\documentstyle [12pt,psfig]{report}
\newcommand{\av}[1]{\langle #1\rangle}

\newcommand{\ket}[1]{|#1\rangle}
\newcommand{\1}{\hspace{1pt}}
\newcommand{\Tr}{{\rm Tr}\,}
\newcommand{\Det}{{\rm Det}\,}

\newcommand{\tto}{{${2\choose1}$}\,}
\newcommand{\uu}{{\sf u}}
\newcommand{\vv}{{\sf v}}
\newcommand{\ww}{{\sf w}}
\newcommand{\one}{\mbox{{\sf 1}\hspace{-0.20em}{\rm l}}}
\newcommand{\beq}{\begin{equation}}
\newcommand{\eeq}{\end{equation}}
\newcommand{\bara}{\begin{eqnarray}}
\newcommand{\ear}{\end{eqnarray}}
\newcommand{\Hi}{\mbox{${\cal H}$}}
\newcommand{\cH}{{\cal H}_}
\newcommand{\Aa}{{\sf A}}
\newcommand{\BB}{{\sf B}}

\title{Quantum Memory in Quantum Cryptography}
\author{TAL MOR \\ D.Sc. work}
\date{16/4/97}

\begin{document}
\maketitle

\setcounter{page}{1}
\renewcommand{\thepage}{\roman{page}}
\tableofcontents
\newpage
\listoffigures
\addtolength{\baselineskip}{2.3mm}
\chapter*{Abstract}
\addcontentsline{toc}{chapter}{Abstract}
\renewcommand{\thepage}{\arabic{page}}
\setcounter{page}{1}

Quantum information theory investigates the information content 
of quantum systems, hence generalizes classical information theory.
The classical bits are generalized into quantum bits (two-level systems).
Unlike classical bits, 
each quantum bit can be in a superposition of its two basis states,
and several quantum bits can be entangled to each other.
Unlike classical bits, quantum bits cannot be duplicated, and furthermore,
the state of a quantum bit cannot be identified, unless more (classical)
information is available. This property of quantum information was used to 
suggest new type of cryptography, {\em quantum cryptography}, which
uses quantum bits to encode classical bits, before providing them
to possible adversaries.

Quantum cryptography suggests various possibilities 
which are beyond the abilities of classical cryptography.
In particular, it suggests schemes for an information-theoretic
secure key distribution,
and an alternative to the existing ``public/secret key distribution'' 
which cannot be information secure, and are 
not proven to be computationally secure.
The main advantage of quantum schemes 
over any classical scheme is the fact that quantum 
information cannot be passively intercepted, hence eavesdropping attempts
induce noise.
Unfortunately, in the presence of noisy channels and devices,
none of the existing schemes for quantum key distribution
is proven secure: the eavesdropper can obtain some small amount of 
information while inducing an acceptable noise;
furthermore, the legitimate users must correct their data using some error
correction code, and the eavesdropper can obtain this information as well.
Classical privacy amplification techniques were suggested to yield a 
secure final key, but there is no complete proof of security yet. 
As is common in cryptography, a promising suggestion might be found
worthless later on;
Future technology, such as
the ability to maintain quantum states for a long time,
could be destructive (or even fatal) to all such schemes.

This thesis investigates the importance of quantum memory
in quantum cryptography,
concentrating on quantum key distribution schemes.

In the hands of an eavesdropper -- a quantum memory is a powerful
tool, putting in question the security of quantum cryptography;
Classical privacy amplification techniques, 
used to prove security
against less powerful eavesdroppers, might not be effective when the
eavesdropper can keep quantum states for a long time.
In this work we 
suggest a possible direction for approaching this problem.
We define strong attacks of this type, and show security against them,
suggesting that quantum cryptography is secure.
We start with a complete analysis regarding the information about a 
parity bit (since parity bits are used for privacy amplification).
We use the results regarding the information on parity bits to prove 
security against very strong eavesdropping attacks, which uses quantum 
memories and all classical data (including error correction codes) to 
attack the final key directly. 

In the hands of the legitimate users, a quantum memory is also a useful
tool. We suggest a new type of quantum key distribution scheme 
where quantum memories are used instead of quantum channels.
This scheme is especially adequate for networks of many users.
The use of quantum memory also allows reducing the error rate to improve 
large scale quantum cryptography, and to enable the legitimate users to
work with reasonable error rate.

\addtolength{\baselineskip}{-2.3mm}
\chapter*{List of Symbols}
\addcontentsline{toc}{chapter}{List of Symbols}

\begin{tabular}{p{2cm}l}\\
BC     & Bit Commitment                                 \\
BCH    & Generalizations of Hamming codes~\cite{MS}     \\
BEC    & Binary Erasure Channel                         \\
BM     & Biham and Mor~\cite{BM1}                       \\
BSC    & Binary Symmetric Channel                       \\
EHPP   & Ekert, Huttner, Palma and Peres~\cite{EHPP}    \\
EP     & Entanglement Purification                      \\
POVM   & Positive Operator Value Measure                \\
OECC   & Obey Error Corretion Code                      \\
OT     & Oblivious Transfer                             \\
QEC    & Quantum Error Correction                       \\
QPA    & Quantum Privacy Amplification                  \\
RDM    & Reduced Density Matrix                         \\
RSA    & Rivest, Shamir and Adleman~\cite{RSA}          \\
qubit  & Quantum bit                                    \\
XOR    & Exclusive OR                                   \\
$i \ldots n$ &  Integers                                \\
$L$, $M$, $N$ & Integers                                \\
$u$, $v$, $x$, $y$ & Binary strings                     \\
$d(u,v)$ & Distortion measure                           \\
$\{v\}$ & A set of binary strings                      \\
$H$    & Entropy                                        \\
$I$    & Information, Mutual information                \\
$p$, $q$, $P$, $Q$  & Error-rates                       \\
$p(x)$  & Parity of a string $x$                        \\
$\hat{n}(x)$  & Number of $1$'s in a string $x$         \\
$U$    & Unitary transformation                         \\
%\end{tabular}
%\newpage
%\begin{tabular}{p{2cm}l}\\
$\otimes$  & A tensor product                           \\
$\odot$  & Bitwise ``AND'' of two strings               \\
$\oplus$ & Bitwise ``XOR'' of two strings               \\
${\cal{H}}_n$, ${\cal{K}}_n$  & Hilbert spaces of $n$ dimensions  \\
\end{tabular}

\begin{tabular}{p{2cm}l}\\
$| 0 \rangle = {1 \choose 0} $ &                        \\
$| 1 \rangle = {0 \choose 1} $ &      
                           Basis vectors in ${\cal{H}}_2$ \\
$|\Phi\rangle$; $|\uu\rangle$  & Quantum pure states    \\
$|\uu'\rangle$ & A state orthogonal to $|\uu\rangle$    \\
$\rho $ & A quantum density matrix                      \\
${\alpha \choose \beta}$; 
${\cos\theta \choose \sin\theta}$& Pure states in ${\cal{H}}_2$ \\
$c_\theta$, etc. & $\cos\theta$, etc.                   \\
$|\uparrow \rangle$, etc. & A spin-up state in ${\cal{H}}_2$, etc. \\
\end{tabular}

\chapter{Introduction to Quantum Cryptography
and Quantum Memory}\label{int}

Classical information theory, originated by Shannon \cite{Shannon1},
quantifies the intuitive concept of information as it appears in a statement
like `this page contains a lot of information'.
It serves as an invaluable tool in communication theory, data processing 
(e.g., compression), computation theory, cryptography and many other areas 
in electrical engineering and computer science.

Information is deeply related to 
physical objects; As such, classical information theory can deal only 
with cases where the carriers of information are classical object 
(even if underneath there is a deeper quantum reality).
In the quantum world it is natural to extend this theory 
to the more general ``quantum information''
theory~\cite{von-N,Kholevo1,Kholevo2,Davies,Levitin,Helstrom,Per93,ben}, 
which considers the information content of a quantum system.
We are mainly interested in cases where
information is encoded into quantum states, and later on decoded
back to classical information. 
The quantum information 
is transformed to classical 
information when a test 
occurs in a laboratory, 
due to the projection postulate of conventional quantum mechanics.
A quantum state (in general, a density matrix) $\rho$, has a practical 
meaning: 
it allows us to predict probabilities regarding various
physical tests. Its expectation value regarding a specific state $\phi$, 
$\langle \phi | \rho | \phi \rangle$, can be measured in the laboratory
(at least, in principle).
In this sense, it is a subjective property of a system:
non-pure states are defined as statistical mixtures of pure states\footnote{
Most books use this definition. Unlike most books Landau and Lifshitz~\cite{LL}
use a different definition which assigns objective meaning to the 
density matrix. 
In this work we 
consider only the subjective 
meaning of the density matrix (namely, 
prediction abilities which can be verified
in the laboratory).
},
hence one considers a state to be non pure due to having partial
knowledge on the preparation procedure. 
If a system is entangled with other systems which are not available
to us, then we can only consider measurements on our system,
and our prediction ability is more limited;
we have to ``trace out'' the degrees of freedom of the other systems,
in order to be able to predict the probabilities of
outcomes of various tests performed in our laboratory.
The state of our system 
depends on classical information which we might obtain
from people who hold the other systems, since such information could
influence our possible predictions.

There are various aspects of quantum information, such as 
entropy of quantum systems~\cite{von-N,Kholevo1}, the coding theorem for 
quantum information~\cite{ben,JS}, optimal identification 
of quantum states~\cite{Davies,Levitin,PW,BrCa,Fuchs}, 
quantum noise in transmission channels~\cite{CD94},
communication using quantum states and many others.
This work deals with quantum communication and cryptography.

In communication theory, 
information is transmitted from one person to another through a channel.
Information theory 
tells us how well a state (say, a string of bits) can be reproduced
after it passed through a noisy channel.
Thus, it quantifies the {\em reliability} of the received state.
In classical communication, the information is carried by distinguishable 
signals, each of which represents a distinct letter of some alphabet  
(usually binary).
Each signal can be duplicated and can also be identified 
unambiguously unless some noise is added. 
When a small amount of noise exists, 
the state could still be identified with excellent 
probability of success if enough redundancy is added to the transmitted data.
In quantum communication, the 
information is carried by signals which are not necessarily 
distinguishable~\cite{Davies,Levitin,BrCa}.
In general, we can 
omit the specification of the carriers and consider only their quantum states
(unless we deal with implementations).
For our purposes, it usually suffices to choose an abstract
two-level system
(with a basis $|0\rangle$ and $|1\rangle$) 
to replace the classical bit. A general state of a 
two-level system $\alpha |0\rangle + \beta |1 \rangle$ 
with $|\alpha|^2 + |\beta|^2 = 1$,
is called a ``qubit''
(quantum bit)~\cite{ben};
any larger system consists of many qubits.
If one uses non-orthogonal states to represent classical signals `0' and `1',
such that the overlap of the two states is not zero,
the encoded data cannot be duplicated nor 
identified unambiguously. 

In cryptography, information must always remain secret from some 
people. In this scientific area, human intentions play an important role,
and we must always assume that people wish to learn this secret data.
For instance, a sender, Alice, wishes to send some data to a receiver, Bob,
in a way that an eavesdropper, Eve, cannot learn that data.
In this case, information theory may tell us 
the maximal amount of information available to Eve, depending on
her abilities.
Thus, it quantifies the {\em security} of the secret data.
An eavesdropper listening to classical communication can (in principle)
remain unexposed, since the data can always be duplicated. 
Quantum communication allows for more.
Using non-orthogonal quantum states to transmit classical bits, 
the signals cannot be duplicated~\cite{WZ}, and any gain of information
about them necessarily induces noise.
Thus, (unlike classical channels) quantum channels 
cannot be passively intercepted, when 
non-orthogonal quantum states are transmitted! 
This fact was used to suggest a new
type of cryptography ---  
quantum cryptography --- where the use of non-orthogonal states 
might permit performing new tasks, which are beyond the ability of 
classical cryptography. 

One of the main goals of cryptography, the secure transmission of messages,
can be achieved using a secret key known only to the sender, Alice,
and the receiver, Bob. 
The problem is how to distribute the key.
Achieving a secure key distribution is among the most important
goals of cryptography, and the current ways this is done are either
very expensive or not proven secure.
The main achievement of quantum cryptography is the design of 
schemes for secure key distribution.
They are using non-orthogonal quantum states,
hence, any eavesdropping attempt induces noise and can be noticed
by the legitimate users\footnote{
For a more accurate analysis of this delicate point see~\cite{GV,Peres-c,GV-r}.}.
The goal of any {\em quantum key distribution} protocol 
is to generate a common key while not leaking any significant information 
to an unrestricted adversary;
If the amount of information leaked to the eavesdropper is much smaller than 
a single bit, the key is said to be 
{\em information-theoretic secure}. 
The classical ``public key distribution'' and many other
protocols in classical cryptography are not designed to withstand adversaries
who have unlimited computation abilities; i.e., they can be at most
{\em computationally secure}. 
This advantage of quantum cryptography would 
not be crucial if indeed classical cryptography was computationally secure.
However, the existing schemes for classical key distribution rely upon 
unproven  
computational assumptions (such as the difficulty of factoring
large numbers).

In addition to key distribution, it is important to
mention a very similar task -- namely ``key expansion''.
In a key expansion protocol some short amount of information 
(say, $m$ bits) 
is shared in advance between Alice and Bob, 
and it is used to derive long common keys
(say, of total length $n$ much larger than $m$).
If such an expansion leaks only an exponentially small amount of information
to an eavesdropper (say, $2^{-m}$ bit) 
the final key is still information-theoretic 
secure.
Quantum key distribution demands, in addition to the quantum channel,
also an ``unjammable'' classical channel in which the transmitted data
can possibly be monitored passively, but not modified.
This requirement~\cite{BBBSS} can be derived in principle using schemes for
broadcasting the classical messages. Alternatively,
it can be fulfilled if the legitimate users can identify each other
through the classical channel. In cryptographic terms, it means that they share
some information in advance, thus the so called ``quantum key distribution''
should be better called ``quantum key expansion''.
Having clarified this point, 
we shall keep using the ``wrong'' term as is common.

Quantum cryptography was invented in 1969 by Wiesner~\cite{Wiesner}
in his paper on ``conjugate coding'' (unpublished by that time).
Ten years later, a collaboration between Bennett and Brassard (based
on Wiesner's idea) was   
the actual starting point of quantum cryptography~\cite{BBBW},
and lead to ``quantum key distribution'' in 1984~\cite{BB84}.
Their basic idea is to use non-orthogonal quantum states for key
distribution since such states cannot be cloned by a potential eavesdropper.
They used four possible states, therefore, we refer to their scheme as the
{\em four-state scheme}.
In 1991 Ekert~\cite{Eke91} proposed (following a suggestion of Deutsch)
to use quantum nonlocality
for cryptographic purposes,
and the connection of his scheme to the original quantum key distribution 
scheme was demonstrated by Bennett, Brassard and Mermin~\cite{BBM}.
We refer to their version of Ekert scheme as the {\em EPR scheme}.
Another important scheme was later proposed by Bennett~\cite{Ben92}, showing
that quantum key distribution can be performed even if one uses only
two non-orthogonal states. 
We refer to this scheme as the {\em two-state scheme}.
Any of these schemes can be used to design a quantum 
key distribution protocol, where a quantum channel and 
an unjammable classical channels are used to transmit a secret key.
If Alice and Bob identify errors (by the discussion over the classical
channel) they quit the protocol. Thus, Eve can 
force them to quit the protocol, but she cannot cheat them into thinking 
she is not listening, and still get information about their key.

The first prototype for quantum key distribution
was built in 1989~\cite{BBBSS}. It provided a 
proof of feasibility by 
performing a distribution over a distance of 30 centimeters.
Since then, much more practical prototypes were built
for a distance of 30 kilometers~\cite{MaTo} in a coil,
a distance of 24 kilometers in underground optical fiber~\cite{LosAl}, 
and 23 kilometers in underwater optical fiber~\cite{muller}.
The implementations are using either    
photon polarization (as originally proposed
in~\cite{BB84}) or phase and interferometry (as~in~\cite{Ben92}).
Unfortunately, they are mainly using weak pulses of coherent light
instead of Fock states of single photons, and this creates
various problems of security, due to the use of a Hilbert 
space which is different 
from the Hilbert space used in the theoretical schemes.
This problem is considered in~\cite{HIGM,Yuen}.
It is yet unclear whether future practical quantum key distribution
will be based upon weak coherent pulses or upon single photons.
The basic theoretical schemes already raise the main fundamental questions
regarding quantum key distribution, thus we shall not consider
here the more complicated practical schemes which use weak coherent pulses.

Most of the fundamental questions in quantum key distribution (and more 
generally, in quantum cryptography) are due to activities which are permitted
by the rules of quantum mechanics, but are not technologically feasible.
The ability to maintain quantum states unchanged for long times was
ignored till recently, but today it strongly influences any suggestion
(or analysis) 
in quantum cryptography.
In this work we investigate the role of quantum memory in quantum cryptography,
and mainly in quantum key distribution. 

The most important question in quantum cryptography is 
to determine how secure
it really~is.  Any practical implementation must take into
account that some errors are inevitable due to errors in creation,
transmission and detection of quantum states.
The legitimate users must be able to withstand some errors,
and furthermore,
must correct them to yield a final (shorter) common key.
Thus, Eve can gain some information from both the quantum states
(if she induce less errors than permitted)
and the error-correction data (transmitted through a classical channel).
Eve's total information depends on the permitted error rate, on her attack
and on the error-correction code used by Alice and Bob.
Clearly, it is not negligible.
To overcome this problem, and enable the derivation of a final key
which is both reliable,
and secure, privacy amplification schemes were 
designed~\cite{BBBSS,BBR,BBCM}.
Their goal is to reduce Eve's
information to be as small as desired, for the price of shortening
the final key even further.
To do so, collective properties of substrings of bits (e.g., the parity
of a substring) are used as the final key.

In~early papers on quantum cryptography,
the security of quantum key distribution was studied under the
assumption that the eavesdropper is performing standard 
(von Neumann) measurements on the photons on their way from 
Alice to Bob.
Such analysis can be found, for instance, in~\cite{HE,BBBSS},
where in~\cite{HE} the issue of Eve's information gain versus the
disturbance to Bob's state was discussed, and in~\cite{BBBSS} Eve's
information on the final string was also considered.
It is rather clear~\cite{BBBSS,BBR} (although, not explicitly proven)
that privacy amplification is effective 
against any attack of that type, implying
the security
of quantum key distribution against such attacks.

Unfortunately, quantum mechanics permits much 
more sophisticated eavesdropping strategies, against which
there are no security
proofs\footnote{
But see Section~\ref{securC}.}.

Quantum gates are currently under 
investigation by both theorists~\cite{Deutsch89,DiV,BDEJ,CZ}
and experimentalists~\cite{Haroche,Kimble,Wineland}. 
Eve can use quantum gates 
to perform {\em translucent attacks}~\cite{EHPP}, which are not
``opaque'' (intercept/resent)
as the standard measurements are.
In translucent attacks Eve interacts with Alice's 
particle as follows:
she attaches an auxiliary particle (a probe)
to each transmitted particle by performing some unitary transformation
on them {\em together}.
After the interaction, Alice's particle 
is sent from Eve to Bob, in a way that Eve does not know its exact state.
If the interaction is weak, 
it results in a very small 
disturbance to the state of the transmitted particle.
Information gain versus disturbance in such attacks is improved compared 
to ``opaque'' attacks~\cite{EHPP} and was 
investigated further~\cite{FP,Chris}. 
When attacking the four-state scheme of~\cite{BB84}, the measurement on the 
probes can be delayed till receiving the data regarding the basis used by Alice.
The main advantage of translucent attack over opaque attack is that
it can be performed on all the transmitted particles. In this case it 
provides very little information on each 
particle, but it might provide much more information  
on a collective property
used as the final key!
Fortunately, 
it is clear~\cite{BBCM} that privacy amplification is still effective 
against such a {\em single particle attack},
and a partial proof was also provided~\cite{DMLS}.
A demonstration of the effectiveness of privacy amplification 
against such attacks is given in Chapter~\ref{secur};
this is done for an explicit but typical example.

Unfortunately, the effectiveness of privacy amplification against 
single-particle attacks is not sufficient.
Eve can do much more than such an attack if she is capable of keeping
the particles in quantum memories.
In this case she can perform {\em joint attacks} which are directed towards
the final key, and the effectiveness of privacy amplification against
them is in question.
Such attacks were considered in some papers~\cite{BBBSS,BBCS},
and were analyzed only in case of error-free channels~\cite{Yao}
(the analysis was done for quantum oblivious transfer, but the proof 
applies to quantum key distribution~\cite{DM3} as well).
In Chapter~\ref{secur} we explain the security problem in details,
and we present 
a restricted class of joint attacks which we call
{\em collective attacks}, in which each transmitted particle is 
attached to a separate probe. 
All probes are measured together {\em after} Eve gets all classical data,
and the attacks are directed against the final key.
Such joint attacks are much simpler than the most general joint attacks,
however, they are still general enough to present the power of joint 
attacks.
Furthermore, we could not find any joint attack which is stronger
than the collective attacks. Thus, we 
suggest a new approach to the security problem: 
\begin{itemize}
\item to prove 
the security against collective attacks.
\item to prove that
collective attacks are the strongest joint attacks. 
\end{itemize}
While we have done a lot of progress regarding the first issue,
we didn't do much progress regarding the second.
We only explain the 
{\em randomization argument} which provides the intuition 
that collective attacks are the strongest joint attacks.

The main achievement of this work is to  
provide various proofs of the security of the final key (the parity bit) 
against a very powerful eavesdropper who has both quantum gates
and quantum memories.
In Chapter~\ref{par} we find the optimal information regarding 
the parity bit, obtained by any type of measurement, and we
show that the best measurement is indeed a joint measurement
in an entangled basis~\cite{BMS}.
In Chapter~\ref{col} we consider various realistic attacks on the parity
bit which use both quantum and classical (error correction) information
and we prove security against them.
Our results provide strong evidence that quantum key distribution
is secure.
We consider various collective attacks which leave the eavesdropper with 
probes in pure states~\cite{BM1} and mixed states~\cite{BM2}, and we
prove security against them.

In parallel to our work, Mayers~\cite{DM2} and a group in England~\cite{oxford}
came up with other approaches to the joint attack, and several works studied 
the single-particle attacks~\cite{LC,FP,lutken}.
The techniques presented
in these papers may well prove useful in advancing the issue of the 
security of quantum key distribution.

In addition to studying the use of a quantum memory  
to the benefit of the eavesdropper in quantum cryptography,
we also study other
applications of quantum memory, to the benefit of the legitimate users.
In Chapter~\ref{net}
we introduce a new scheme for quantum key distribution which is 
implemented with quantum memory and does not require quantum channels!
It is a time-reversed EPR scheme and correlations between particles
are introduced by projecting their states onto entangled states.
Our scheme is especially suited for building a quantum cryptographic network
of many users.
A quantum memory plays an important role in the security problem 
also when the legitimate users are permitted to use it (ignoring the
practicability of the scheme).
In Chapter~\ref{QECQPA} 
we discuss new possibilities related to quantum 
memories: first, they become more available due to suggestions for
quantum error correction~\cite{Shor1}, and fault-tolerant error 
correction~\cite{Shor2}. Second, new applications 
of quantum memories, namely the possibility 
to purify singlets~\cite{purification},
can be used to design quantum privacy amplification~\cite{oxford}.
We connect the two issues together: we suggest 
quantum privacy amplification based upon quantum error correction,
and discuss their implications for
quantum cryptography. This discussion corrects wrong impressions 
that the security problem has already been solved by quantum privacy
amplification~\cite{oxford}. 
Our new quantum privacy amplification suggests the surprising possibility
of using ``quantum repeaters'' for improving secure transmission of 
non-orthogonal quantum states over 
a long distance.

The practical interest in quantum cryptography is clear.
Furthermore, quantum cryptography raises very interesting 
physical questions: 
\begin{enumerate}
\item How is information split between two 
observers (Bob and Eve)? 
\item Is secure transmission of data possible under the laws of Physics?
\item How do (reduced) density matrices (thus, predictions regarding the 
probabilities of various outcomes in any possible
test) change due to additional classical information?
\end{enumerate}
The intensive research on quantum information, quantum cryptography
and quantum computing already yielded (or helped improve)
various surprising advances in
quantum mechanics 
such as teleportation, generalized measurements, fast quantum computation,
purification 
of singlets, and preservation of qubits. 

The protocols considered here
and the attacks 
against them are discussed in mathematical terms, and are required 
to be consistent only with the mathematical foundations of quantum mechanics.
We prefer to simplify the discussion as much as possible,
thus we present simplified models (such as two-level systems, and pairs of
two level systems), and not the more realistic models which are 
implemented in laboratories (such as weak coherent states).
The only exception is the case of quantum memory. Due to its special
role in this work, we give a brief overview
of possible implementations. 
We shall frequently use the terms {\em existing} vs.\ {\em future}
technologies, where we actually mean {\em feasible in the near future}
vs.\ {\em infeasible in the near (foreseeable) future}.
We follow the common literature of both theorists and experimentalists
in order to decide how to classify each operation,
without investigating further into the feasibility question. 
For instance, creating or measuring a singlet state 
(or any state of two two-level
systems), and applying any 
single unitary transformations
onto two two-level systems,
are considered as {\em existing} although performing them in practice
is still a problem.  
On the other hand, combining several such transformations on more
than two two-level systems, and keeping quantum states for long times
are considered as future technology.

The distinction between existing and future technologies has a special
role in quantum cryptography: the legitimate users are normally permitted to
use existing technology only, while the opponents are permitted to use
any operation consistent with the rules of quantum mechanics.
Thus, a proof of security must hold against a ``quantum'' opponent
and a practical protocol is performed by existing technology.
In cases we permit the legitimate users to use future technology
we explicitly say it, and we explain the motivation for it.

In our work, we shall use alternately 
Dirac's bracket notation, 
vector notation and spin notation  
$|0\rangle = {1 \choose 0} = | \uparrow \rangle $, 
$|1\rangle = {0 \choose 1} = | \downarrow \rangle $, 
$ \cos \theta |0\rangle + \sin \theta |1\rangle 
= {\cos \theta  \choose  \sin \theta} = \cos \theta |\uparrow \rangle
+ \sin \theta | \downarrow \rangle$, etc.
In case of spin in $x$ direction we also denote 
$| \rightarrow \rangle = (1/\sqrt{2}) (| \uparrow \rangle 
+ | \downarrow \rangle) = \frac{1}{\sqrt 2} {1 \choose 1}$, 
etc.

The rest of this introduction is organized as follows:
The basics of classical information are described in Section~\ref{int.cla-inf}.
Quantum information theory, and mainly the use of quantum states
for transmitting classical bits, are described in Section~\ref{int.qua-inf}.
The principles of classical cryptography
are described in Section~\ref{int.cla-cry}.
We review the basic quantum key distribution schemes in
Section~\ref{int.qua-cry}, where 
we also explicitly present a key distribution protocol with all its details,
and a detailed explanation of basic
eavesdropping attacks and the security against them.
We introduce quantum memory in Section~\ref{int.mem}, in which we 
also discuss basic uses of quantum memories
in quantum cryptography, and 
suggest implementations for a quantum memory.
Recent progress in quantum computing~\cite{Deutsch85,Shor} are deeply
related to this work, although indirectly.
A brief introduction to the exciting new
developments in quantum computing is given in
Section~\ref{int.qua-com}.

\section{Classical Information}\label{int.cla-inf}

Classical information theory~\cite{Shannon1,Ash} 
quantifies the idea of information.
If one tells us that the temperature on June 25 went up to about 30 degrees 
(Celsius) in Haifa, we would not gain much new information due to our 
a priori knowledge about the weather in that season. 
But if he tells us that it was snowing that day he surely provides 
us a lot of new information.
To quantify this, some definitions are required.
The self information of an event $u$ is 
$ I(u) = \log (1/P(u)) $,
where $P(u)$ is the probability of the event, and 
where we always use $\log_2$ so the basic unit is a bit.
A rare event (e.g., snowing in June) has a large amount of self information. 
A multiplication of probabilities 
yields summation of information.
The entropy of an ensemble $A$ of events (letters)  
$a_1,\ldots , a_n$ where each letter 
$a_i$ appears with probability $P(a_i)$ is  the expected value of the
self information
\beq  H(A) = E\big[\log_2 (1/P(A))\big] = - \sum_{i=1}^n P(a_i) \log (P(a_i)) \eeq 
and can be thought of as a measure of our 
lack of information on the unknown event in $A$. 
For instance, for a binary alphabet,  
$ H = - \sum_{i=1}^{2} \frac{1}{2} \log \frac{1}{2} = 1 $
if 
the letters appear with equal probability,
and $H = 0$ if one letter appears
with certainty.
The entropy of a binary alphabet with probabilities
$p$ and $(1-p)$ is denoted by $H(p)$ 
\beq H(p) =  -p\log p - (1-p)\log(1-p) \ .\eeq 

Two ensembles of letters $A$ and $B$ might have dependent probability 
distributions.
The trivial example is when A is the input alphabet for some information 
channel and B is the output.
The source is assumed to be memoryless and stationary.
An average distortion between input vector and output vector is a measure 
for the channel quality.
Usually an additive distortion measure is used 
$d(\uu,\vv)=  \sum_{i=1}^m d(u_i,v_i)$ 
where $d(u_i,v_i)$ is a one letter distance.
It is called the Hamming distance if 
$d(u_i,v_i) = 0$ for $u_i=v_i$ and $d(u_i,v_i)=1$ for $u_i \ne v_i$.
For instance, the Hamming distance between the strings
$000$ and $111$ is three.
This measure can also be used as a measure of 
distinguishability of two input vectors
(code words).
In channel coding we increase the number of bits which encode each code word
so that the additive Hamming distance between any two legitimate code 
words is increased.
Hence, they can be distinguished even when the channel is noisy.
Such techniques (known as error-correction techniques) enable
receiving a reliable output when the channel is noisy.

The relation between input and output of a transmission 
channel is described in terms of information
\cite{Shannon1,Ash}, and the Hamming distance can be a measure for
the distortion of a string transmitted through the channel.
The conditional probability of $B=b_i$ 
given that $A=a_j$
is $P(B=b_i|A=a_j)$ 
[we write $P(b_i|a_j)$ for convenience].
The joint probability for both $a_j$ and $b_i$
is $P(b_i,a_j)$.
In case of independent probability distributions it is equal to $P(b_i) P(a_j)$,
and in general it is given by the Bayes probability law
\beq P(b_i|a_j) P(a_j) = P(b_i,a_j) = P(a_j|b_i) P(b_i) \label{Bayes} \ .\eeq  
The Bayes probability law  
presents the symmetry of input and output.
The conditional entropy is defined by
\beq  H(B|A) = - \sum_j \sum_i P(a_j,b_i) \log P(b_i|a_j) \label{conditH} \eeq 
where the sum is over the size of the input and the output alphabets.
When $A$ and $B$ are independent, $H(B|A) = H(B)$, 
meaning that we gain no information on $B$ by being told what $A$ is.
Otherwise $H(B|A) < H(B)$ 
meaning that our lack of knowledge about $B$ is decreased when $A$ is given.
The gain in information when $A$ is given  is called the 
mutual information and is among the most important notions in  
information theory.
The mutual information  $I(A;B)$ is symmetric in $A$ and $B$.
\bara I(A;B) = 
H(B) - H(B|A) = 
H(A) - H(A|B) \nonumber \\ = 
\sum_i \sum_j P(a_j,b_i) \log \frac{P(a_j,b_i)}{P(a_j)P(b_i)} \label{mutI}\ear 

In a perfect channel the output signals are completely determined by the input 
signals. 
The channel introduces no errors.
A Binary Symmetric Channel (BSC) with input $A$ and output $B$ is a channel 
with binary alphabet where the signals are transported with 
symmetric distortion (i.e., equal error probabilities $P_e$).
The optimal mutual information in this case is achieved when the input 
probabilities are also equal (hence also the output probabilities).
It is 
\bara I(A;B) &=& H(B) - H(B|A) = 1 - H(P_e) \nonumber \\ 
&=&  1 + P_e \log P_e + (1-P_e) \log (1-P_e)\nonumber  \\ 
&=& \frac{1}{2} \Big[ 2  P_e \log (2 P_e) + 2 (1-P_e) \log (2(1 - P_e))\Big]\ ,
\label{MI}\ear 
sometimes denoted as $I_2(P_e)$.

For a given channel, the optimal mutual information 
(over all possible input probabilities) is 
called the channel capacity.
The channel capacity ${\cal{C}}$ is a measure of the minimal distortion 
caused by the channel. 
For a binary channel ${\cal{C}} \le 1$.
The `Channel Coding Theorem' states that it is possible 
to derive an error-free code using a non-perfect channel \cite{Shannon1,Ash}:
for any positive $\epsilon$, we can choose an integer $m$ 
and $N$ vectors, 
$N < 2^{m{\cal{C}}}$,
of length $m$ (to be used as code words), such that all 
code words can be 
distinguished with error probability smaller than $\epsilon$.

Another important channel is the Binary Erasure Channel (BEC) which is actually
not a binary channel since its output contains, in addition to `0' and `1' also
the letter `?' which stands for an inconclusive result.
This channel introduces no error but the output of either inputs is 
inconclusive with probability $P_?\ $.
Channel capacity is derived with equal input probabilities and is calculated 
using eq.(\ref{Bayes}), (\ref{conditH}) and (\ref{mutI}) to be 
\bara 
I &=& H(A) - H(A|B) = 1 + \sum_j P(b_j) \sum_i P(a_i|b_j) \log P(a_i|b_j) 
\nonumber \\ &=& 1 - P_? \ \ .\ear 
Note that the error probability $P_e$ and the probability $P_?$ (of an 
inconclusive result) have a different weight in their contribution  
to the mutual information.

\section{Quantum Information}\label{int.qua-inf}

Contrary to a classical state, a quantum  
state cannot be identified unless additional information is provided.
Suppose that a spin measuring device 
is aligned in the $z$ direction,  
so that it can identify the spin of an electron unambiguously 
if it is prepared along the $z$ direction.
An electron with some arbitrary (pure) spin state, say  
\beq \ket{\psi} = \alpha \ket{0} + \beta \ket{1} \eeq
(with $|\alpha|^2 + |\beta|^2 = 1$),
will yield
a result `0' (electron went up in a measuring device) or `1' (electron went
down in that measuring device), unpredictably.
The only clue that the state was as stated is that
the probability for each result can be calculated.
These probabilities are
$|\alpha|^2$ for the result `0' and $|\beta|^2$ for the result `1'.
Only if the physical state is taken from a set of 
orthogonal states in a known 
basis can it be identified in one measurement as in the classical case.

The $n$ letters of a quantum communication channel can be any different
(not necessarily orthogonal nor pure) quantum states 
$\rho_1 \ldots \rho_n$
on an $N$-dimensional Hilbert space $\cH{N}$.
The mutual information of such a channel is a function not only of 
the input probabilities, but
also of the choice of measurement performed by the receiver.
Even in case of two orthogonal states, 
the receiver may select a bad choice of
measurement direction which will tell him nothing.
Therefore, one of the main 
questions is  how to maximize the mutual information.
For given signals and given input probabilities of the source, that is, 
for a given density matrix 
\beq \rho = \sum_{i=1}^n p_i \rho_i \ , \eeq 
what is the optimal measurement,
and what is the mutual information in this case
(called accessible mutual information)? 
In classical information theory, the only parameter controlled by
the legitimate users is the input probabilities:
optimizing the mutual information over all possible input probabilities
yield the ``channel capacity''.
In quantum information theory, the users control both the input probabilities
and the measurement, and optimizing both to maximize the mutual information
yields the channel capacity.
We usually deal with given (and equal) 
input probabilities, 
and care only about the accessible
mutual information and not about the channel capacity.

The issue of finding the accessible information
is frequently considered in the literature.
One way to attack this problem is to find upper \cite{Kholevo1}  
and lower \cite{JRW}  bounds on this information where the entropy 
of a quantum system, 
\beq S = \Tr (\rho \log \rho ) - \sum_i \Tr (\rho_i \log \rho_i ) \label{entropy} \eeq 
serves as the upper bound.
Another way is to investigate and solve 
some special cases (subclasses)~\cite{Levitin,Davies,BrCa} or concrete 
examples~\cite{Davies,Kholevo2}.
Levitin~\cite{Levitin} discussed two density matrices 
with equal determinants 
in 2-dimensional
Hilbert space, and Braunstein and Caves~\cite{BrCa} considered density matrices 
which are close to each other.
Various cases of pure states are also solved;
Kholevo~\cite{Kholevo2} and Davies~\cite{Davies} presented examples 
(with $n>N$) where the optimal mutual information is derived with 
a non-standard measurement (i.e.\ non von Neumann measurements) 
since non-standard measurements~\cite{JP-DL} can result in 
more than $N$ results. 
On the other hand, it seems~\cite{Levitin} (but has not been proven) 
that if the Hilbert space spanned by $n$ pure states
is $n$-dimensional, the measurement which optimizes mutual information
is a standard 
(projection) measurement.
With some additional conditions, it probably holds for 
density matrices as well.

Let us consider the simple case of two non-orthogonal pure states
in $\cH2$.
A generalization of this to $\cH{N}$ is trivial since
two pure states always span some 2-dimensional subspace of $\cH{N}$. 
Suppose that the sender of the information, 
Alice, emits particles in one of the two
non-orthogonal states $|\uu\rangle$ and $|\vv\rangle$, which represent bits 0
and 1, respectively (the state of each particle is known to Alice).
The states are sent with equal probabilities.
Since these two states are not orthogonal, the receiver, Bob,
cannot identify with certainty the state of a given photon.
By a suitable choice of basis, the two non-orthogonal states $|\uu\rangle$
and $|\vv\rangle$ can be written as
\beq |\uu\rangle={\cos\alpha\choose\sin\alpha}\qquad\qquad
 {\rm and}\qquad\qquad
 |\vv\rangle={\cos\alpha\choose -\sin\alpha},\vspace{2mm} \label{two-state-input} \eeq
where \,$0<\alpha<45^{\circ}$. 
The angle between the two states 
is $2\alpha$.
Note that the overlap between the two states is 
\beq C=\langle\uu|\vv\rangle = \cos^2\alpha - \sin^2\alpha = \cos2\alpha\  .\eeq
A standard measurement in an orthogonal basis 
symmetric to the two states optimizes~\cite{Levitin} the mutual information;
it is analogous to a BSC.
This measurement introduces an error\vspace{-2mm}
\beq 
\label{BSC-err}
P_e = \sin^2(\frac{\pi}{4}-\alpha)= \frac{1 - \sin 2 \alpha}{2}\ , \eeq
hence,\vspace{-2mm}
\beq I_{{}_{BSC}} = 1 - H(P_e) = I_2(P_e)\ . \eeq

Although a standard measurement is optimal in this case  
a non standard measurement~\cite{IvPe} may be better for certain purposes.
The receiver can build a device which gives a definite answer in a
fraction of the cases, and an inconclusive answer otherwise. 
A measurement of this type (with a perfect channel) creates an analogy 
to the BEC.
It lets the receiver 
derive the optimal {\em deterministic } results.
It is useful 
if the receiver is permitted to use a subset of 
selected bits and throw undesired result. 
In such a case, a measurement which optimizes the mutual information
regarding the selected bits 
is preferable to the one which optimizes the average mutual information
regarding all bits.
A simple way to perform a deterministic measurement 
is to perform a standard measurement in the basis of one of the vectors.
We define two vectors,
\beq 
|\uu'\rangle={-\sin \alpha \choose \cos\alpha }
\qquad\qquad{\rm and}\qquad\qquad
|\vv'\rangle={\sin \alpha  \choose \cos \alpha }
\ , \eeq
which are orthogonal to $|\uu\rangle$ and $|\vv\rangle$ respectively. 
Suppose Bob measured in the $|\uu\rangle$ $|\uu'\rangle$ basis:
if the result is $|\uu'\rangle$ which is orthogonal
to $|\uu\rangle$, it must have been the other state $|\vv\rangle$ prior
to the measurement; if the result is $|\uu\rangle$, 
it is inconclusive, since it could be either of the two
possibilities prior to the measurement.

The above can also be described in terms of non standard measurements,
a description which is sometimes more appropriate for such analysis.
The generalization of the standard projection measurement is 
called a {\em positive
operator valued measure\/} (POVM) \cite{JP-DL,Helstrom,Per90,Per93}. 
A POVM is a set of positive operators $A_1 \ldots A_k$ 
which sum up to the unit matrix \one.
Each operator corresponds to a possible outcome of the measurement.
The probability that this generalized quantum measurement yields the
$\mu$'th element of the POVM, if the system was prepared in a pure state
$\psi$, is $\langle \psi|A_\mu|\psi\rangle$. More generally, for a preparation
represented by a density matrix $\rho$, this probability is
\,$\Tr(\rho A_\mu)\1$. 
Contrary to the two positive operators which define a standard 
(projection) measurement,
the operators which build up the POVM do not have to fulfill the condition that 
$     A_i A_j = 0   $  
for $i \ne j$, nor that $A_i A_i = A_i$.

To derive the optimal POVM in our case, we use the vectors 
$|\uu'\rangle$, $|\vv'\rangle$ and a third vector 
\beq |\ww\rangle={1\choose 0}, \eeq equidistant from
$|\uu \rangle$ and $| \vv \rangle$.
It is easy to verify that the three positive operators
\beq A_\vv=|\uu'\rangle\langle \uu'| /(1+C), \qquad
  A_\uu=|\vv'\rangle\langle \vv'| /(1+C), \qquad
  A_\ww=2C\, |\ww\rangle \langle \ww| /(1+C), \label{POVM1} \eeq
(where $C=\cos 2\alpha$) sum up to the unit matrix \one, 
and that $A_\uu$ and $A_\vv$ are maximal (increasing them further will 
violate the conditions for a POVM). 
A quantum test yielding $A_\vv$ rules out the initial state $\uu$ which is 
orthogonal to it.
The same is true for $A_\uu$ while $A_\ww$ yields an inconclusive result.
The probability of an inconclusive result is
\beq \av{\uu|A_\ww|\uu}=\av{\vv|A_\ww|\vv}=C,\eeq
hence, 
\beq
I_{{}_{BEC}} = 1 - C \ .
\eeq

It is always
possible \cite{Per90} to extend the Hilbert space of states, \Hi, in which
this POVM is defined, in such a way that there exists, in the extended
space, ${\cal K}$, a set of {\em orthogonal\/} projection operators,
$P_\mu\1$, summing up to the unit operator, and such that each one of
the $A_\mu\1$ in Eq.~(\theequation) is the projection of one of these
$P_\mu\1$ from ${\cal K}$ to \Hi. This is Neumark's theorem~\cite{Per93}. The
physical interpretation of this theorem is that the extended space
${\cal K}$ corresponds to a combination of the system to be measured
with another system, called {\em ancilla\/}, prepared in a 
known state \cite{IvPe,Per90}.

Another POVM, less efficient but simpler to realize than the above one,
is the one we described earlier (a standard measurement in one of the basis).
This POVM is made of the positive operators:
\beq A_\vv=(|\uu'\rangle \langle \uu'|)/2, 
\qquad A_\uu=(|\vv'\rangle \langle \vv'|)/2,
\qquad A_\ww=(|\uu\rangle \langle \uu| + |\vv \rangle \langle \vv|)/2
\ .
\label{POVM2} \eeq
In this
case, the ancilla merely registers which basis was used.

The optimal mutual information for the two non-orthogonal pure states 
is derived by a standard projection measurement.
The optimal basis is an orthogonal  basis symmetric
to the two 
states\cite{Levitin,Levitin2}. 
An unproven claim of Levitin is that the optimal measurement is a standard measurement also for
the case of two density matrices in $\cH2$.
Levitin finds the accessible information for the simple case of two 
density matrices with
equal determinant and input probabilities, where, again, the optimal basis is 
symmetric. 
Exact solutions for more complicated cases are not known.

\section{Classical Cryptography}\label{int.cla-cry}

The art of cryptography is almost as old as wars are.
First, the transmission of a secret message
was done by trusted messengers.
Then, people started to use substitution codes which `prevented' the enemy
from reading the messages even if he has got them.
In this cryptographic method each letter is replaced by 
another letter using some secret permutation,
whose choice is chosen from all the 
$26!$ possible permutations.
The permutation must be known to the receiver,
thus the sender and the receiver must share some secret
data ({\em a key}) in advance.
However, statistical properties of the language 
can be used to crack such codes.
More recently (16th century), a new technique was replacing 
the substitution codes.
In this technique, the data (which for our purposes is always represented
by a string of bits) is divided into strings of equal length $k$,
and each string is combined 
(say, by a bitwise XOR operation) with a random key 
of length $k$ known to the sender and the receiver. 
The combined data is sent through the communication channel.
The usage of a random key reduces the statistical
characteristics, since a letter is not always
replaced by the same letter.
The {\em one time pad} version of the key encoding (also called {\em Vernam}
cipher), where the random key $K$ is as long as the message $M$, 
was invented at the beginning
of this century. In this encoding the $i$'th bit of the message
is XORed with the $i$'th bit of the key 
to yield the $i$'th bit of the encoded message $C$,
$ C_i = M_i \oplus K_i$ (so that $C_i$ is zero if $K_i = M_i$ and  
one otherwise). For example,  
\begin{equation} \begin{array}{cccccc}
 M \quad & 001110 & 101011 & 100101 & 001001 & 110111  \\
 K \quad & 100010 & 111010 & 010100 & 001101 & 011010  \\
 C \quad & 101100 & 010001 & 110001 & 000100 & 101101   \end{array}\ .
\end{equation}
This cipher
cannot be broken at all since the key randomizes the message completely.
It is easy to see that knowing the transferred message $C = M \oplus K$
gives no information about $M$ to an eavesdropper. 
This is formally proven using information theory;
The eavesdropper's 
information is not increased since the probability of guessing the message
$M$ is independent of $C$: $ P(M|C) = P(M) $.

The safety of the transmission  
depends on the safety of the key, which has to 
be uniformly random, secret 
and shared only by the legitimate users. Moreover, safety can 
be guaranteed only if the key is used only once.
The problem is therefore how to distribute the random key
between users in a secure way. 
Classically, the only 
possibility is either through personal meetings, or through 
trusted couriers.  Therefore, one time pads are very expensive,    
and are impractical for many
applications. 

Most practical cryptographic systems nowadays rely 
on different principles, i.e.,
{\em computational} security.
This means that the system can be broken in principle, but
that the computation time required to do so is too long to 
pose a real threat. 
Unfortunately, none of the existing schemes is proven to be
computationally secure, and they rely on various computational assumptions.
On one hand, they can be adjusted to yield any desired
level of security 
once these assumptions will
be proven. On the other hand, 
the existing schemes might be broken in the future due 
to technological progress (faster computers),
mathematical advances (faster algorithms or future 
theoretical progress in computation theory), or new type of computers
(such as quantum computers) which are not equivalent to the standard
computers. 

The existing public key distribution schemes are based on 
the very reasonable (but unproven) 
complexity assumption that there are 
problems which can be solved in non-deterministic
polynomial time but cannot be solved in polynomial time (see explanation in
Section~\ref{int.qua-com}),
combined with additional less solid assumptions, such as that factoring
large numbers is difficult.
While the first assumption is very reasonable,
the additional assumption is not.
Indeed, some public key cryptosystems have already been broken
in the past. 
The most popular public key cryptosystems nowadays are based on the 
assumptions that the discrete logarithm 
and factorization problems are difficult.
These assumptions might be found wrong in the future. 
Furthermore, recent  
developments in quantum computation 
enable to use quantum   
computing devices to crack (at least in principle) all 
public key cryptosystems which are based  
on the discrete logarithm and factorization problems
(and maybe many others) in polynomial time.

The principle of public key cryptography was invented by
Merkle~\cite{Merkle} and Diffie and Hellman~\cite{DH}.
Later on, this principle had lead to the RSA crytposystem~\cite{RSA}
which is the basis for the most important public key schemes.
The idea of public key cryptography 
is that the receiver chooses a pair of mutually inverse
transformation $E$ and $D$, one used for encryption and one for decryption.
The sender does not need to know the receiver's secret key. 
Instead, the receiver publishes the encryption method $E$ so that 
any user can use it (by calculating $C=E(M) \ $)
to send the receiver any message.
The decryption algorithm $D$ remains secret, 
so that only the receiver  
can compute it and read the message $M=D(C)$.
As we mentioned previously, public key cryptography
is not information-theoretic secure, but, based on some computational
assumptions,
it is designed to be computationally secure. 
It is based on the presumed existence of `one way functions', $E$, 
which are easy to calculate while 
it is very difficult to calculate their inverse.
This would still be fine if we could prove that a huge computing power
is indeed required. Unfortunately, this is only assumed,
and none of the suggested transformations is proven to yield 
a reasonable and useful one way function.
To be useful for public key cryptography
a one way function must have a ``trapdoor'' which is used by the 
legitimate users, and sometimes, the same trapdoor helps the adversary
in reversing the function.

Modern cryptography deals with many other protocols, aside from
the secure transmission of messages,
such as authentication, identification, electronic signatures,
contract-signing, zero-knowledge proofs, 
and more.
Some basic protocols (from which more complicated protocols are built)
are coin-tossing, bit commitment and
oblivious transfer. 
Public key cryptography provides beautiful solutions to many
such tasks, but usually these implementations 
suffer from security problems similar to those we previously explained.

\section{Quantum Cryptography}\label{int.qua-cry}

A different technique for key distribution, which might provide 
an information-theoretic 
secure key distribution 
is quantum cryptography.
The legitimate users of a quantum key distribution scheme 
use non-orthogonal quantum states as the information carriers.
They cannot prevent Eve
from listening to their information exchange, but 
due to the ``no-cloning'' principle they will notice if she does,
and in such a case, they 
will not use the non-secret information.
The no-cloning principle assures us that any eavesdropping attempt induces 
noise.
This noise can be detected by Alice and Bob during the second stage
of the transmission, which includes discussion over a classical channel.

Let us present the no-cloning principle in more details.
The ``no-cloning theorem'' tells us that a
single quanta cannot be cloned \cite{WZ,Dieks}:
Assuming that (any) quantum state can be cloned contradicts
the unitarity of quantum mechanics.
Let $\uu$ and $\vv$ be two non-orthogonal (and non-identical)
quantum states, let 
$\Aa$ be the initial state of 
an auxiliary system used for the cloning process, and
let $U$ be a general unitary transformation.
A cloning process which duplicates 
both states is \cite{Per93}:
\begin{eqnarray} |\Aa \rangle |\uu \rangle  \longrightarrow 
 U(|\Aa \rangle |\uu \rangle) &=&      
|\BB_\uu \rangle  |\uu \rangle |\uu \rangle  \nonumber \\
 |\Aa \rangle |\vv \rangle  \longrightarrow 
 U(|\Aa \rangle |\vv \rangle) &=&      
|\BB_\vv \rangle  |\vv \rangle |\vv \rangle  \label{nc1} \ \end{eqnarray}
with all states normalized.
This process
is impossible since unitarity implies
$  \langle \uu | \vv \rangle \langle \BB_\uu | \BB_\vv \rangle = 1  $
(note that the states $\BB$ live in a smaller dimensional Hilbert space 
than $\Aa$).
Moreover, even a less ambitious cloning process 
\begin{eqnarray}   |\Aa \rangle |\uu \rangle   \longrightarrow 
     U(|\Aa \rangle |\uu \rangle ) &=&   
|\Aa_\uu \rangle |\uu \rangle  
\nonumber \\
   |\Aa \rangle |\vv \rangle   \longrightarrow 
     U(|\Aa \rangle |\vv \rangle ) &=&
|\Aa_\vv \rangle |\vv \rangle 
\label{nc2} \ , \end{eqnarray}
(with $|A_v \rangle \ne |A_u \rangle$)
is impossible; 
In this process
one attempts to gain an imprint of the received state
($|\uu\rangle$ or 
$|\vv\rangle$), without disturbing the original.
Unitarity implies  
$ \langle \Aa_\uu | \Aa_\vv \rangle = 1 $,
meaning that no information at all can be gain on the
states while keeping them unchanged~\cite{BBM}.
There is no duplicating attempt here and we adopt Peres' 
suggestion to name this ``an {\em imprint} process''.
Thus the second version of ``no cloning'' is referred to
as the ``no clean imprint'' theorem:
one cannot get an imprint of a quantum state without dirtying the origin.

In quantum cryptography Alice and Bob use a quantum channel to transmit
the non-orthogonal quantum states.
Any attempt by Eve to learn anything about these states induces noise.
The information transmitted in a quantum channel 
may be modified, hence Alice and Bob use also,
in addition to the quantum channel, an unjammable 
classical channel which is not secured against eavesdropping.
The quantum channel is 
used for the distribution of the secret key, 
while the classical
channel is used in particular
to verify that there are no errors (hence no eavesdropping),
and later on, to send the encoded message.
In principle, this is sufficient to ensure the safety of the transmission:
Alice and Bob exchange a larger string than required for the key and use
the extra part to test for eavesdropping. If they find no errors the channel
is secure hence also the transmitted key.
This can be verified to any degree of safety by checking a larger 
fraction of the bits.
If they do find errors they know that Eve  
was listening.

Unfortunately, perfect quantum channels do not exist and Alice
and Bob may find errors even if no eavesdropper is listening.
As a result, a much more complicated protocol is required,
and the eavesdropping analysis also becomes much more complicated.
At first it seemed that quantum key distribution can still be secure
in this case, but it was recently understood that security in noisy
channels (and devices) is not guaranteed, if Eve is equipped with 
a quantum memory!  
When we use the terms ``secure'' or ``secret'' in the rest of the 
work we refer to being information-theoretic secure (unless stated otherwise).

Key distribution is only one of the implementations of quantum cryptography.
Many other tasks (and their implementations) have been considered in 
the literature, based on 
the four-state scheme.
We shall also discuss two tasks which have been the subject of very interesting
works in quantum cryptography: oblivious transfer and bit commitment.

\subsection{The Four-State Scheme and a Protocol for
Quantum Key Distribution}\label{int.qua-cry.BB84}

The first quantum key distribution scheme is 
the Bennett--Brassard (BB84) scheme~\cite{BB84} which we call 
the {\em four-state scheme}. 
We describe it using the terminology 
of spin 1/2 
particles, but it can use any two-level system.
A classical two-level system, such as a bistable device,
can only be found in one of the two possible states, hence it encodes
one bit. In contrast, a quantum two-level system can be prepared in any
coherent superposition of the two basis states, which creates a much
richer structure.  Such a system is now known as a
``qubit''~\cite{ben}
(i.e. quantum bit).
For each qubit, Alice chooses at random whether 
to prepare her state along the $z$ or the $x$ axis, i.e., 
in one of  the two eigenstates of either 
$\hat{S}_z$ or $\hat{S}_x$.
This state, denoted by: $|\! \uparrow \rangle$, $|\! \downarrow
\rangle$, 
$|\! \leftarrow \rangle$ or $|\! \rightarrow \rangle$ is then sent to
Bob. 
It is agreed that the two states 
$|\! \uparrow \rangle$  and $|\! \leftarrow \rangle$ stand for bit
value
`0', 
and the other two states, $|\! \downarrow \rangle$ and $|\!
\rightarrow
\rangle$
stand for `1'.   
Bob chooses, also at 
random, whether to measure $\hat{S}_z$ or $\hat{S}_x$.  When his 
measurement is along the same axis as Alice's preparation 
(e.g., they both use $\hat{S}_z$), 
the measured value should be the same as hers, whereas when 
they use conjugate axes, there is no correlation between 
his result and Alice's original choice. 
In addition to the quantum channel, the legitimate users also use 
a classical channel as previously explained.
By discussing over this channel Alice and Bob agree
to discard all the instances where they did not use the same
axes. The result should be two strings of perfectly correlated bits.
These two strings shall be identical in case no natural noise occurs and
no eavesdropper interferes.
We shall call this resulting string {\em the raw data}.
As the choice of axis 
used by Alice is 
unknown to Eve, any interaction by her will 
unavoidably modify the 
transmission and introduce some errors. 

In practice however, the transmission will never be perfect 
and there will 
be some errors, even in the absence of an eavesdropper. 
Alice and Bob use the classical channel 
to compare some portion of their data and calculate the error rate.
They decide about some {\em permitted} error rate, $P_e$, which they accept,
and determine this error rate according to the properties of the channel
and their devices.
Eve could take advantage of that, and attack some of the bits
to gain some information, as long as the error rate she induces, $p_e$, 
plus the natural errors, do not exceed the permitted error rate.
For all further discussions we assume that Eve is powerful
enough to replace the channel by a perfect channel so she can 
induce as many errors as accepted by the legitimate users $p_e\approx P_e$
(or more precisely -- a little less):
Eve can detect the particles
very close to Alice's cite (where not many channel errors are yet
induced) and release her particles very close to Bob's cite.

Assume Alice and Bob used the same basis.
When Eve eavesdrops on a fraction $\eta$
of the transmissions, performing a standard measurement
in one of the bases (as Bob does),
the error rate created is 
$p_e = \eta/4$: when Eve uses the correct basis, she does not
introduce any error, while she creates a 50\% error rate when 
she uses the wrong basis. 
Eve knows the permitted error rate (agreed using the classical channel),
thus she chooses $\eta = 4 P_e$.
The average mutual information she 
obtains is $\eta/2$: she has total information when 
she used the correct basis, and none when she used the wrong 
one. Note that the scheme is completely symmetric, so that Eve
shares the same information with Alice and with Bob. Therefore,
we can write the mutual information shared by Alice and Eve and
shared by Eve and Bob as a function of the error rate:
$ I_{AE} (P_e) = I_{EB} (P_e) = 2 P_e$.
More complicate attacks are discussed in Chapter~\ref{secur}.

As long as $P_e$ 
is not too high, Alice and Bob might be able to 
use classical information processing techniques, such as
error correction and privacy amplification~\cite{BBBSS,BBR,BBCM},
to obtain a reliable and secure final key.
It is called reliable if they succeed to reduce the error rate to zero,
and it is called secure if they succeed to reduce the information
obtained by Eve to zero as well
(by ``zero'' we mean close to zero as they wish). 
Both techniques are based on the use of parity bits of long strings
where the parity of a string is zero if it contains an even number
of $1$'s and the parity is one if the string contains an odd number
of $1$'s. For a two-bit string, this is exactly the XOR 
of the two bits.

The simplest error-correction code is the repetition code from one bit
to three bits: $0$ is encoded as $000$ and $1$ as $111$.
It is not efficient (since the amount of transmissions is increased 
by a factor of three) 
and it can correct only one error in the three bits; and if two
errors occur the ``corrected'' final bit is wrong.
This code uses two additional bits to correct a possible error in one
bit.
A more efficient code~\cite{MS} is the Hamming code $H_3$ which uses three
parity bits
to correct one error in four bits: 
Alice add to her four bits three parity bits so 16 different words
are written using 7 bits. Since this example is explicitly used
in Chapter~\ref{col}, let us present it in more details. The words are
easily calculated by taking the fifth bit to be the parity of bits
1, 2, and 3;
the sixth bit to be the parity of bits 1, 2, and 4; and the seventh bit
to be the parity of bits 2, 3, and 4. For instance:
\begin{equation}
\begin{array}{cccccccc}
  0 & 0 & 0 & 0 & \quad & 0 & 0 & 0 \\
  0 & 0 & 0 & 1 & \quad & 0 & 1 & 1 \\
  0 & 0 & 1 & 0 & \quad & 1 & 0 & 1 \\
  0 & 0 & 1 & 1 & \quad & 1 & 1 & 0 \\
  0 & 1 & 0 & 0 & \quad & 1 & 1 & 1 \\
  \ & . & . & . & \quad & \ & \ & \ \\
  1 & 1 & 1 & 1 & \quad & 1 & 1 & 1 \\
\end{array} \ . \end{equation} 
Let us call these seven-bit words $w_i$ where $i$ goes from 1 to 16.
It can be easily checked (an also proven) that any two words
differ by at least three bits, thus any single error can be corrected.
(We say that any two words are at a Hamming distance of at least
three).
When Alice sends a word $w_i$ and Bob receives it with a single error
he can correct it and re-derive $w_i$.
In practice Alice will send Bob seven random bits.
If the seven-bit word is in the above set, she (later on) sends $000$.
If it is $0000\ 001$ or any word which is the bitwise XOR of
$w_i$ with $0000\ 001$ Alice sends (later on) the string $001$.
If it is any word which is the bitwise XOR of
$w_i$ with $0000\ 010$ Alice sends the string $010$, etc.
Note that Alice actually sends the parity bits of $(1,2,3,5);(1,2,4,6);
(2,3,4,7)$.
In this case we say that Alice sends 
the parities of the parity substrings $(1110100; 1101010; 0111001)$.
Doing so she also tells the parities of all their linear combinations, such as
$1110100 \oplus 1101010 = 0011110$.
This Hamming code has generalizations to codes $H_r$ of
length $2^r-1$ and $r$ parity bits,
and generalizations to BCH codes which correct
more than one error~\cite{MS}.

Assume that Eve had no knowledge on their string apart of the error-correction
data.
After performing that type of error correction Alice and Bob remain 
with a corrected word of the same length (e.g., of length of seven bits 
in the above example), and Eve knows that it is taken from
a smaller set of words (16 words, in this example).
Alternatively, they can distill a shorter common string 
(of length four, in the above example), which is completely unknown to
Eve (simply by keeping only the four first bits in the above example).

In reality, Eve has both the error-correction data, and some data
she gained by eavesdropping on the transmitted particles.
Alice and Bob perform a privacy amplification technique to reduce
Eve's information to be as small as they wish. 
The simplest privacy amplification uses the parity bit
of a string as the secret bit. 
The analysis of Eve's information could be very complicated.
Let us start from the simplest case.
In case the eavesdropper
does not know one (or more) bit(s) of the string,
she does not have any information on the parity bit.
In other simple cases Eve has either full information or zero information on 
each bit, and we can calculate the probability $P(n)$ 
that Eve knows all the bits 
thus we can calculate her average mutual information 
$I = 1 [P(n)] + 0 [1- P(n)] = P(n)$.
If $P(n)$ is exponentially small with $n$ we say that the final 
bit is {\em exponentially secure}\footnote{
When Eve has a total information which is much smaller than one bit, the
key is already information-theoretic secure. An exponential security where
$I \approx 2^{-n}$ is much better than
security with $I\approx 1/n$.
}.
In general, the situation is more complicated, 
and Eve's information on each bit is neither zero nor one.
When Eve has some probabilistic information on the bits, 
her information on the parity bit is not a simple function as in the example
above.
Analysis of the privacy amplification in such cases
were done in~\cite{BBCM,DMLS} using the concept of
{\em collision entropy}.
Alternatively, one could calculate the probability that Eve 
guesses the parity bit
correctly. 
When privacy amplification is done after error correction, its analysis
is even more complicated since 
Eve has both the information on the bits and the error-correction data.
Furthermore, we might wish 
to obtain more than one final bit, so that the parity of the whole
string is not a good choice, and the final key 
is usually constructed from parities of several substrings
of length of about half the string.

Fortunately, for particular symmetric cases,
where Eve has the same error probability for each bit,
the analysis of privacy amplification
is still possible.

Let us return to the key distribution scheme:
All these operations,
error estimation, error correction and privacy amplification,
waste many bits (say, $l$). 
In order to remain with a final key of $L$ bits 
Alice should send  $L' > 2(L + l)$ qubits (where the factor 2
is due to the two bases).
In more details, Alice and Bob start with $L'$ qubits, 
compare basis, and use some bits (of the raw data) to estimate
$P_e$.
As a result, $n$ bits remain, from which about $P_e n$ are 
erroneous. 
Alice and Bob randomize the order of the bits 
and correct the errors
using an appropriate error-correction code~\cite{MS}.
This process generates parity bits which 
Alice sends to Bob (using the classical channel), 
who uses them to obtain an $n'$-bit string ($n'$ could be
smaller than $n$)
identical to Alice's, 
up to some very small error probability. 
By that time Eve has both the information she gained in 
her measurements,
and the additional information of the error-correction data.
Alice and Bob amplify the security of
the final key by using privacy amplification: 
by choosing some parity bits
of substrings to be the final key. 
Both the information available to Eve on the final key, and 
the probability $p_f$ of having error in the final string must be 
close to zero as much as desired by Alice and Bob.
Various eavesdropping techniques, and the efficiency
of privacy amplification against them, are discussed in details
in the next chapters.

Another fact which could be useful to Eve is that some of the transmitted
particles do not reach Bob. 
The easiest way to deal with them is to consider them as errors
(but this might affect the practicality of the scheme).

By the time we started this work, it was assumed that the privacy 
amplification of~\cite{BBR} and its extension in~\cite{BBCM},
indeed provide the ultimate security.
Only after intensive discussions with Charles Bennett, Bruno Huttner, 
Dominic Mayers and Asher Peres, 
(initiated in Torino, 1994, at the ISI conference),
it became clear to us that privacy amplification might not work
against a ``quantum'' eavesdropper equipped with a quantum memory,
and the possibility to do any unitary transformation on the qubits. 
The security problem  
is explained in Chapter~\ref{secur}, together with our suggestion for  
a way to approach the security problem.
We develop it further in Chapters~\ref{par} and~\ref{col}, 
providing strong evidence
that quantum key distribution is secure, even against a ``quantum 
eavesdropper'' equipped with any possible devices, and restricted
only by the rules of quantum mechanics.

Let us review other schemes for quantum key distribution, and concentrate only 
on the parts which are different from the above scheme.

\subsection{The EPR Scheme}\label{int.qua-cry.EPR}

The second basic quantum key distribution scheme is  
based on EPR~\cite{EPR} correlations. 
It was suggested by Deutsch~\cite{Deutsch85},
formalized by Ekert~\cite{Eke91} and modified by 
Bennett, Brassard and Mermin~\cite{BBM}. 
We describe here the modified version which we call the {\em EPR
scheme}.
In this scheme Alice creates pairs of spin 1/2 particles in the 
singlet  state, 
and sends one particle from each pair to Bob. 
When the two particles are measured
separately the results obtained for them are correlated. 
For example, if they are measured along the same axis, 
the results are opposite, 
regardless of the axis. 
Alice and Bob use the
same sets of axes as for the four-state scheme,
say $\hat{S}_z$ and $\hat{S}_x$, and keep
the results only when they used the same axis.   
It is noteworthy that, in the EPR scheme,  the pairs could 
be created by any other party, including Eve herself. 

Let us discuss this point  
in more details.  The singlet state may be written in
infinitely many ways. We write it as:
\begin{eqnarray}               
\Psi^{(-)} &=& \sqrt{ \frac{1}{2} }
 \left(|\!\uparrow_A \downarrow_B \rangle -
       |\!\downarrow_A  \uparrow_B
\rangle
\right)  
 \nonumber \\ &=& 
\sqrt{ \frac{1}{2} }
 \left(|\! \leftarrow_A \rightarrow_B \rangle - 
        |\! \rightarrow_A \leftarrow_B \rangle \right)
 \label{singlet-pair} \ ,
\end{eqnarray}           
where the equality follows from 
$|\!  \rightarrow \rangle = \frac{1}{\sqrt 2}(|\!  \uparrow \rangle + |\! 
\downarrow \rangle) \ $ 
and $|\!  \leftarrow \rangle = \frac{1}{\sqrt 2}(|\!  \uparrow \rangle -
|\! 
\downarrow \rangle) \ $,
and where the subscripts `A' and `B' stand for Alice and for
Bob. 
When Alice and Bob use  the same axis, 
either $z$ or $x$, we use the first or the second equation
respectively,
to see that their measurements always yield opposite results.
The singlet state is the only state with this property.
Therefore, as Alice and Bob may measure either of these two options, 
any deviation from the protocol by Eve (i.e., any attempt
to create another state), will induce errors.
So Eve must   
create the required singlet state, from
which she cannot extract any information about Alice's
and Bob's measurement (see~\cite{Eke91,BBM} for more details).

\subsection{The Two-State Scheme}\label{int.qua-cry.two}
The last elementary scheme is the {\em two-state scheme}~\cite{Ben92}.
In this scheme, Alice chooses one of 
two non-orthogonal
states, 
\begin{equation}
\phi_0 = {\cos \theta \choose \sin \theta}
\quad {\rm or} \quad 
\phi_1 = {\cos \theta \choose - \sin \theta}
\end{equation}
and sends it to Bob to encode 0 and 1 respectively. 
As these states are not orthogonal, 
there is no way for Bob to decode them deterministically.

Bob must not use the optimal 
projection measurements (in the symmetric basis)
because then the protocol is unsafe: Eve also measures in that basis and
sends Bob photons (according to her results) in the orthogonal 
symmetric basis; 
hence her string and his string are the same.
Bob expects a certain amount of errors which is exactly
the amount of errors in Eve's string, 
Therefore Eve introduces no additional errors to what Bob would expect. 

Bob must obtain deterministic information on Alice's state 
in order to notice the presence of Eve.
We have already seen how to analyze this system 
in Section~\ref{int.qua-inf}, and we saw that partial deterministic
information is obtained by 
means of a POVM. When the best possible POVM (\ref{POVM1}) is used, 
Bob's mutual information is equal
to $1 - P_? = 1 - \cos 2\theta$ where $\cos 2 \theta$ is the overlap 
of the two states.
When Bob uses the optimal POVM, Eve cannot avoid introducing errors.  
The safety of the protocol against 
eavesdropping is ensured by the fact that Eve cannot always get
deterministic results either. Therefore, in many instances where 
she intercepts the transmission she will have to guess 
Alice's choice. 
If the channel is very lossy, Eve could take advantage of it and 
learn all information: she performs the optimal deterministic attack,
and throws away the bits whenever she got an inconclusive result.
Thus, in order to use this scheme we must restrict also the losses in 
addition to the restriction on the error rate,
or alternatively use a reference signal~\cite{Ben92}.
To simplify further discussion of this scheme we redefine $P_e$ to include the 
losses, thus it is better for Eve to send something
(which might not be counted as error) than to keep the particle.

In our further discussions we consider a scheme in which Bob does not
perform the optimal POVM.
Instead, Bob performs a less optimal test which also 
provides him with a conclusive or inconclusive
result. This test was described in Section~\ref{int.qua-inf}
by a less efficient POVM (\ref{POVM2}),
but it can also be described using ordinary measurements:
Bob performs one of two possible tests with equal probability.
In half the cases he tests whether a specific particle is in a state
$\phi_0$ or a state orthogonal to it ${\phi_0}'={\sin \theta \choose
-\cos \theta}$; A result $\phi_0$ is 
treated as inconclusive (since it could also be $\phi_1$ sent by Alice), 
but a result ${\phi_0}'$ is always identified as $\phi_1$.
In the other half he tests whether a specific particle is in a state
$\phi_1$ or a state orthogonal to it ${\phi_1}'={\sin\theta \choose
\cos \theta}$; A result $\phi_1$ is 
treated as inconclusive (since it could also be $\phi_0$ sent by Alice), 
but a result ${\phi_1}'$ is always identified as $\phi_0$.
All the conclusive results are the raw data.
Note that Bob can get a correct conclusive result only if he chooses a basis
which does not contain Alice's state (in contrast to the situation 
in the four-state scheme).
\subsection{Weak Coherent Pulses}\label{int.qua-cry.coh}

As single photons are 
difficult to produce experimentally, a ``slight''
modification of the four-state scheme, 
using weak pulses of coherent light instead of
single photons, was the first one to be implemented in
practice~\cite{BBBSS,tow93,mul93}. 
Such schemes are sensitive to losses in the channel
and to the existence of two photons in a pulse.
In an earlier work~\cite{HIGM}, not included in this thesis,
we focused on quantum cryptographic schemes 
implemented with weak pulses of coherent light. 
We compared the safety of the four-state scheme 
and the two-state scheme, 
and presented a new system, which is a
symbiosis of both, and for which the safety can be 
significantly increased.  
In that work we also provided a thorough analysis of  
a lossy transmission line.
A recent work of Yuen~\cite{Yuen} suggests that most of the practical
schemes implemented in laboratories based on weak coherent pulses
are not ``slight'' modifications since they use a completely
different Hilbert space. Yuen actually shows that these implemented
schemes are insecure due to that fact.

In this thesis we do not deal with weak coherent pulses.
Whenever we discuss the four-state or the two-state schemes
we refer to the schemes 
as previously described, which are using two level systems.

\subsection{Quantum Bit Commitment and 
Oblivious \mbox{Transfer}}\label{int.qua-cry.BCOT}

Let us consider two other
possible applications of quantum cryptography,
Bit Commitment (BC) and Oblivious Transfer (OT).
These are defined as follows:
\begin{description}
\item  BC: Assume Alice has a bit $b$ in mind.
She would like to be committed to this bit towards Bob, in a way that she 
cannot change it, but yet, Bob will have no information on that bit.
In a bit commitment scheme Alice provides Bob some evidence through a procedure
$commit(b)$ which tells him that she is committed while following 
these two requirements.
At a later time Alice can reveal $b$ to Bob through another procedure 
$unveil(b)$, which provides Bob, in addition to the value of the bit,
with some information from which he can make sure she did not change the bit
she was committed to.
\item \tto OT \cite{Wiesner,EGL}: Alice has two bits in mind.
She sends them to Bob in a way that he receives only one and knows nothing 
on the other bit.
Alice does not know and can have no control on whether Bob gets 
the first or the second bit.
\item OT \cite{Rabin81}: Assume Alice has a bit $b$ in mind. 
She would like to transmit this bit to Bob so that 
he receives this bit with probability 50\%.
Bob knows whether he received the bit or not, but Alice
must be oblivious to this.
Neither of them can control the probability of reception.
\end{description}

In the BC protocol, a
cheating Alice would like to change her commitment at the $unveil$
stage, and a cheating Bob would like to learn the bit from her
commitment.
In the \tto OT a cheating Alice would like to control or to
know which bit Bob received, and 
a cheating Bob would like 
to learn more than one bit.
In the OT protocols a cheating Alice would like to control the
probability that Bob receive the bit in OT or to learn whether Bob got the bit
or not, and 
a cheating Bob would like to learn the bit with
probability better than 50\%.

The two versions of OT are equivalent\cite{CK}:
OT is derived from \tto OT by considering only one bit of the two.
A protocol for \tto OT is derived from OT as follows: 
Alice and Bob perform OT several times.
Bob arranges the bits in two strings --- a string 
of bits he knows and a string of bits he doesn't;
He tells Alice the groups but not which group is known to him.
Then, the parity bits of each string are used for \tto OT.

BC can be derived easily from 
\tto OT: 
Suppose Alice sends a string of $2n$ bits to Bob such that each pair
of bits $b_{2i} , b_{2i+1}$ are transmitted to Bob via \tto OT scheme.
Alice wants to commit to a bit $b$, so she chooses the pairs of bits to
fulfill $b = b_{2i} \oplus  b_{2i+1}$.
Bob knows nothing on the value of $b$.
Procedure $unveil(b)$ includes the value of $b$ and the value of all 
the $2n$ bits.
If Alice wants to change the value of $b$ she needs to change one bit in 
each pair.
Unless Bob has accidentally received (in $commit(b)\ )$ 
only the bits she does not 
change, she will be caught.
Her chance to cheat successfully is $2^{-n}$, so that this procedure 
provides exponential security.
It is generally believed that OT cannot be derived from BC.

The best protocols which are available today for quantum oblivious
transfer and quantum bit commitment are severely influenced by the possibility
to use quantum memories.
On one hand, we do not wish to explain these complicated protocols 
in details since they are not 
relevant for our work. On the other hand, their sensitivity to 
the existence of quantum memory is remarkable, and we wish to show it.
As a compromise, we present two toy-models which exhibit the same sensitivity
to quantum memories. 

The four-state protocol for key distribution  
can be modified to produce a naive \tto OT protocol~\cite{CK}:
Alice sends particles as in the four-state scheme and Bob
measures them in one of the two bases. If he measures 
in the same basis Alice has used he gets a bit, hence he receives each bit 
with probability 50\%.
Alice then tells him the basis she used for each photon, so he knows
which bits he has received as required.
Bob organizes the bits in two sets, one set for those he identified
and one for those he didn't. He tells Alice the members of each set
but not which is the identified set. Alice sends him error-correction 
data for the two sets so that he can correct one of the strings only,
but not the other. Finally Alice uses the parity bit of one of
the strings to transmit him her secret bit.

A naive bit commitment scheme based on using the 
four-state scheme can also be easily 
implemented by choosing the basis as representing the secret bit. 
Alice sends to Bob the bit `0' using either 
$ | \rightarrow \rangle $ or
$ | \leftarrow    \rangle $, and `1' using 
$ | \uparrow  \rangle $ or
$ | \downarrow  \rangle $.
In this case, of course,  Alice does not tell Bob anything 
at the $commit(b)$ stage. In the $unveil$ stage she tells him both the
bit and the state she used. Bob cannot learn the bit since 
both possible values are described by the same density matrix
\beq (1/2) ( | \rightarrow \rangle \langle \rightarrow| +
 | \leftarrow \rangle \langle \leftarrow| ) = (1/2) 
 ( | \uparrow \rangle \langle \uparrow| +
 | \downarrow \rangle \langle \downarrow| ) \ . \label{unpollar} \eeq
Alice can cheat with probability of $1/2$; for instance she sends 
$\uparrow$ to commit to `1', and in order to change her commitment she
later on claims that she has sent $\leftarrow$. 
Thus the protocol demands
that she commit to the same bit $n$ times, so the probability of
successful cheating is reduced to $2^{-n}$.

Both these protocols are insecure, since the parties can cheat using 
quantum memories as we explain below.

\section{Quantum Memory}\label{int.mem}

In this section we present the quantum memory\footnote{
Parts of this section are original.}:
its definition, basic use, and its possible implementations.

\subsection{Defining a Quantum Memory}\label{int.mem1}

A classical bit or a classical string of bits can be kept in
a memory, where their value is preserved for a long time.
The main physical feature required for keeping a bit is the huge redundancy
used for the two possible bit values.

When it comes to the quantum world things become less obvious,
but still we can provide an appropriate definition:
\begin{itemize}  
\item
A quantum memory is a device where a 
quantum state can be kept for a long time 
and be fetched when desired with excellent fidelity. 
\end{itemize} 
Clearly, both ``a long time'' and ``excellent fidelity'' are determined
by the desired task for which the state is kept.

A quantum state develops in time according to some unitary
operation, and is exposed to interactions with the environment.
If redundancy is added in a simple way as in the classical case above,
we might lose all the advantages of using quantum states.
Thus, it is not easy to keep a quantum state unchanged for a long time.
The same problem appears when we want to transmit a state 
over a long distance.
However, with existing technology, one can talk about transmissions
of quantum states (e.g., sending photons' polarization states
to a distance of more than 10 kilometers),
while it does not make much sense 
to discuss memories where a state can be kept for 
$10^{-3}$ seconds (which is much shorter than a phone talk).
Consider a quantum bit (a two level system) in a state
\begin{equation} \psi =  {\alpha \choose \beta} \ ; \end{equation}  
If it changes according to some unitary transformation to a state
\begin{equation} \psi^1 =  {\alpha^1 \choose \beta^1} \end{equation}  
we may still be able to use it if we know the transformation,
but if it decoheres due to interactions with other systems,
in a way which we cannot reverse in time, the state is lost.
As in the case of transmission over a long distance --- one can choose 
some acceptable error rate $P_e$, and agree to work with
the experimental system as long as the estimated
error rate $p_e$ does not exceed
$P_e$.
To estimate the error rate, one first does all effort to re-obtain the
desired state (e.g., take the unitary transformation into consideration),
and then one compares the expected state $\psi^1$ with the obtained state 
$\rho$ and defines the error rate as the percentage
of failure.
In theory, if we consider some irreversible change to the state
(due to environment, or an eavesdropper, or any other reason), and we
can calculate the obtained state, the error rate is  
\begin{equation} p_e = 1 - \langle \psi^1 | \rho | \psi^1 \rangle
\ . \end{equation}

The definition of a quantum memory contains more that just the
ability to preserve a quantum state for a long time.
Other necessary conditions are the input / output abilities:
One must be able to produce a known quantum state;
One must be able to input an unknown state
into the memory;
One must be able to measure the state in some well defined basis,
or furthermore, to take it out of the memory (without measuring it)
in order to perform 
any unitary transformation
on it (alone, or together with other particles).

\subsection{Basic Uses of a Quantum Memory}\label{int.mem2}

We shall describe here only simple uses known by the time
we have started this work. The rest of this work concentrates on other uses
of quantum memory, which are much more advanced.
While doing this work more fascinating uses of quantum memory
were suggested by others, 
and we will shortly review them in the relevant chapters (mainly
in Chapter 6).

\begin{itemize}
\item
Doubling the efficiency of
the four-state scheme for QKD: Bob keeps the particles in a memory,
and measures them after Alice tells him the bases.
\item
Improving the EPR scheme: instead of measuring the states
of their particles, writing the results on a paper and hiding the paper,
Alice and Bob keeps them in a quantum memory.
This way they are assured that no one else had access to the paper, and
therefore, no one else has the key.
They perform a measurement only when they want to use the key~\cite{Eke91}.
\item
Teleporting~\cite{BBCJPW}
a state using singlet pairs
prepared in advance. 
\item
Suggesting new schemes 
for quantum key distribution which relies upon 
the existence of short time quantum memory~\cite{BW,GV}.
\item
Cheating the naive \tto OT scheme presented in the previous section. 
Bob delays his measurements till after he receives the classical information 
regarding the basis used by Alice; In this case he gets full information.

To overcome this problem, the \tto OT protocol was modified~\cite{BBCS}
to contain a step where Bob commits to the measurement he performed,
thus he cannot avoid performing the measurement.
However, this ``solution'' creates a tremendous   
complication to the protocol, and even worse --
it adds an undesired dependence on the existence
of secure bit commitment scheme!
\item
Cheating the naive 
bit commitment scheme. 
Alice use photons taken from EPR 
singlet pairs,
keeps one particle of a pair in a memory,
and sends the other to Bob;
she chooses the basis just right before she performs 
$unveil(b)$; 
She may thus change her commitment as she wishes to~\cite{BC91}. 

A better (and much more sophisticated) 
bit commitment was later suggested~\cite{BCJL},
but most amazingly -- 
it was recently broken by Mayers~\cite{DM4} due to similar (but deeply
hidden) flaws!
Mayers' attack is based on a theorem regarding the 
classification of quantum ensembles~\cite{HJW93}.
In terms relevant to quantum cryptography this theorem states that
two ``different'' quantum systems (in Bob's hands)
which are described by the same density matrix 
[e.g. as in eq.~(\ref{unpollar})]
can always be created from one entangled state shared by Alice and Bob, 
by measurements performed at Alice's site.
Thus, Alice can postpone the decision of whether a state comes from the
first or the second system, and change her commitment at the $unveil$ stage.  
The consequence to bit commitment is that, either the two density matrices are 
different and Bob can partially learn the commitment, 
or they are similar and Alice
can change her commitment.
\item
Breaking the classical {\em logical chain} of achieving 
a secure protocol by using
secure building blocks.
Thus, any protocol which is based on secure building blocks must be also checked
as a whole.
As a result, the importance of the building blocks 
(such as OT and BC) is reduced.

There is a very surprising consequence of breaking the logical chains:
Due to the facts that (1) BC can be built upon OT and (2) secure BC probably
does not exist~\cite{DM4} one could induce (by reduction) that secure 
OT cannot exist. However, secure OT
might still exist, despite the fact that the BC protocol built upon
it is insecure.
\end{itemize}

\subsection{Implementations}\label{int.mem3}

By the time we began this work, the possibility of implementing
a quantum memory was hardly considered. 
The following discussion is taken from our work~\cite{BHM}, and was done with
the help of David DiVincenzo.

We demand the possibility to program, 
store and manipulate quantum bits.
Any 2-dimensional Hilbert space system can be considered and this 
opens a variety of possible implementations.
Fortunately, apart from a different timing requirement,
the same requirements appeared recently in
quantum computing, and are being thoroughly
investigated
by both theorists 
\cite{Deutsch89,DiV,BDEJ,CZ},
and experimentalists
\cite{Haroche,Kimble,Wineland}. 
The difference in the timing requirements is that the quantum bits in
most applications of a quantum memory 
are subjected only once (or very few times)
to a unitary operation of calculation, while the quantum bits in quantum 
computation are subjected to a huge number of computations;
On the other hands, the common tasks where quantum memory is used 
take long times.
As a result, the problem
of decoherence is different. For quantum memories
we don't need to consider
the switching time versus the decoherence time, but we must require a
decoherence time on the time scales of minutes, days, and even years,
depending on the applications.

It may be possible to implement a working prototype of 
such a memory within a few years. 
Such a prototype should enable to keep quantum states for
a few minutes (or even hours), and to  
perform two-bit operations on them. 
At the moment, the best candidates for combining these two operations 
are ion traps.
In ion traps the quantum bits can be kept in internal degrees of
freedom 
(say, spin) of the ions, and
in phononic degrees of freedom of few ions together.
It is already possible to keep quantum states in the spin 
of the ions  
for more than 10 minutes~\cite{Wineland2} 
and in principle it is possible 
to keep them for years.
These ion traps are thus good candidates for implementing quantum 
memories. Moreover, they are also among the best candidates for quantum 
manipulations, since
there are ways to use the phononic degrees of freedom to
perform two-bit operations~\cite{CZ}, such as the
very important controlled-NOT gate.
Barenco et al.~\cite{BDEJ} realized that
a single quantum controlled-NOT logical gate would be 
sufficient to perform the Bell measurement, a measurement
which appears in almost all applications of quantum memories. 
Thus, using ion traps it is possible to (partially) perform the Bell
measurement, and this was shown 
both in theory~\cite{CZ} and in an experiment
\cite{Wineland}.
Combining together the two experiments\footnote{
Which were done by the same group in NIST.} to have
both long lived quantum states and the possibility to manipulate them
will create a useful quantum memory.

The way to establish a real working quantum memory which suits for
all purposes is still long.
The main obstacle is that it is currently impossible to transfer
a quantum state from one ion trap to another.
A fully operating quantum memory must allow us to put quantum states
in a register (that is, a separate ion trap), and pull it out 
to another register when we want to perform operations on it,
together with other qubits (possibly held in other registers).
Recently, the possibility of doing this arose from the
idea of combining
ion-traps (where the ions are well controlled) and QED-cavity
together,
and to use the same internal degrees of freedom for 
both~\cite{DD,Pellizz}. 
We shall call this combination {\em cavitrap} for convenience.
In QED cavity \cite{Haroche} the internal degrees 
of freedom are coupled to photons and not to phonons.
Recently, another group \cite{Kimble}
has shown that it is possible to use polarization
states of
photons instead of using the $| 0\rangle$ and $| 1\rangle$ Fock
states.
If such photon states are used in a cavitrap, it may be possible to
use 
them to transmit a quantum state from
one cavitrap to another~\cite{DD}.
In some sense, this will be an implementation of the nuclear spins
based
``quantum gearbox'' suggested by DiVincenzo~\cite{DiV}.

All this discussion would be considered nothing but a fantasy  
just few years ago. 
However, similar (and more complicated) ideas 
are required for using 
quantum gates
in both quantum computing and quantum information,
and a lot of effort is invested in both the theory and application 
of quantum gates.  Also, 
quantum memory is considered more and more seriously, for various purposes,
and the fantasy becomes closer to reality every year.

\section{Quantum Computation}\label{int.qua-com}

A classical computer can solve certain problems in a time which 
is polynomial with the size of the input. Such problems are called 
{\em easy}, and other problems, which are solved only in exponential
time are called {\em difficult}. 
Public key cryptography uses one way functions with trapdoors,
i.e., functions which are easy to compute while their inverses are difficult
to compute unless some secret key is known.
Easy problems are not useful for public key cryptography.
An important class of problems (which are useful for public key
cryptography) is those solvable in non-deterministic 
polynomial time, for which a given solution can be verified in
polynomial time.
It is well accepted that the most difficult problems in this class
are difficult to solve, although this assumption is not proven.
Unfortunately, the popular cryptosystems are not based 
upon this assumption alone, but on much less solid assumptions.
For instance, on the assumption that factoring large numbers 
and computing discrete logarithms are difficult.

More than 20 years ago Feynman raised an important 
question:
Can quantum devices, based on quantum principles, simulate physics 
faster than classical computers?
Deutsch \cite{Deutsch85} proposed a Turing-like model for quantum computation 
in 1985, called a quantum Turing machine.
He constructed a universal quantum computer which can simulate any given
quantum machine (but with a possible exponential slowdown).
The quantum Turing machine 
is different from classical Turing machine in that, in addition to
standard abilities, it can also get into superpositions of states.
Bernstein and Vazirani~\cite{BV} 
constructed an efficient universal quantum computer
which can simulate a large class of
quantum Turing machines with only polynomial slowdown.
Yao~\cite{Yao1} 
developed a complexity model of quantum circuits showing that any
function computable in polynomial time by a quantum Turing machine
has a polynomial 
size quantum circuit.
Quantum circuits and networks were defined by Deutsch \cite{Deutsch89}
on the basis of reversible classical logic \cite{Bennett73}.
Yao~\cite{Yao1} 
also developed a theory of quantum complexity which may be useful
to address the question whether quantum devices can perform computations
faster than classical Boolean devices.

There were early attempts of solving problems on quantum computers 
more efficiently 
than it is possible on classical computers~\cite{BerBra,DJ,Sim},
focusing on problems which are not known to be solved in polynomial time 
on a classical computer.
These works were the groundwork for the 
recent breakthrough of Shor \cite{Shor}.
Shor succeeded to solve in polynomial time the problems of discrete logarithm
and factorization on a quantum computer! 
As we previously said, 
these two problems are extremely important since most existing classical 
cryptosystems rely on the assumption that they are not easy (not polynomial).

The interest in quantum cryptography also raised a lot due to the interest 
in this breakthrough.
In addition, many experimentalists are now trying to make quantum gates
in their labs, and many implementations for quantum cryptography
which looked as fantasy
three years ago, are now very close to practice.

The question of building practical quantum computers is not only
a technological problem, since decoherence and other environmental
interceptions could limit the computing abilities very much.
However, it seems that quantum error-correction codes~\cite{Shor1,Steane1}
and their extension to fault-tolerant quantum error correction~\cite{Shor2}
can overcome the decoherence problem. This issue is still under 
thorough investigation. We discuss quantum error correction in 
Chapter~\ref{QECQPA} but we concentrate on its relevance to 
quantum memories and quantum cryptography.

\section{Structure of the Thesis}

Chapter 2 describes the security problem in details and our
approach to it.
Chapter 3 discusses the information on the parity bit in various cases,
and proves that it reduces exponentially with the length of the string in the
case relevant to eavesdropping analysis.
Chapter 4 extends the analysis regarding the information on a parity bit 
to deal with error correction, 
and applies the analysis 
to show the security of quantum key distribution schemes against
various attacks.
Chapter 5 describes a new scheme for quantum key distribution
which is based on the use of a quantum 
memory. This scheme is especially adequate for building networks of many users.
Chapter 6 discusses quantum error correction, quantum privacy amplification 
and suggests a new quantum privacy amplification 
based upon quantum error correction.
It also develops the surprising possibility of having ``quantum repeaters''.
Chapter 7 provides our conclusions in brief.

\newpage

\chapter{On the Security of Quantum Key Distribution Against 
a `Quantum' Eavesdropper}\label{secur} 

The security of quantum key distribution in 
noisy channels and environment is still   
an open question\footnote{
But see Section~\ref{securC}.} 
despite  
the use of privacy amplification; 
no bound has yet been found on the amount of information obtained by 
an adversary equipped
with {\em any} technology consistent with 
the rules of quantum mechanics
(to be called {\em a quantum eavesdropper}).
In particular, neither of the suggested schemes is proven secure
against sophisticated {\em joint} attacks, which 
use quantum memories and quantum gates to attack directly the final
key.
Note that in order to make quantum key distribution a practical tool
one must present a scheme which 
uses {\em existing} technology, and is
proven secure against an adversary equipped
with {\em any} technology.

The main achievement of this thesis is to develop a new approach to attack
this problem. In this chapter we explain the security problem
in more details and we present our  
approach to it.
We separate the complicated security problem 
into several simpler steps. 
We define a class of strong attacks which we call
{\em collective attacks}, and we conjecture that 
the problem of security against any attack can be replaced 
by the simpler problem of security against collective attacks.
In the next two chapters we discuss security against collective attacks,
providing some important security results.

Our approach tries to prove the efficiency of privacy amplification 
against a quantum eavesdropper,
by analyzing the density matrices available to her, and bounding 
the information which can be extracted from them.

The security of quantum cryptography is a very complicated and tricky problem.
Several security claims done in the past were found later on to contain 
loopholes. 
Recently, in parallel to our work,
two other attempts were made to solve the 
security issue~\cite{DM2,oxford}, based on different interesting
approaches. 
The first one~\cite{DM2} considers the efficiency
of privacy amplification as we do 
(but without analyzing Eve's density matrices);
we discuss it in the last section of this 
chapter.
The other one~\cite{oxford} assumes 
non realistic perfect devices\footnote{This
is not clear from~\cite{oxford}. 
The assumption appears on page 2819, col.\ 1, line 18, and it is never removed.}
with perfect accuracy,
hence provides only a limited
security proof. Furthermore, it 
suggests a {\em quantum privacy amplification} hence is not
applicable with existing technology. We discuss it in Chapter 6,
where we suggest another quantum privacy amplification scheme.

\section{Privacy Amplification and Security Against Existing Technology}\label{securA}

We have seen that, in a realistic protocol, Eve might gain some information
on the raw data:
she obtains some information on the transmitted data, as long as she
induces less errors than permitted\footnote{
We assume that the permitted error rate is reasonably small, 
but not extremely small
(say, around 1\%).};
furthermore,
she obtains also the error-correction data, since it is transmitted via 
a classical channel which is accessible to her.
Thus, the key common to Alice and Bob is insecure at this stage.
Actually, even in an error-free channels, and even when no errors are found
at the error-estimation stage, the error correction is still required;
for instance, Eve could eavesdrop only on two qubits hence be unnoticed
with large probability.

To overcome these problems, privacy amplification techniques~\cite{BBR,BBCM}
were suggested.  
The simplest one
uses the parity bit of the full string as the secret bit,
so that the final key contains $L=1$ bits.
Such techniques are designed to reduce Eve's information on the final key to be
exponentially small with the length of the initial string.
Currently, they are shown to work only against very restricted attacks. 

Eve can measure some of the particles and gain a lot of information on them,
but this induces a lot of error. Hence, if she performs such an attack,
she must restrict 
herself to attack only small portion of the particles, and thus, 
her information
on the parity of many bits is reduced to zero.
Let us explain this in more details: 
When Eve measures only a small portion of the particles, the probability
that no other bit will be used for creating the final key
is exponentially small.
If, to avoid inducing too much noise, she measures less 
particles than used for the final string
her information on the parity of this string is zero. 
Alternatively, she can measure
all the particles. For instance, 
when she attacks the four-state scheme and measures all particles, each 
particle in one of the bases, she gains full information if she guessed
the basis correctly for all bits which shall be used for the final key.
However, the probability of success is exponentially small
with the length of the final key. 
Furthermore, the probability that she induces less than 
the permitted error rate
when she attacks all particles is exponentially small as well.
Analysis of such attacks were done, for instance, in~\cite{HE} for
the four-state scheme, and in~\cite{EHPP} for the two-state scheme.

Eve can do much better than measure some of the particles --- she can obtain 
information on each bit without inducing too much noise.
Of course, in this case, she must induce only a small amount of 
noise on each bit.
Eve can achieve this task by performing an incomplete measurement:
She lets another particle interact with the particle sent by Alice,
and performs a measurement on that additional particle.
Such {\em translucent attacks} were defined in~\cite{EHPP}.
They are done using quantum gates and the additional particle is
called {\em a probe}.
The analysis in \cite{EHPP} considered the mutual information available
to the eavesdropper on a single transmitted particle,
and did not deal with 
the processes of error correction and privacy amplification.
Fortunately, it is rather clear \cite{BBCM} 
that privacy amplification will reduce Eve's information 
exponentially to zero also 
in this case, hence,
such an {\em individual} translucent attack is ineffective.

Let us show in detail that privacy amplification can reduce
Eve's information on the final key (the parity bit of the string)
to be exponentially small, when a particular translucent attack is used.
Let us consider
the ``translucent attack without entanglement'' of \cite{EHPP}, 
which is applied onto the two-state scheme, and 
which leaves Eve with probes in a pure state. 
It uses a two-dimensional probe in an initial state ${1 \choose 0}$,
and transforms the states $\phi_0$ and $\phi_1$ 
(defined in Section~\ref{int.qua-cry} in the introduction) so that  
\begin{equation} {1 \choose 0}
                 {\cos \theta \choose \pm \sin \theta} \longrightarrow
                 {\cos \alpha  \choose \pm \sin \alpha } 
                 {\cos \theta' \choose \pm \sin \theta'} \end{equation}
with `$+$' for $\phi_0$, and `$-$'  
for $\phi_1$. 
As a result, $\theta'$ is the angle of the states received
by Bob, and $\alpha$ is the angle of the states in Eve's hand. 
Using the basis 
\begin{equation}
{1 \choose 0}
{1 \choose 0} = 
\left(\begin{array}{c}
1 \\ 0 \\ 0 \\ 0
\end{array}\right) \ ; 
{1 \choose 0}
{0 \choose 1} = 
\left(\begin{array}{c}
0 \\ 1 \\ 0 \\ 0
\end{array}\right) \ ; 
{0 \choose 1}
{1 \choose 0} = 
\left(\begin{array}{c}
0 \\ 0 \\ 1 \\ 0
\end{array}\right) \ ; 
{0 \choose 1}
{0 \choose 1} = 
\left(\begin{array}{c}
0 \\ 0 \\ 0 \\ 1
\end{array}\right) \ , 
\end{equation}
the transformation can be written as 
\begin{equation}
\left(\begin{array}{cccc}
 \frac{c_\alpha c_{\theta'}}{c_\theta} &     0                              &
                      0               &-\frac{s_\alpha s_{\theta'}}{c_\theta} \\
                      0               & \frac{c_\alpha s_{\theta'}}{s_\theta} &
-\frac{s_\alpha c_{\theta'}}{s_\theta} &     0                              \\
                      0               & \frac{s_\alpha c_{\theta'}}{s_\theta} &
 \frac{c_\alpha s_{\theta'}}{s_\theta} &     0                              \\
 \frac{s_\alpha s_{\theta'}}{c_\theta} &     0                   &
                      0               & \frac{c_\alpha c_{\theta'}}{c_\theta}
\end{array}\right) \ , 
\label{EHPPtrans} \end{equation}
with $c_\theta = \cos \theta$ etc.

The error rate is defined as the probability of wrong identification;
The probability that Bob measured $\phi_0'={\sin\theta \choose -\cos\theta}$  
(which is identified as $\phi_1$) when Alice sent $\phi_0$ 
is 
\begin{equation} p_e = {\rm Tr} \left[  \left(
\begin{array}{cc} {c_{\theta'}}^2 & c_{\theta'} s_{\theta'} \\
               c_{\theta'} s_{\theta'} &  {s_{\theta'}}^2
\end{array}\right)  \left(
\begin{array}{cc} {s_{\theta}}^2 & -c_{\theta} s_{\theta} \\
               -c_{\theta} s_{\theta} &  {c_{\theta}}^2
\end{array}\right)
\right] = \sin^2 (\theta - \theta') \ . \end{equation}
The connection between this induced error rate and the angle $\alpha$
is calculated using the unitarity condition~\cite{EHPP} (preservation
of the overlap)  
\begin{equation} \cos 2 \theta = \cos 2 \theta' \cos 2 \alpha \ .
\end{equation}  

In order to analyze the effect of privacy amplification, 
suppose that Eve performs this translucent attack  
on all the bits (with identical transformation for each bit). 
As result, she gets  
$n$ probes (for the $n$ bits left from the raw data
after error estimation), each in one of the two states 
$\psi_0 = {c_\alpha \choose  s_\alpha}$ or  
$\psi_1 = {c_\alpha \choose -s_\alpha}$.    
Since she attacks all the bits, she must attack them
weakly, so that $p_e$ and $\alpha$ are small.
For weak attacks which cause small error rate 
we use the approximation 
$1/ \cos 2\alpha \approx
1 + 2 \alpha^2 +O(\alpha^4)$  
in the unitarity condition, to get
$ 2\theta' \approx \arccos (\cos 2 \theta [1 + 2 \alpha^2 +O(\alpha^4)]) 
\approx 2\theta - 2 \alpha^2
\cot 2 \theta + O(\alpha^4)$. 
Thus, the error rate is 
$p_e = \sin^2(\theta - \theta') \approx \alpha^4 \cot^2 2\theta + O(\alpha^6)$.
The angle of Eve's probe satisfies 
\begin{equation}
\alpha = (p_e \tan^2 2\theta )^{1/4} \ ,
\end{equation}
and from this expression we shall obtain the information on the parity bit
as a function
of the error rate.
It is important to note\footnote{ 
We shall see in Chapter~\ref{col} that Alice and Bob can use other states
(in addition to the two states of the protocol) which are much more
sensitive to this attack, so that $\alpha \approx {p_e}^{1/2}$ for a 
slightly modified scheme.
However, this shall become relevant only when we compare
the quality of different possible
attacks.}
that $\alpha = O[(p_e)^{1/4}]$ since other attacks usually give
$\alpha= O[(p_e)^{1/2}]$.

We have seen in the introduction (Section~\ref{int.qua-inf}) that
for these two possible pure states of each probe,
$\psi_0$ and $\psi_1$ (called there, $\uu$ and $\vv$), 
a standard measurement in an orthogonal basis symmetric 
to the two states 
optimizes the mutual information.
The angle between one basis vector 
and these polarization state is
$\frac{\pi}{4} \pm \alpha$. The measurement results in an error 
with probability
\beq  Q_e  
= \frac{1 - \sin(2 \alpha)}{2} \label{one-part} \ ,  \eeq
and with the same error probability for both inputs, 
thus, leading to a binary symmetric channel (BSC).
The optimal information of such a channel is
$I_{{}_{BSC}} = I_2(Q_e) $, but we care only about the
fact that the mean error probability is also optimized 
(minimized) by this measurement.
We denote this error-probability by $r\equiv Q_e$ for convenience.
Let us calculate Eve's information on the parity bit by calculating
the probability that she guesses it correctly.
(Another approach could use the ``collision entropy'' on each
bit~\cite{BBCM} to find the optimal mutual information on the parity bit).
The probability of deriving the wrong parity bit is equal to the
probability of having an odd number of errors on the individual
probes 
\bara  Q_e^{(n)}  =
   \sum_{j = odd}^n {n \choose j} r^j (1-r)^{n-j} \ .\nonumber \ear
To perform the sum only over odd $j$'s we use the formulae
\bara  (p+q)^n = 
\sum_{j = 0}^n {n \choose j} p^{n-j} q^j \ \ {\rm and}\
\ (p-q)^n = 
\sum_{j = 0}^n {n \choose j} p^{n-j} (-q)^j \ ,\nonumber \ear
to derive 
\beq 
\sum_{j = odd}^n {n \choose j} p^{n-j} q^j = 
\frac{(p+q)^n - (p-q)^n}{2} 
\ . \eeq
Assigning $q=r$ and $p=1-r$ we get
\beq   Q_e^{(n)}  = \sum_{j = odd}^n {n \choose j} r^j (1-r)^{n-j}
= \frac{1^n - (1 - 2r)^n}{2} = \frac{1}{2} - \frac{(1-2r)^n}{2}  
       \label{n-part} \ , \eeq
and finally
\beq   Q_e^{(n)} = \frac{1}{2} - \frac{\sin^n(2\alpha)}{2}
\label{n-errors}   \eeq
using $1 - 2 r = \sin 2 \alpha$.
The mutual information $I_S$ in this single-particle measurement is
thus \beq  
I_S = I_2(Q_e^{(n)}) = I_2 \left(
\frac{1}{2} - \frac{\sin^n(2\alpha)}{2} \right)
\ . \eeq
This is the optimal information obtained by a single particle measurement,
performed on $n$ particles each in either of the states $\psi_0$ and $\psi_1$,
and it was calculated by us in~\cite{BMS}.

A lot of useless side-information is obtained by this measurement 
(e.g., on the
individual bits). This fact indicates that Eve might be able to do
much better by concentrating on deriving only useful information.
This will be shown explicitly in Chapter~\ref{par}.

In case of small angle $\alpha$, 
let us calculate the 
optimal information obtained by individual measurements.  In that 
case, equation (\ref{n-errors}) yields 
\beq Q_e^{(n)}  \approx \frac{1}{2} - \frac{(2 \alpha)^n}{2} \ .\eeq
For small $\eta$ the logarithmic function is approximated by
\beq 
\log(\frac{1}{2} \pm \eta ) = \frac{\ln(\frac{1}{2} \pm \eta)}{\ln 2}
\approx
 -1 \pm \frac{2}{\ln 2} \eta - \frac{2}{\ln 2} \eta^2 \ , \eeq
from which the mutual information  
\bara I_2 (\frac{1}{2} - \eta) &=& 1 - H(\frac{1}{2} - \eta) = 1 + 
                (\frac{1}{2} + \eta)  \log(\frac{1}{2} + \eta) +
       (\frac{1}{2} - \eta)  \log(\frac{1}{2} - \eta) \nonumber \\ 
           &\approx& \frac{2}{\ln 2} \eta^2 \label{Iapprox}  \ear
is obtained.  Using this result and assigning $\eta = (2\alpha)^n/2$,
the information (to first order)
obtained by the optimal single-particle measurement is
\beq I_S = \frac{2}{\ln 2} \frac{(2 \alpha)^{2n}}{4}
= \frac{(2 \alpha)^{2n}}{2 \ln 2}. \label{Iindi} \eeq
We see that it is exponentially small with the length of the string
$n$.
This result is the maximal information Eve can obtain 
if she has  
$n$ particles, each in either of 
the states $\psi_0$ and $\psi_1$,
and she uses single-particle attack.

Using $\alpha = (p_e \tan^2 2\theta )^{1/4}$
we finally get the information as a function of the error rate 
\beq I_S = O \left( 
(16 p_e \tan^2 2\theta )^{n/2}
\right) \ . \end{equation}

\section{Can a `Quantum' Eavesdropper Learn the Key?}\label{securB1}

Eve can do much more than an individual attack if she keeps the 
probes in a quantum memory!
At a later time, 
she can attack the {\em final key}
directly using both 
the classical information regarding the error correction 
and privacy amplification, and the quantum  
information contained in the quantum states of her probes.
In such a case, she might be able to gain only useful information regarding 
the desired parity bits. 
Privacy amplification techniques were not designed to stand against such 
attacks, hence their efficiency against them is unknown.

The most general attack allowed by the rules of quantum mechanics
is the following:
Eve lets all 
particles pass through a huge probe (a probe which lives in
a very large Hilbert space) 
and performs transformations
which create correlations between the state of the particles and the 
state of the probe.
Eve waits till receiving all classical information (for instance: 
1. the relevant
bits; 2. the error-correction data; 3. the privacy amplification technique.) 
Eve measures the probe to obtain the optimal information on the final key 
using all classical data.
Such an attack is called a {\em coherent}
or a {\em joint} attack~\cite{BBBSS}.
(We shall use the term {\em joint} attack in this thesis,
and keep the term coherent measurement for describing measurements
which are done in some entangled basis).
Joint attacks are the most general attacks 
consistent with the rules of quantum mechanics.
Doing several measurements in the middle can always be simulated by that
system as well, by using {\em measurement gates}.   
The analysis of joint attacks is very complicated, and although security
against them is commonly believed, it is yet unproven~\footnote{
But see Section~\ref{securC}.}.
For instance, it could well be that Eve would obtain much more information
on parity bits by measuring all bits together. 
If that information is not reduced sufficiently
fast with the length of the string,
the known quantum key distribution becomes insecure! 

\section{The Collective Attack}\label{securB2}

In this thesis we suggest a new approach to the security problem.
It started as a practical question: if we cannot prove security against 
the most general attack, what type of restricted
security proofs can we achieve? 
It continued as a very promising direction for approaching
the security problem.
The results we present here strongly suggest 
that quantum key distribution is secure.
We hope that the steps which still remain open will be soon solved
as well.

Consider the following attack
which we call~\cite{BM1} the {\em collective} attack:
\begin{enumerate}
\item Eve attaches a \underline{separate, uncorrelated} 
probe to each transmitted
particle using a translucent attack.
\item Eve keeps the probes in a quantum memory till 
receiving {\em all} classical 
data including error-correction and privacy amplification data.
\item Eve performs the optimal measurement on her probes in order to learn
the maximal information on the {\em final} key.
\end{enumerate}
The case in which Eve attaches 
one (large-Hilbert-space) probe to all transmitted particles  
is the joint attack.
No specific joint attacks were  
yet suggested; the collective attack defined above 
is the strongest 
joint attack suggested so far,
and there are good reasons
to believe that it 
is the strongest possible joint attack. 
If this is indeed the case, then
security against it establishes an ultimate security.

In order to devise a proof of security against collective attacks
we shall provide a solution for an example based on
the ``translucent attack without entanglement'' of \cite{EHPP}, 
which leaves Eve with probes in a pure state, and prove security 
against it.
As before we assume that this translucent attack is 
performed on all the bits. After throwing inconclusive results,
and after estimating the error rate, Alice and Bob are left
with $n$ bits.
As a result, Eve holds an $n$-qubit state which corresponds to 
Alice's string $x$.
For simplicity, we choose the final key to consist of one bit, 
which is the parity of the
$n$ bits.
Eve wants to distinguish between two density matrices corresponding
to the two possible values of this parity bit.
Our first aim is to calculate the optimal mutual information she can 
extract from them. 
A priori, all strings are equally probable 
and Eve needs to distinguish between the two density matrices
describing the parities. In the next chapter 
we show that the optimal information is obtained by a coherent measurement
and not by individual measurement.
We calculate this information and we prove that it is exponentially
small. Thus, we provide the first proof of efficiency of
privacy amplification 
against coherent measurements.
In real protocols an error correction must be done.
Since Eve is being told what 
the error-correction code is,
all strings consistent with the given error-correction 
code (the given $r$ subparities) 
are equally probable, while other strings are excluded.
Full analysis of this case is given in Chapter~\ref{col},
leading to a proof that privacy amplification is still effective when 
the error-correction data is available to Eve.
In Chapter~\ref{col} we also
extend this result to various other collective attacks, 
strongly suggesting security against any collective attack. 

\section{Collective Attacks and the Ultimate Security}\label{securB3}

Our approach solves interesting problems regarding security against
collective attacks, but the 
more puzzling question regarding our approach is whether
it can also provide a route to the proof of the ultimate security.
Unfortunately, the proof is not available yet, 
but we have intuitive arguments suggesting that it exists, based on a 
{\em randomization argument}.

The randomization procedure is a very important step in any
protocol for quantum key distribution.
Bob chooses randomly the basis in which he performs his measurements
(in either of the schemes).
Alice and Bob choose randomly the qubits which shall be used and thrown away
in the error-estimation step. Also, they randomize the bits before choosing
any error-correction code, and they choose randomly which parity bits 
will form the final key. 
At the time Eve holds the transmitted particles she has no knowledge of the 
relevant qubits for each of these steps.
A correct guess would surely help her, putting her in a position more 
similar to Bob's. 
However, the probability of guessing each of these random steps
(i.e., Bob's basis, the bits left after
error estimation, the error-correction code for a randomized string and 
the substrings used for privacy amplification)
is exponentially small with the length of the string.
Hence, the average information which result from such a guess is
exponentially small.

Thus, we conjecture that 
she cannot gain information by searching or by creating correlations between 
the transmitted particles; it is better for her to keep 
one separate probe for each 
particle, and to perform 
the measurements after obtaining the missing 
information as is done in the collective attacks. 
Any attempt of creating such coherent correlations at the first
step of the attack induces errors, while it cannot lead 
to a noticeable increase in the resulting average information; 
Searching for exponentially many such correlations could help Eve,
but it would also cause an appropriate induced error rate.

Unfortunately, proving this intuitive argument is yet an open problem.
 
A simplified approach to prove this can 
be based on providing Eve with most
of the missing details by choosing in advance (and telling her)
all relevant data
(the bits used for error estimation, 
the error correction
code, and 
the privacy amplification code)  
except for the basis used by Alice and Bob.
Furthermore, we can also let her have all 
the particles together. 
Despite all this,
as long as she does not know the basis used by Alice and Bob,
she cannot know which bits will form the raw data,
and the collective attack
will probably be her best attack.

\section{An Alternative Approach}\label{securC}

As we were working on this research, 
Yao~\cite{Yao} and Mayers~\cite{DM2} suggested another
approach to the security problem.

Yao~\cite{Yao} claims to prove security of OT (assuming secure BC)
in case of error-free channel,
and this proof applies to key distribution 
(even if secure BC does not
exist)
based on reduction from OT, suggested by Mayers~\cite{DM3}.
This important result is certainly not surprising.
It translates the ``no clean imprint'' theorem to a calculation,
which takes error correction and privacy amplification into account:
if Eve is not permitted to induce errors, she can gain only exponentially
small information.
The main achievement of this proof is that it does not limit Eve in
any way, thus it is the first work which considers a quantum eavesdropper.
It considers the quantum state Eve obtains versus the quantum state
she sends further to Bob, 
and not the density matrices
available to Eve.
Since the state obtained by Bob must be very 
similar to the one sent by Alice, Eve gain only exponentially small
amount of information.

Going to real channels the situation is different, since Eve can certainly
gain information on each transmitted particle. 
When Eve attacks in a way which induces constant (but permitted) error
rate, the probability that she will be noticed does not approaches one
in the limit of large strings, but remains zero.
In this case, we believe that 
it does not suffice to consider the difference between
Bob's state and 
Alice's state, since it shall not be negligible anymore.
Instead, one have to investigate Eve's 
density matrices as a function of that difference.
Surprisingly, 
Mayers claims~\cite{DM2} that it is possible 
to extend Yao's result to real channels, 
without analyzing Eve's density matrices.
We do not understand Mayers' proof well enough and at the moment
we believe that it cannot
be correct. 
It is important to state that various security proofs claimed in the
past were later on found to be wrong (e.g., to be based on a hidden
assumption).
Mayers' result is not yet accepted by the general community, and
furthermore, it was not published in a refereed journal. 
Till this claim is clarified, 
we must assume that the problem is still
open. 

Even if Mayers' proof will be found correct and complete
in the future (or will lead
to a complete proof), our approach might still be crucial:
it is possible that their approach is limited to deal with extremely
small error rates only, or that it requires much longer strings compared 
to our approach.

\chapter{The Parity Bit}\label{par}

A major question in quantum information theory 
\cite{Kholevo2,Davies,Helstrom,Per93,Levitin,Fuchs}
is ``how well can two
quantum states, or more generally, two density matrices $\rho_0$ 
and $\rho_1$, be distinguished?''
In terms of a communication scheme this question is 
translated to an identification task: 
A sender (Alice) sends a bit $b=i$ ($i=0;1$) 
to the receiver (Bob) by sending the 
quantum state $\rho_i$, and the receiver does his best to identify
the value of the bit, i.e.,~the quantum state.
The two-dimensional Hilbert space $\cH2$ is usually used to 
implement such a binary channel. The transmitted signals can be 
polarization states 
of photons, spin-states of spin-half particles, etc.
The transmitted states may be pure states or density
matrices, and need not be orthogonal.
Usually, the mutual information $I$ is used to describe 
distinguishability, 
such that $I=0$ means indistinguishable, and $I=1$ 
(for a binary channel) means perfect distinguishability.
The ensemble of signals is agreed on in advance, and  
the goal of Alice and Bob is to optimize the average mutual 
information over the different possible measurements at the 
receiving end. For example, 
two orthogonal pure states transmitted through an error-free
channel are perfectly distinguishable; The optimal mutual 
information ($I=1$) 
is obtained if Bob measures in an appropriate basis.
Finding the optimal mutual information is still an open 
question for most ensembles.
Some cases with known analytic solutions are the case of two pure 
states, the case of two density matrices in two dimensions with
equal determinants~\cite{Levitin,Fuchs} and the case of commuting
density matrices~\cite{BCJL}. 
There are no known analytic solutions for two non-trivial density
matrices in dimensions higher than two.
In this chapter we find a solvable case which has very important
implications to quantum cryptography.
This part of the work was done together with Charles Bennett and John 
Smolin~\cite{BMS}, and with the help of Asher Peres.

Suppose that a source produces binary string $x$ of length $n$ with equal 
and independent probabilities for all the digits.
Let the string be encoded into a quantum-mechanical channel, in which the 
bits `0' and `1' are represented by quantum states (density matrices)
$\rho_0$ and $\rho_1$ of independent two-state quantum systems.
These can be either pure states
or density matrices with equal determinants.
Suppose Bob wants to learn 
the parity bit (exclusive-OR) of the $n$-bit string 
and not the specific value of each bit.
The parity bit 
is described by one of two density matrices 
$\rho_0^{(n)} $
and  
$\rho_1^{(n)} $ which lie in a $2^n$-dimensional Hilbert 
space $\cH{2^n}$.  
These parity density matrices $\rho_p^{(n)}$ are the average density matrices,
where the average is taken over all strings the source might produce, which 
have the same parity $p$.
Since the parity bit of the source string $x$ is encoded by $\rho_p^{(n)}$, 
information about which of the two density matrices was prepared is 
information about the parity of $x$.

Let $x$ be any classical string of $n$ such bits, and  
$\rho_x = \rho_{(1^{st} \ {\rm bit})} \ldots  
          \rho_{(n^{th} \ {\rm bit})}$  
be the density
matrix made up of the tensor product of the signaling 
states $\rho_{(i)}$ corresponding to the $i^{th}$ bit of $x$.
Formally, we distinguish between the two density matrices:
\begin{equation}
\rho_0^{(n)}=\frac{1}{2^{n-1}}\sum_{x\,|\,p(x)=0}\!\!\!\!\!\rho_x\ 
\quad {\rm and} \quad
\ \rho_1^{(n)}=\frac{1}{2^{n-1}}\sum_{x\,|\,p(x)=1}\!\!
\!\!\!\rho_x \ ,
\end{equation}
where the sum is over all possible strings with the same parity
[each sent with equal probability $(1/2^n)$]
and $p(x)$ is the parity function of $x$.
We show a simple way to write the parity density matrices.
We find that they are optimally distinguished by a non-factorizable coherent
measurement, performed on the composite $2^n$ dimensional quantum system, 
and we calculate the optimal mutual information which can be 
obtained on the parity bit.

We concentrate on the special case where the two
signaling states have large overlap, which is important in the analysis
of the security of quantum key distribution against powerful 
multi-particle eavesdropping attacks.
We show that the optimal obtainable information decreases exponentially 
with the length $n$ of the string.
This result provides a clue that classical privacy amplification 
is effective against coherent measurements, limiting the ability of 
an eavesdropper to obtain only negligible information on the final key.

The first sections deal only with the case where each bit is encoded
by a pure state.
In Section \ref{par1} we find a simple way to write the density 
matrices of the parity bit for any $n$ when the signaling states
are pure; We show that the parity matrices can be 
written in a block diagonal form and explain the importance of 
this fact. In Section \ref{par2} we investigate the 
distinguishability of the parity
matrices; The optimal
measurement that distinguishes them is found to be a standard
(von Neumann) measurement in an entangled
basis (a generalization of the Bell basis of two particles);
We calculate exactly 
the optimal mutual information 
on the parity bit (derived by performing the optimal measurement). 
In Section \ref{par3} we obtain the main result of this
chapter: for two
almost fully overlapping states, the optimal mutual information $I_M$
decreases exponentially with the length of the string.  
While exponentially small, this optimal information is nevertheless
considerably greater than the information that would have been 
obtained by measuring each bit separately and classically combining 
the results of these measurements. 
We are also able to calculate the maximal deterministic (conclusive) 
information
on the parity matrices obtained in Section \ref{par1};
This is done in Section \ref{par4} where we also confirm a result 
previously obtained by Huttner and Peres \cite{HP} for two bits.
In Section \ref{par5} we repeat the calculation
of the optimal mutual information for the more general case where 
the bits are represented by non-pure states (in $\cH2$ and 
with equal determinants).
In Section \ref{par6} we summarize the conclusions of this chapter.
In the next chapter we study  
the implications of these
results to the security of quantum cryptography. 

\section{Density Matrices for Parity Bits}\label{par1}

Let Alice send $n$ bits.  The possible values of a single bit
($0$ or $1$) are represented by 
\beq   
\psi_0={\cos\alpha\choose\sin\alpha} \quad \ {\rm and } \ \quad 
\psi_1={\cos\alpha\choose -\sin\alpha} \label{pure-state}\eeq
respectively.
In terms of density matrices these are:
\beq \rho_0^{(1)} = 
  \left(\begin{array}{rr}   c^2 &  sc  \\  sc &  s^2     
  \end{array}\right) 
\quad \quad {\rm and } \quad \quad 
 \rho_1^{(1)} = 
  \left(\begin{array}{cc}   c^2 &  -sc  \\  -sc &  s^2     
  \end{array}\right),\label{rho^1} \eeq
where we use a shorter notation 
$s \equiv \sin\alpha$; $c \equiv \cos\alpha$,
for convenience, and the superscript $[]^{(1)}$ is explained in the 
following paragraph.

The parity bit of an $n$-bit string is the exclusive-OR of all 
the bits in the string. In other words, the parity is 1 if there 
are an odd number of
1's and 0 if there are an even number.
The parity density matrices of $n$ bits will be denoted as
$\rho_0^{(n)}$ and $\rho_1^{(n)}$ in case the parity is 
`0' and `1' respectively.
Using these density matrices we define also the {\em total} density 
matrix
$\rho^{(n)} \equiv  \frac{1}{2} (\rho_0^{(n)} + \rho_1^{(n)})$ 
and the {\em difference} density matrix
$\Delta^{(n)} \equiv \frac{1}{2} (\rho_0^{(n)} - \rho_1^{(n)})$, so 
that 
\beq \rho_0^{(n)} = \rho^{(n)} + \Delta^{(n)} 
 \quad \quad {\rm and } \quad \quad 
\rho_1^{(n)} = \rho^{(n)} - \Delta^{(n)} \ . \label{rho^n} \eeq
 
The one-particle density matrices 
(equation \ref{rho^1}) also describe the parities of one 
particle, 
and therefore we can calculate 
\beq \rho^{(1)} = \frac{1}{2} (\rho_0^{(1)} + \rho_1^{(1)}) = 
  \left(\begin{array}{cc}   c^2 &  0  \\  0 &  s^2     
  \end{array}\right),\label{rho} \eeq
\beq \Delta^{(1)} = \frac{1}{2} (\rho_0^{(1)} - \rho_1^{(1)}) = 
  \left(\begin{array}{cc}   0 &  sc  \\  sc &  0     
  \end{array}\right) \label{Delta} \ . \eeq
The density matrices of the parity bit
of two particles are:
\bara \rho_0^{(2)} &=& \frac{1}{2} (\rho_0^{(1)} \rho_0^{(1)} + 
                     \rho_1^{(1)} \rho_1^{(1)})  \nonumber \\ 
     \rho_1^{(2)} &=& \frac{1}{2} (\rho_0^{(1)} \rho_1^{(1)} + 
                             \rho_1^{(1)} \rho_0^{(1)})  \ear
where the multiplication is a tensor product.
The total density matrix is 
\bara 
\rho^{(2)} &=& \frac{1}{2} (\rho_0^{(2)} + \rho_1^{(2)}) \nonumber \\
      &=& \frac{1}{4} [\rho_0^{(1)} (\rho_0^{(1)} + \rho_1^{(1)})
       + \rho_1^{(1)} (\rho_1^{(1)} + \rho_0^{(1)})] \nonumber \\
      &=& \rho^{(1)} \rho^{(1)} \ , \nonumber \ear 
which, by using the basis  
\bara |b_0\rangle \equiv {1 \choose 0}_1 
      {1 \choose 0}_2 = \left(  \begin{array}{c} 
                    1 \\ 0 \\ 0 \\ 0 \end{array} \right)
  \ ; \quad  |b_1\rangle \equiv 
      {1 \choose 0}_1 
      {0 \choose 1}_2 = \left(  \begin{array}{c} 
                    0 \\ 1 \\ 0 \\ 0 \end{array} \right)
  \ ; \nonumber \\ |b_2\rangle \equiv
      {0 \choose 1}_1 
      {1 \choose 0}_2 = \left(  \begin{array}{c} 
                    0 \\ 0 \\ 1 \\ 0 \end{array} \right)
  \quad   {\rm and} \quad |b_3\rangle \equiv  
      {0 \choose 1}_1 
      {0 \choose 1}_2 = \left(  \begin{array}{c} 
                    0 \\ 0 \\ 0 \\ 1 \end{array} \right)
\label{basis}
\ear   
in $\cH4$, can be written as 
\bara \rho^{(2)} = 
       \rho^{(1)} \rho^{(1)} =  
  \left(\begin{array}{cccc}    c^4   &    0    &    0    &    0   \\ 
                                0    & c^2 s^2 &    0    &    0   \\ 
                                0    &    0    & c^2 s^2 &    0   \\ 
                                0    &    0    &    0    &   s^4  \\ 
  \end{array}\right). \ear
The difference density matrix is
\bara 
\Delta^{(2)} &=& \frac{1}{2} 
(\rho_0^{(2)} - \rho_1^{(2)}) \nonumber \\
      &=& \frac{1}{4} [\rho_0^{(1)} (\rho_0^{(1)} - \rho_1^{(1)})
 + \rho_1^{(1)} (\rho_1^{(1)} - \rho_0^{(1)}) ] \nonumber \\
    &=& \Delta^{(1)} \Delta^{(1)} = 
\left(\begin{array}{cccc}     0    &    0    &    0    & c^2 s^2 \\ 
                              0    &    0    & c^2 s^2 &    0    \\ 
                              0    & c^2 s^2 &    0    &    0    \\ 
                           c^2 s^2 &    0    &    0    &    0    \\ 
\end{array} \right) \ . \ear 

The density matrices of the parity bit
of $n$ particles can be written recursively:
\bara \rho_0^{(n)} = \frac{1}{2} (\rho_0^{(1)} \rho_0^{(n-1)} + 
                          \rho_1^{(1)} \rho_1^{(n-1)})  \nonumber \\
      \rho_1^{(n)} = \frac{1}{2} (\rho_0^{(1)} \rho_1^{(n-1)} + 
                             \rho_1^{(1)} \rho_0^{(n-1)}) \ , \ear
leading to 
\beq \rho^{(n)} = \frac{1}{2} (\rho_0^{(n)} + \rho_1^{(n)}) 
      = \rho^{(1)} \rho^{(n-1)}   \ , \nonumber  \eeq
and
\beq \Delta^{(n)} = \frac{1}{2} (\rho_0^{(n)} - \rho_1^{(n)})
      = \Delta^{(1)}  \Delta^{(n-1)}   \ .\nonumber   \eeq
Using these expressions recursively we get
\beq \rho^{(n)} = 
       (\rho^{(1)})^n   \label{rho-n} \eeq
which is diagonal, and
\beq \Delta^{(n)} = 
      (\Delta^{(1)})^n      \label{Delta-n} \eeq
which has non-zero terms only in the secondary diagonal.
The density matrices  
$  \rho_0^{(n)}$
and $\rho_1^{(n)}$ are now immediately derived for any $n$
using equation (\ref{rho^n}):
\bara \rho_0^{(n)} = 
       (\rho^{(1)})^n +
      (\Delta^{(1)})^n      \nonumber  \ear
and
\bara 
     \rho_1^{(n)} = 
       (\rho^{(1)})^n -
      (\Delta^{(1)})^n    \ . \label{rho_p^n} \ear
As an illustrative example we write $\rho_0$ and $\rho_1$ for two
particles: 
\bara \rho_0^{(2)} = 
 \left(\begin{array}{cccc}    c^4   &    0    &    0    & c^2 s^2 \\ 
                               0    & c^2 s^2 & c^2 s^2 &    0    \\ 
                               0    & c^2 s^2 & c^2 s^2 &    0    \\ 
                            c^2 s^2 &    0    &    0    &   s^4   \\
  \end{array}\right) \ ; 
      \rho_1^{(2)} = 
 \left(\begin{array}{cccc}    c^4   &    0    &    0    &-c^2 s^2 \\ 
                               0    & c^2 s^2 &-c^2 s^2 &    0    \\ 
                               0    &-c^2 s^2 & c^2 s^2 &    0    \\ 
                           -c^2 s^2 &    0    &    0    &   s^4   \\
  \end{array}\right) \ . 
\ear
The only non-zero terms in the parity density matrices are the 
terms in the diagonals for any $n$, thus the parity density matrices
have an {\sf X}-shape in that basis.

\medbreak

The basis vectors can be permuted  
to yield block-diagonal matrices built of $2\times 2$  blocks. 
The original basis vectors [see, for example, equation~(\ref{basis})], 
$|b_i\rangle$, are simply $2^n$-vectors where 
the $i^{\rm th}$ element of 
the $i^{\rm th}$ basis vector 
is $1$ and all other elements are $0$ ($i$ ranges from $0$ 
to $2^n-1$).
The new basis vectors are related to the old as follows:  
\bara
|b'_i\rangle = |b_{i/2}\rangle\ {\rm for\ even}\ i \quad
{\rm and}\ |b'_i\rangle=|b_{2^n-(i+1)/2}\rangle\ {\rm for\ odd}\ i
\ \ .\ear
The parity density matrices are now, in the new basis (we omit 
the $'$ from
now on as we will never write the matrices in the original basis):
\beq \rho_p^{(n)} = 
      \left(  \begin{array}{cccc} 
         B_p^{[j=1]} & 0 & \ldots & 0 \\
           0 & B_p^{[j=2]} & \ldots & 0 \\
           0 & 0 & \ldots & B_p^{[j=2^{(n-1)}]} 
                                                \end{array} \right)
\label{block-diag} \eeq
where the subscript $p$ stands for the parity ($0$ or $1$). Each of 
the 2$\times$2 matrices has the form 
\beq   B_p^{[j]} = \left(\begin{array}{cc}   
         c^{2(n-k)} s^{2k}    &  \pm c^n s^n             \\ 
         \pm c^n s^n          &  c^{2k} s^{2(n-k)} 
  \end{array}\right), \label{Bpj} \eeq
with the plus sign for $p=0$ and the minus sign for $p=1$, and
$0 \le k \le n$, 
and all these density matrices satisfy
$ \Det B_p^{[j]} = 0 $. 
The first block ($j=1$) has $k=0$; there are 
${n \choose 1}$ blocks which have  
$k=1$ or $k=n-1$; there are ${n \choose 2}$ $j$'s which 
have $k=2$ or $k=n-2$, etc.  
This continues until $k=(n-1)/2$ for odd $n$.  For even $n$ the 
process continues up to $k=n/2$ with the minor adjustment that 
there are 
only $\frac{1}{2}{n \choose n/2}$ $j$'s of $k=n/2$.  This enumeration
groups blocks which are identical or identical after interchange
of $k$ and $n-k$ and accounts for all $2^n/2$ blocks.  We will see 
later that blocks identical under interchange of  $k$ and $n-k$ will
contribute the same mutual information about the parity bit, thus 
we have grouped them together.

With the density matrices written in such a block-diagonal form 
of $2\times 2$ blocks the problem of finding the optimal mutual information
can be analytically solved.  It separates into two parts:
\begin{itemize}
\item Determining in which of $2^n/2$ orthogonal 2d subspaces
(each corresponding to one of the $2\times 2$ blocks) the system lies. 
\item Performing the optimal measurement within that subspace.
\end{itemize}
The subspaces may be thought of as $2^n/2$ parallel channels,
one of which is probabilistically chosen and used to encode the 
parity
by means of a choice between two equiprobable pure states within that
subspace (these two states are pure because the $B_0$ and $B_1$ 
matrices each
have zero determinant).  We shall present in the next section the 
optimal
measurement that yields 
the optimal mutual information transmissible
through such a two-pure-state quantum channel. The channel then 
corresponds to a classical binary symmetric channel
(BSC), i.e.,~a classical one-bit-in one-bit-out channel whose output
differs from its input with some error probability $p_j$ independent
of whether the input was 0 or 1.  
The optimal mutual information in each subchannel is the optimal
mutual information
of a BSC with error
probability $p_j$ and
is $I_2(p_j)=1-H(p_j)$,  
with $H(x)=-x\log_2x-(1-x)\log_2(1-x)$,    
the Shannon
entropy function. 
The optimal mutual information $I_M$ for
distinguishing $\rho_0^{(n)}$ from $\rho_1^{(n)}$ can thus be 
expressed
as an average over the optimal mutual information of the subchannels:
\beq I_M = \sum_{j=1}^{2^n/2} q_j I_2(p_j),\label{Imutual}
\eeq where $q_j=\Tr
B_0^{[j]}=\Tr B_1^{[j]}$ is the probability of choosing the $j$'th
subchannel.
The BSC error probability $p_j$ for the
$j$'th subchannel depends on the subchannel's $2\times 2$ renormalized 
density matrices $\hat{B}^{[j]}_p=B^{[j]}_p/q_j$, 
and is easily calculated once the optimal measurement is found.    
For each
subchannel the $q_j$ and renormalized $2\times 2$ matrices look like
\beq
q_j=c^{2(n-k)} s^{2k} + c^{2k} s^{2(n-k)} \label{qj}
\eeq
and
\beq
\hat{B}_p^{[j]}=
\left(\begin{array}{cc}   
\frac{c^{2(n-k)} s^{2k}} 
                        {c^{2(n-k)} s^{2k} + c^{2k} s^{2(n-k)}} &
\frac{\pm c^n s^n} 
                  {c^{2(n-k)} s^{2k} + c^{2k} s^{2(n-k)})} \\
\vspace{0.2cm} 
\frac{\pm c^n s^n} 
                  {c^{2(n-k)} s^{2k} + c^{2k} s^{2(n-k)})} &
\frac{c^{2k} s^{2(n-k)}} 
                        {c^{2(n-k)} s^{2k} + c^{2k} s^{2(n-k)}} 
        \end{array}\right) \  \label{blocks} \ .
\eeq
Since they correspond to pure states we can find an angle 
$\gamma^{[j]}$ for each block such that the density matrices are
\begin{equation} \left(  \begin{array}{cc}
         \cos^2 \gamma^{[j]} &   \pm \cos_\gamma^{[j]} \sin \gamma^{[j]} \\
      \pm  \cos \gamma^{[j]} \sin \gamma^{[j]}   &  \sin^2 \gamma^{[j]}  
\end{array}\right)   \ . \label{pure-block} \end{equation}

In our previous example of $n=2$ the matrices are put in a block 
diagonal form:
\bara \rho_0^{(2)} = 
 \left(\begin{array}{cccc}    c^4   & c^2 s^2 &    0    &    0    \\ 
                            c^2 s^2 &   s^4   &    0    &    0    \\ 
                               0    &    0    & c^2 s^2 & c^2 s^2 \\ 
                               0    &    0    & c^2 s^2 & c^2 s^2 \\ 
  \end{array}\right) \ ; 
      \rho_1^{(2)} = 
 \left(\begin{array}{cccc}    c^4   &-c^2 s^2 &    0    &    0    \\ 
                           -c^2 s^2 &   s^4   &    0    &    0    \\ 
                               0    &    0    & c^2 s^2 &-c^2 s^2 \\ 
                               0    &    0    &-c^2 s^2 & c^2 s^2 \\
  \end{array}\right) \ , 
\ear
so that 
$q_{j=1} = c^4 + s^4$; $q_{j=2} = 2 c^2 s^2$; and 
\beq\label{bnormal} \hat{B}_p^{[j=1]} = 
\left(\begin{array}{cc}   
               \frac{c^4}{c^4+s^4} & \pm\frac{c^2 s^2}{c^4+s^4} \\ 
                   \pm \frac{c^2 s^2}{c^4+s^4} & \frac{s^4}{c^4+s^4}
       \end{array}\right) \ ;
\quad  \hat{B}_p^{[j=2]} = 
  \left(\begin{array}{cc}   1/2 & \pm 1/2  \\ 
                        \pm 1/2 &  1/2
      \end{array}\right) \ . \label{ex_n=2} \eeq

\section{Optimal Information in a Parity Bit}\label{par2}

Two pure states or two density matrices in $\cH2$ with equal 
determinants can always be written (in an appropriate basis) in 
the simple form
\beq\label{amatrix}  \rho_0 = \left(\begin{array}{cc}   
                      a_1   &  a_2   \\ 
                      a_2   &  a_3   
     \end{array} \right) \ ; \quad
       \rho_1 = \left(\begin{array}{cc}   
                      a_1   & -a_2   \\ 
                     -a_2   &  a_3   
     \end{array} \right) \eeq
with $a_i$ real positive numbers such that 
$\Tr \rho_p = a_1 + a_3 = 1$.
For two pure states, say 
\begin{equation}
\left( \begin{array}{cc}
{c_\gamma}^2         &   \pm c_\gamma s_\gamma \\
\pm c_\gamma s_\gamma &   {s_\gamma}^2 
\end{array} \right) \ ,  \end{equation} 
we already said 
that a standard measurement in an orthogonal basis symmetric 
to the two states 
optimizes the mutual information.
The density matrices of pure states  
can be written as 
$\rho_i = (\one + \sigma \cdot {\bf r_i} )/2$ with the $\sigma$ being
the Pauli matrices and ${\bf r} = (\pm \sin 2 \gamma , 0 , \cos 2
\gamma)$ being a three dimensional vector
which describes a spin direction.
Using this notation any density matrix is described by a point in a 
three dimensional unit ball, called the Poincar\'e sphere
(also called the Bloch sphere).
The pure states are points on the surface
of that sphere 
and the above two pure states have $x = \pm 2 a_2 = \pm \sin 2 \gamma$.
With the density matrix notation  
the optimal basis for distinguishing the states
is the $x$ basis. Note that all angles are doubled in this spin notation,
so that the angle between the two states is $4 \gamma$,
and the angle between the basis vector 
and the state is $\frac{\pi}{2} - 2 \gamma$.
The measurement of the two
projectors
\beq      A_{\rightarrow} =  1/2 \left(\begin{array}{cc}   
                      1   &  1  \\ 
                      1   &  1 
     \end{array} \right) \quad \quad {\rm and} \quad \quad  
       A_{\leftarrow} =   1/2 \left(\begin{array}{cc}   
                      1   &  -1  \\ 
                     -1   &   1 
     \end{array} \right) \eeq
yields 
\beq P_e = \Tr \rho_1   A_{\rightarrow}  = \frac{1}{2} - a_2  
= \frac{1}{2} - \frac{\sin 2 \gamma}{2}
\label{Err-Prob} \ , \eeq
which recovers the result of equation (\ref{one-part}) in case of 
pure states obtained in Chapter~\ref{secur}.
However, the treatment of density matrices is more general
and this is the optimal measurement also 
in the case of deriving block-diagonal matrices
for the parity of string of non-pure states with equal 
determinants \cite{Levitin,Fuchs},  
when $\rho_i$ of equation (\ref{rho^1}) 
are replaced by $\rho_i^{dm}$ of 
equation (\ref{rhod_0}) and (\ref{rhod_1})
of Section \ref{par5},
and this case is also described by a BSC.  
The only difference between the matrices 
is that $\Det \rho_p = 0$ for pure states
and $ 0 \le \Det \rho_p \le \frac{1}{4} $ for density matrices. 

Instead of measuring the density matrices in the $x$ direction 
we perform the following unitary transformation  
on the density matrices 
\beq U =   1/\sqrt2 \left(\begin{array}{cc}   
                      1   &   1  \\ 
                      1   &   -1 
     \end{array} \right) \eeq
to obtain $\rho'=U \rho U^\dagger$ which is then 
measured in the $z$ basis.  
Note that the transformation transform the original $z$-basis 
to $x$-basis
(the motivation for this approach will be understood when we 
discuss the $2\times 2$ blocks of the parity matrices).
The new density matrices are 
\beq {\rho_0}' =\left(\begin{array}{cc}   
           \frac{1}{2} + a_2  &         \frac{a_1 - a_3}{2}    \\ 
           \frac{a_1 - a_3}{2}      &   \frac{1}{2} - a_2 
     \end{array} \right)    \ ; \   
     {\rho_1}' =\left(\begin{array}{cc}   
           \frac{1}{2} - a_2  &         \frac{a_1 - a_3}{2}    \\ 
           \frac{a_1 - a_3}{2}      &   \frac{1}{2} + a_2 
     \end{array} \right)  \label{dm_tag}   
\eeq
and their measurement yields the probability
$ \frac{1}{2} \pm a_2  $
to derive the correct (plus) and the wrong (minus) answers 
(as we obtained before), leading to 
optimal mutual information of 
\beq I_2(\frac{1}{2} - a_2)\ \label{BSC-info} ,\eeq
which depends only on $a_2$.
Note that the same information is obtained in case 
$a_1$ and $a_3$ are interchanged.
Now, instead of considering the information as a function of
$\gamma$ we have it as a function of $a_2$, and we can 
apply it directly to the blocks $\hat{B}_p^{[j]}$ of the previous section.

Let us now consider the information one can obtain on the parity bit.
The naive way to derive information on a parity bit 
of $n$ particles [each in one of the states
${\cos \alpha \choose \pm \sin \alpha}$] is to derive the
optimal information on each particle separately and calculate the
information on the parity bit.  This individual (or {\em
single-particle} measurement) is the best Bob can do in case he
has no quantum memory in which to keep the particles (which, usually
arrive one at a time) or he has no ability to perform more advanced
coherent measurements.  The optimal error-probability for each particle
is $ Q_e = \frac{1 - \sin 2 \alpha}{2} $, and the 
probability of deriving the wrong parity bit, derived in the 
previous chapter~(eq.~\ref{n-errors}) is
$Q_e^{(n)}  = 
             \frac{1}{2} - \frac{(\sin 2\alpha)^n}{2}  $,
and the mutual information in this case was found to be 
\beq  
I_S = I_2 \left(
\frac{1}{2} - \frac{(\sin 2\alpha)^n}{2} \right)
\ .\eeq

As we previously said, the derivation of 
useless side-information on the
individual bits indicates that Bob might be able to do
much better by concentrating on deriving only useful information.
The optimal measurement for finding mutual information on the 
parity bit
is not a single-particle measurement, but is instead a measurement on
the full $2^n$-dimensional Hilbert space of the system.  In general,
optimizing over all possible measurement is a very difficult task
unless the two density matrices in $\cH{2^n}$ are pure states.
However, in the preceding section we have shown how to reduce the
problem to that of distinguishing the $2\times 2$ blocks of our
block-diagonal parity matrices.  We now have only to apply the
optimal single-particle measurement to the 2{\sf x}2
$\hat{B}^{[j]}$'s of
equation~(\ref{blocks}) and use the result in equation~(\ref{Imutual}).

The error probability (equation~\ref{Err-Prob}) for distinguishing 
the $\hat{B}^{[j]}$'s of equation~(\ref{blocks})
is seen to be:
\beq p_{j} = \frac{1}{2} - \frac{c^n s^n}{ c^{2(n-k)} s^{2k} 
+  c^{2k} s^{2(n-k)} }
\ ,\label{pj}  \eeq
from which the information $I_2(p_j)$ in each channel is obtained.
Plugging the error probability $p_j$ (equation \ref{pj}) 
and the probability of choosing the j'th subchannel $q_j$ 
(equation~\ref{qj})
into (\ref{Imutual}), the 
optimal information on the parity bit is now:
\beq I_M = \sum_{j=1}^{2^n/2} 
\big( c^{2(n-k)} s^{2k} + c^{2k} s^{2(n-k)} \big)
\   I_2\left( \frac{1}{2} - \frac{c^n s^n}
   {c^{2(n-k)} s^{2k} + c^{2k} s^{2(n-k)}}  \right) \ . 
\label{I-mutual} \eeq
For various simple cases we can see the implications of this result,
and furthermore, we can compare it to the best result obtained
by individual measurements. 
In the simple case of orthogonal states ($\alpha = \frac{\pi}{4}$) 
all these density matrices are the same and we get
$q_j = \left(\frac{1}{2}\right)^{n-1}$, $\ p_j = 0$ and $I_M = 1$
as expected.
\begin{itemize}
\item A brief remark is in order at this stage.
The transformation to the $x$ basis for each 2 by 2 matrix, 
$\hat{B}_p^{(n,k)}$ is actually a transformation from a product basis
to a fully entangled basis of the $n$ particles.
That basis is a generalization of the Bell basis of \cite{BMR}.  
\beq  {1 \choose 0}_1 
      {1 \choose 0}_2 
       \ldots       
      {1 \choose 0}_{n-1}
      {1 \choose 0}_n  \pm  
      {0 \choose 1}_1 
      {0 \choose 1}_2 
       \ldots       
      {0 \choose 1}_{n-1}
      {0 \choose 1}_n 
 \ ; \nonumber \eeq   
\beq  {1 \choose 0}_1 
      {1 \choose 0}_2 
       \ldots       
      {1 \choose 0}_{n-1}
      {0 \choose 1}_n  \pm  
      {0 \choose 1}_1 
      {0 \choose 1}_2 
       \ldots       
      {0 \choose 1}_{n-1}
      {1 \choose 0}_n 
\eeq   
etc.
The Bell basis for two particles is frequently used and its basis 
contains the EPR singlet state and the other three orthogonal fully 
entangled states.
\end{itemize}

For large $n$, the number of blocks is exponentially large
and performing the summation required in equation (\ref{I-mutual}) 
is impractical,
since all the $2^{n-1}$ matrices must be taken into account.
However, that problem can be simplified by realizing that all blocks
with a given $k$, as well as all blocks with $k$ and $n-k$ 
interchanged,
contribute the same information to the total.  This is easily seen
in equation~(\ref{I-mutual}) where both the weight and the argument
of $I_2$ are symmetric in $k$ and $n-k$.
The optimal mutual information for even $n$ is then
\beq  
I_M^{even} = \sum_{k=0}^{\frac{n}{2}-1} {n \choose k} q_k I_2(p_k) + 
\frac{1}{2} 
{n \choose \frac{n}{2}} q_{\frac{n}{2}} I_2(p_{\frac{n}{2}})\ , 
\label{Ieven} \eeq
and for odd $n$
\beq I_{M}^{odd} =  \sum_{k=0}^{\frac{n-1}{2}} 
{n \choose k} q_k I_2(p_k) \ .
\label{Iodd}  \eeq

As an example we calculate $I_M$ for $n=2$ (of course, 
the counting argument is not needed in that case).  This particular 
result complements the result in \cite{HP} where the deterministic 
information 
of such a system is considered (see also Section \ref{par4}).  
In the new basis (\ref{dm_tag}) the density matrices of 
(equation~\ref{ex_n=2})
become
\bara  \hat{B}_0^{'(k=0)} =
  \left(\begin{array}{cc}  1/2 + \frac{c^2 s^2}{c^4 + s^4}  & 
                         \frac{1}{2}\frac{c^4-s^4}{c^4+s^4}  \\ 
        \frac{1}{2}\frac{c^4-s^4}{c^4+s^4} &      
                           1/2 - \frac{c^2 s^2}{c^4 + s^4}   
        \end{array}\right) \ ; \quad
 \hat{B}_1^{'(k=0)} =
  \left(\begin{array}{cc}  1/2 - \frac{c^2 s^2}{c^4 + s^4}  &
 \frac{1}{2}\frac{c^4-s^4}{c^4+s^4}  \\ 
        \frac{1}{2}\frac{c^4-s^4}{c^4+s^4} &      
                           1/2 + \frac{c^2 s^2}{c^4 + s^4}   
        \end{array}\right)
\nonumber \ear and 
\beq   \hat{B}_0^{'(k=1)} =
  \left(\begin{array}{cc}                1    &     0 \\ 
                                         0    &     0 
      \end{array}\right) \ ; 
\quad  \hat{B}_1^{'(k=1)} = 
  \left(\begin{array}{cc}                0    &     0 \\ 
                                         0    &     1 
      \end{array}\right) \ . \nonumber \eeq
We use the notation $S = 2sc = \sin 2\alpha$; 
$C = c^2 - s^2 = \cos 2 \alpha$
(hence,  $c^4 - s^4 = C$ and  $ c^4 + s^4 = 
\frac{1+C^2}{2}$)
to obtain
$q_1 = 2c^2 s^2 = \frac{S}{2}$, $p_1 = 0$,    
$q_0 = \frac{1}{2} (1 + C^2)$ and  
$p_0 = \frac{C^2}{1+C^2}$ (the $q_j$'s were obtained in the previous
section).
The mutual information of the parity of two bits is obtained using
equation (\ref{Imutual})
\bara I_M &=& q_0 I_2(p_0) +  q_1 I_2(p_1)  \nonumber \\ &=&  
      \frac{1}{2} (1 + C^2) 
 I_2 \left(\frac{C^2}{1+C^2}\right) + \frac{S^2}{2} \ . \ear

\section{Information on the Parity Bit of Almost Fully Overlapping 
States}\label{par3} 

The case of almost fully overlapping states is extremely important to
the analysis of eavesdropping attacks on any quantum key distribution
scheme.  In this case the
angle $\alpha$ is small so $s\equiv\sin \alpha \simeq \alpha$ and
$c\equiv\cos \alpha \simeq 1 - \frac{\alpha^2}{2}$.  To observe the 
advantage of the coherent measurement, let us first recall that the 
optimal information obtained by individual measurements
[with $Q_e^{(n)} \approx \frac{1}{2} - \frac{(2 \alpha)^n}{2}\ $], 
calculated to first order for small $\alpha$, is (see~\ref{Iindi}) 
\beq I_S \approx 
 \frac{(2 \alpha)^{2n}}{2 \ln 2}.  \label{Iindiv} \eeq

We use the same approximations used in Chapter~\ref{secur}, 
and equations (\ref{pj}) 
and (\ref{qj}) to calculate the leading terms in the 
optimal mutual
information (\ref{Ieven}) and (\ref{Iodd}).
For $k = \frac{n}{2}$ ($n$ even) we get $p_k = 0$ 
(regardless of the small angle) and 
\beq I_2(p_{\frac{n}{2}}) = 1 \ . \eeq
For $k < \frac{n}{2}\  $ we get 
$p_k \approx \frac{1}{2} - \frac{s^n}{s^{2k}} \approx
 \frac{1}{2} - \alpha^{n-2k} $ 
which yields 
[using equation~(\ref{Iapprox}) 
with $\eta = \alpha^{n-2k}$]
\beq  I_2(p_k) \approx
 \frac{2}{\ln 2} \alpha^{2n - 4k} \ . \eeq 
The coefficient $q_k = \alpha^{2k}$ for $k< \frac{n}{2}$ and
$q_k = 2 \alpha^{2k}$ for $k= \frac{n}{2}$, 
so that 
\beq q_k I_2(p_k) \approx
 \frac{2}{\ln 2} \alpha^{2(n-k)} \  \eeq
for $k < \frac{n}{2}$, and  
\beq q_k I_2(p_k) \approx  2 \alpha^n \  \eeq
for $k = \frac{n}{2}$. 
The dominant terms are those with the largest $k$, that is, $k$ 
closest to $\frac{n}{2}$.
The next terms are smaller by two orders in $\alpha$.
The number of density matrices with these $k$'s are also the largest
(up to a factor of 2 in case of even $n$). 
Therefore, the terms
$k= \frac{n}{2}$ for even $n$ and $k= \frac{n-1}{2}$ for odd $n$
are the dominant terms in the final expression.
Thus, for almost fully overlapping states, the mutual information is
\bara 
I_M^{even} \approx \frac{1}{2} {n \choose \frac{n}{2}  } 2 \alpha^n 
 = { n \choose \frac{n}{2} } \alpha^n \nonumber \ear 
for even $n$,
and 
\beq 
I_M^{odd} \approx {n \choose \frac{n-1}{2} }
         \frac{2}{\ln 2} \alpha^{n+1}  \label{Iopt} \eeq
for odd $n$.
 
These expressions can be further simplified.
The number of density matrices of any type is bounded (for large $n$)
using Stirling formula (see \cite{MS} in the chapter on 
Reed-Solomon codes)  
\beq  
{ n \choose k} < \frac{2^{nH(k/n)}}{\sqrt{2\pi (k/n)(1-k/n) n}} \ .
\eeq
For $k$ near $\frac{n}{2}$,  
$\eta \equiv \frac{1}{2} - \frac{k}{n}$ is small, and the 
standard approximation (\ref{Iapprox}):
$H \approx 
1 - O\big(\eta^2\big) = 
1 - O\left((\frac{1}{2} - k/n)^2 \right) < 1 $
is used to derive 
${ n \choose k} < \frac{2^n}{\sqrt{2\pi (k/n)(1-k/n) n}}$.
Using also $k/n(1-k/n) \approx \frac{1}{4} -  
  \eta^2$, we derive 
\beq
{ n \choose k} < \frac{2^n}{\sqrt {\frac{\pi}{2} n}} (1+O(\eta^2))\ .
\eeq

Thus the leading term in $I_M$ is
\beq I_M < \frac{2^n}{\sqrt{\frac{\pi}{2} n}} \alpha^n = 
             (2\alpha)^n /\sqrt{\frac{\pi}{2} n}  \eeq
for even $n$
and 
\beq 
I_M < \frac{2^n}{\sqrt{\frac{\pi}{2} n}} 
         \frac{2}{\ln 2} \alpha^{n+1} = 
         \frac{2}{\ln 2} \alpha
                 (2 \alpha)^n  /\sqrt{\frac{\pi}{2} n}  
             <   (2 \alpha)^n  /\sqrt{\frac{\pi}{2} n}  \eeq
for odd $n$ (using $\alpha < \ln2 / 2$). 
We see that we could keep a better bound for odd $n$ but for
simplicity we consider the same bound for both even and odd $n$'s.

We can now compare the optimal information $I_M$ from a coherent 
measurement on all $n$ particles to the optimal information $I_S$
from separate measurements (cf. eq.~\ref{Iindiv}): 
\beq\begin{array}{ccl} I_M & = & O(1)\times (2\alpha)^n/\sqrt{n} \\
I_S & = & O(1)\times (2\alpha)^{2n}. \\
\end{array}\eeq Since $\alpha$ is a small number (corresponding to
highly overlapping signal states), the coherent measurement is 
superior to the individual measurement by an approximate factor of 
$(2\alpha)^{-n}$. 
However, it is only superior by a
polynomial factor, since 
\beq I_M \approx \sqrt{I_S} \ . \eeq

\section{Deterministic Information on the Parity Bit}\label{par4}

For a single particle Bob can perform a different kind of individual
measurement which is not optimal in terms of average mutual 
information but is sometimes very useful \cite{Ben92,Per93}.
It yields either a conclusive result about the value
of that bit or an inconclusive one, and Bob will know which of the
types of information he has succeeded in obtaining.
Such a measurement corresponds to a binary erasure channel
\cite{Per93,HP,EHPP}.
With probability $p_?$ of an inconclusive result, 
the mutual information is 
$I_{p_?} = 1 - p_?$.
The minimal probability for an inconclusive result is
$\cos 2\alpha$  leading to 
$I_{p_?} = 1-\cos 2\alpha$ \cite{Per93}. This result is obtained
by performing a generalized measurement (Positive Operator Value 
Measure \cite{Per93,Helstrom,JP-DL})
on the system or a standard measurement performed on a larger 
system which contains the system and an auxiliary particle 
\cite{Per93,IvPe}. Note that this results in less mutual information 
than the optimal measurement for one-particle mutual information.
If Bob uses this type of measurement on each particle separately 
his deterministic single-particle information about
the parity bit is $(1-\cos 2\alpha)^n$.

We now use the block-diagonal density matrices derived in 
Section~\ref{par1}
to derive also the optimal {\em deterministic} 
information on the parity bit.  We note that each of the $2 \times 2$ blocks 
in the block-diagonal
density matrices is the density matrix of a pure state, so we
may replace the optimal measurement in each subchannel with 
the optimal deterministic measurement and proceed as before.
The total optimal deterministic 
information is easily calculated by replacing $I_2(p_k)$ in 
(\ref{Ieven}) and (\ref{Iodd}) by 
$I(p_{?_k}) = 1 - p_{?_k}$.
To find the minimal 
$P_{?_k}$ we recall that each of the normalized density 
matrices $\hat{B}_p^{(n,k)}$ corresponds to pure states (\ref{pure-block})
with some angle $\gamma$. 
Comparing with equation (\ref{blocks}) we can write the overlap
as the difference between the diagonal terms  
of the density matrices 
\beq p_? = \cos(2 \gamma) = 
\cos^2\gamma - \sin^2\gamma = 
\frac{c^{2(n-k)} s^{2k} - c^{2k} s^{2(n-k)}} 
                  { c^{2(n-k)} s^{2k} + c^{2k} s^{2(n-k)}} \ , \eeq
hence
\beq I(p_{?_k})  = 1 - 
\frac{c^{2(n-k)} s^{2k} - c^{2k} s^{2(n-k)}} 
                  {c^{2(n-k)} s^{2k} + c^{2k} s^{2(n-k)}} \ .\eeq
The total information is 
\beq  
I_{D}^{even} = \sum_{k=0}^{\frac{n}{2}-1} {n \choose k} 
q_k I(p_{?_k}) 
+ \frac{1}{2} {n \choose \frac{n}{2}} q_{\frac{n}{2}} 
I(p_{?_{n/2}})\ , 
\label{Ideteven} \eeq
for even $n$, and
\beq 
I_{D}^{odd} = 
\sum_{k=0}^{\frac{n-1}{2}} {n \choose k} q_k I(p_{?_k}) 
\label{Idetodd}  \eeq
for odd $n$.

For $n=2$ we 
recover a result previously obtained by Huttner and Peres 
\cite{HP} by performing the optimal POVM on the first pair of
density matrices of 
equation (\ref{ex_n=2}), 
and a measurement in the entangled basis (as before) on the second.
The probability of an inconclusive result is
$\cos^2\gamma - \sin^2 \gamma = \frac{c^4 - s^4}{c^4 + s^4} 
= \frac{2C}{1+C^2}$, 
hence the optimal deterministic information is
$1 -  \frac{2C}{1+C^2}$, 
leading to the total deterministic information
\beq I_D = q_1 I_D(p_?) + q_2 I_2(p_2) =  
      \frac{1}{2} (1 + C^2) 
          (1 - 2\frac{C}{1+C^2}) + \frac{S^2}{2} = 1 - C \ , \eeq
which is exactly the result obtained by Huttner and Peres (note, 
however,
that they used an angle which is 
$\frac{\pi}{4}-\alpha$ hence derived $1-S$ for the 
deterministic information).

For almost overlapping states (small $\alpha$) the dominant terms
are still the same as in the case of optimal information.
The term $q_k$ is as before
and the information in each port is \beq I(p_{?_k})=1 \eeq 
for $k = \frac{n}{2}$ and 
\beq I(p_{?_k}) = 1 - \frac{c^{2n-4k} - s^{2n-4k}}{
                         c^{2n-4k} + s^{2n-4k}} =
           1 - (1 - \alpha^{2n-4k})^2  = 2 \alpha^{2n-4k}         
   \eeq
for $k<\frac{n}{2}$.
Taking into consideration only the dominant term we get 
\bara I_D^{even} 
\approx  {n \choose \frac{n}{2}  }   \alpha^n
\nonumber \ear 
for even $n$ which is the same as the optimal information, 
and 
\beq 
I_D^{odd} \approx {n \choose \frac{n-1}{2} }
         2 \alpha^{n+1}   \eeq
for odd $n$ which is smaller 
than the optimal mutual information by a factor of $\frac{1}{\ln 2}$.

\section{Parity Bit for Density Matrices of Mixed States}\label{par5}

The previous discussion assumed that $\rho_p^{(1)}$ are pure states.
The generalization to the case of density matrices with equal 
determinants is straightforward.
Let the bit `0' and the bit `1' be represented by  
\beq \rho_0^{dm} = 
  \left(\begin{array}{cc}   c^2 &  sc-r  \\  sc-r &  s^2     
  \end{array}\right),\label{rhod_0} \eeq
and 
\beq \rho_1^{dm} = 
  \left(\begin{array}{cc}   c^2 &  -(sc-r)  \\  -(sc-r) &  s^2     
  \end{array}\right),\label{rhod_1} \eeq
(with $s=\sin \alpha$ etc., and $r<sc$)
which contains the most general density matrices of the desired type.
On the Poincar\'e sphere 
these density matrices have the same $z$ components
as the previously written pure states but smaller $x$ components
(hence smaller angle $\alpha'$).
We could choose other ways of representing these density matrices,
e.g., with identical $x$ components and smaller $z$ components.
Such representations are appropriate for comparison with pure
states (since they yield the same mutual information for a single
particle) but are less convenient for showing that the previous result
is easily generalized.

Clearly 
\beq \rho^{(1)_{dm}} = \frac{1}{2} (\rho_0^{(1)} + \rho_1^{(1)}) = 
  \left(\begin{array}{cc}   c^2 &  0  \\  0 &  s^2     
  \end{array}\right),\label{rhod} \eeq
\beq \Delta^{(1)_{dm}} = \frac{1}{2} (\rho_0^{(1)} - \rho_1^{(1)}) = 
  \left(\begin{array}{cc}   0 &  sc-r  \\  sc-r &  0     
  \end{array}\right),\label{Deltad} \eeq
The total density matrix doesn't change and the difference density 
matrix
has terms $(sc - r)^n$ instead of $(sc)^n$.
Reorganizing the basis vectors we again get the block diagonal 
matrices
where each of the 2 by 2 matrices has the form 
\beq   B_p^{(n,k)} = \left(\begin{array}{cc}   
         c^{2(n-k)} s^{2k}    &  \pm (cs-r)^n             \\ 
         \pm (cs-r)^n          &  c^{2k} s^{2(n-k)} 
  \end{array}\right). \label{Bpnk-dm} \eeq
When normalized, these density matrices have the form of 
equation~(\ref{amatrix}) and are optimally distinguished by measuring 
them in the $x$ direction. Transforming to the $x$ basis as before 
we get the same 
\beq q_k = 
       c^{2(n-k)} s^{2k} 
                 +  c^{2k} s^{2(n-k)} \eeq
as before, and 
\beq  p_k = \frac{1}{2} - \frac{(cs-r)^n}{ 
                                          c^{2(n-k)} s^{2k} 
                                    +  c^{2k} s^{2(n-k)}} \ . \eeq
The total information can now be calculated as before by assigning 
these $p_k$ and $q_k$ into equations (\ref{Iodd}) and (\ref{Ieven}).
Thus, the case of mixed states is also analytically solved for any 
number of bits, and the influence of mixing on the optimal mutual 
information is through the $p_k$'s.

Calculating the optimal information for small $\alpha$ and any 
$r$ is possible but complicated.
A much simpler alternative is to find a bound on the 
optimal
information using pure states with the same angle, $\alpha'$, 
on the Poincar\'e sphere, using
\beq 
\tan 2 \alpha' = \frac{\sin 2\alpha - 2 r}{\cos 2\alpha} \ ,\eeq  
or using an alternative form for the mixed 
states (\ref{rhod_0}) and (\ref{rhod_1}).

\section{The Connection to Quantum Cryptography}\label{par6}

In this chapter we solved a problem in quantum information:
suppose Bob is given $n$ particles, each in one of two possible states
(either pure or mixed in 2-d with equal determinants). 
How much mutual information
can he extract on the parity bit and what is the optimal measurement?
The density matrices describing the parity bit are very complicated
and have $2^n$ dimensions. 
Thus, it is
very surprising that the optimal information can be calculated, since previous
results consider much simpler case (e.g. density matrices in 2-d).

This result is important to the analysis of privacy amplification,
and has crucial impact on quantum cryptography.
Previously, it was known that privacy amplification 
is effective when particles are not measured together,
but its effectiveness against coherent measurements
was in question.
Our result provides the optimal measurement which can be done to 
find a 
parity bit.
In particular cases, when almost fully overlapping states are used,
we proved two complementary results regarding that optimal 
measurement:
\begin{itemize}
\item The optimal information $I_M$ obtained by allowing all possible 
measurements is
much larger than the one ($I_S$) obtained by measuring 
each bit separately.
\item The optimal information $I_M$ is still 
exponentially small with the length of the string.
This is an {\em effectiveness result}: classical 
privacy amplification techniques are effective
against {\em any} quantum measurement.
\end{itemize}
Our ``effectiveness result'' leads to the derivation of strong
security results against sophisticated attacks directed against the final
key. However, such analysis is much more complicated since
it involves the key distribution protocol in real channels, and because
Eve is exposed to classical information in addition to the quantum 
information.  This is the subject of the next chapter.

\chapter{Security Against Collective Attacks}\label{col}

The discussion of the previous chapter
treats an imaginary scenario which is very 
general 
but is not general enough to deal with the common situation in
quantum cryptographic protocols. Still, it provides some
basic tools required for such an analysis.

The basic differences from the situation previously described are:
\begin{itemize}
\item  Eve 
probes states transmitted
from Alice to Bob 
rather than fed by Alice; 
Thus, she 
has some freedom in choosing the states she obtains.
\item In addition to the quantum information (the possible
states of the probes) Eve obtains classical data; 
The state of Eve's probes 
is information-dependent, and this fact must 
be taken into account in calculating the optimal information,
and also in the optimization process.
\end{itemize}
These differences create
a more complicated situation where the final measurement
must be optimized, not by itself, but {\em together} with
the probing process. 

As far as quantum cryptography is concerned, this chapter provides strong
evidences that quantum key distribution is secure.
As far as physics is concerned, this chapter contains interesting 
examples for information-dependent quantum states.
Another achievement is the development of new type of
bounds on the information which can be extracted from quantum states.

To emphasize the importance of the ``effectiveness result'' 
obtained in the previous chapter 
let us consider the scenario which is common in quantum key
distribution schemes:
Alice and Bob (the legitimate users) try to establish a secret
key.
They use any binary scheme and Alice sends $L'$ particles
(see Section~\ref{int.qua-cry}) through a noisy
channel to Bob.  
An adversary, Eve, is trying to learn information on their key.
She gets the particles one at a time, 
interacts with each one of them weakly (on average), 
and sends it forward to Bob.
She must interact weakly with each particle
if she wants to induce only
small error rate (or 
alternatively, she could attack strongly only a few of the particles, 
but privacy amplification
is already proven effective against such type of attack \cite{BBCM}).
Although the specification of the privacy amplification technique
is announced after the transmission is 
over, 
Eve can keep a quantum state of a system which has
interacted with all the transmitted particles, and use it after all
the specification is announced.
Security against such joint attacks in error-free channels is shown in
\cite{Yao}.
Our main goal is to obtain a concrete bound on Eve's information for a
given error rate, and show that it reduces to zero, if large strings 
are used.
Typically, one would like to consider the case where Alice and
Bob use existing technologies, while Eve is restricted only by
the laws of physics.  

In the first sections we concentrate on {\em symmetric collective attacks} 
in which
the same translucent attack is applied to each transmitted
particle, and 
the attack is symmetric to any of the possible quantum states 
of each particle.
Such an attack induces the same probability of error to each transmitted
bit.
It must be weak, or else it would induce 
a non acceptable error rate.
Thus, the possible states of Eve's probe cannot differ much.

In Section \ref{colA} we design a specific case in quantum key distribution 
where each of Eve's probes is in one of two pure states
(with small angle between them), so the efficiency
result of the previous chapter can be applied. 
We also incorporate error correction as is done in any realistic protocol
and thus the analysis becomes more complicated. This part was done
together with Eli Biham and Dominic Mayers and with the help of Gilles Brassard
and John Smolin.
It was published with Eli Biham
in~\cite{BM1}. 
Sections~\ref{colB},~\ref{colC} were 
done with Eli Biham~\cite{BM2}.
In Section \ref{colB} we present new bounds on the information
which can be extracted from mixed states  
in case the result regarding pure states is known.
In Section \ref{colC} we combine 
the results of the previous sections and
the fact that density matrices in Eve's hands are information-dependent
and we calculate Eve's optimal information for various symmetric
collective attacks. We show how to
prove security against any symmetric collective attack which 
uses 2-dimensional probes.
In Sections \ref{colD} and \ref{colE}
we discuss the relevance of the previous result to
non-symmetric collective attacks and to the optimal collective attack.

\section{The Simplest Case -- Eve's Probes are in One Out Of Two
Possible Pure States}\label{colA}

In this section we consider the two-state scheme and devise a specific attack
against it.
In the two-state~\cite{Ben92} scheme 
the classical bits `0' or `1' are represented by 
two non-orthogonal pure states, which can be written as  
\begin{equation}  \psi_0 = {\cos \theta \choose \sin \theta}\quad {\rm and}
\quad \psi_1 = {\cos \theta \choose -\sin \theta} \end{equation} respectively. 
Alice sends $L'$ such qubits.
Bob performs a test which provides him with a conclusive or inconclusive
result. For instance, he can test whether a specific particle is in a state
$\psi_0$ or a state orthogonal to it ${\psi_0}'={\sin \theta \choose
- \cos \theta}$; A result $\psi_0$ is 
treated as inconclusive (since it could also be $\psi_1$ sent by Alice), 
and a result ${\psi_0}'$ is identified as $\psi_1$.
Alternatively, he could test if a particle is in a state 
$\psi_1$ or a state orthogonal to it ${\psi_1}'={\sin \theta \choose
 \cos \theta}$, and identify a result $\psi_1'$ conclusively as $\psi_0$.
Bob informs Alice  
which bits were identified conclusively, using the 
unjammable classical channel, and these bits are the raw data.
Later on, they check the error rate on the raw data,
and if it is reasonable, they use the remaining $n$ bits 
to obtain a (hopefully secure) final key.
The errors are corrected using any error-correction protocol.
Finally, the parity of the full $n$-bit string
is used as their final one-bit key
(Of course, for practical purposes our approach must be generalized
to yield a longer final string).

Our first example of a collective attack is based on
the ``translucent attack without entanglement'' of \cite{EHPP}, 
which leaves Eve with probes in a pure state. 
The translucent attack without entanglement uses the unitary transformation
\begin{equation}  {\cos \theta \choose \pm \sin \theta} \longrightarrow
                 {\cos \theta' \choose \pm \sin \theta'} 
                 {\cos \alpha \choose \pm \sin \alpha} \end{equation}
where $\theta'$ is the angle of the states received
by Bob, and $\alpha$ is the angle of the states in Eve's hand.
The error rate is the probability of wrong identification
and it is $p_e = \sin^2 (\theta - \theta')$.  
Using the unitarity condition 
$ \cos 2 \theta = \cos 2 \theta' \cos 2 \alpha $ we get for
weak attacks that 
\begin{equation}
\alpha = (p_e \tan^2 2\theta )^{1/4} \ .
\end{equation}
This translucent attack is 
performed on all the bits, 
and it
leaves Eve with $n$ relevant probes, 
each in one of the two states ${c \choose \pm s}$, 
with $c = \cos \alpha$ and $s = \sin \alpha$.
As a result, 
Eve holds an $n$-bit string $x$ which is concatenated from its bits
$(x)_1 \ (x)_2 \ldots (x)_n$.
For simplicity, we choose the final key to consist of one bit, 
which is the parity of the
$n$ bits.
Eve wants to distinguish between two density matrices corresponding
to the two possible values of this parity bit.
Our goal is to calculate the optimal mutual information she can 
extract from them. 

For our analysis we need some more notations. 
Let $\hat{n}(x)$ be the number of $1$'s in $x$, and $p(x)$ be the 
parity of $x$. 
For two strings of equal length $x \odot y$ is the bitwise ``AND'', so that 
the bit $(x \odot y)_i$ is one if both $(x)_i$ and $(y)_i$ are one,
and zero otherwise.
Also $x \oplus y$ is the bitwise  ``XOR'', so that 
$(x \oplus y)_i$ is zero if $(x)_i$ and $(y)_i$ are the same, and one otherwise.
For $k$ (independent) strings,
$v_1, \ldots , v_k$, of equal length let the set $\{v\}_k$
contain all the $2^k$ linear combinations 
of $v_1, \ldots , v_k$ (including the string $v_0 = 00 \ldots 0$): 
$(v_0),(v_1), \ldots, (v_k), (v_1 \oplus v_2), (v_1 \oplus v_3), \ldots,
(v_{k-1} \oplus v_k), (v_1 \oplus v_2 \oplus v_3), \ldots  
(v_1 \oplus v_2 \ldots \oplus v_k)$.
Since the (original) $k$ strings are linearly independent   
these $2^k$ strings are all different. 
The quantum state of a string is the tensor product  
\begin{equation} \psi_x = {c \choose \pm s}{c \choose \pm s} \ldots
{c \choose \pm s} 
 = \left( \begin{array}{c}
      c c c  \ldots   c c c \\
 \pm  c c c  \ldots   c c s \\
             \ldots               \\
 \pm  s s s  \ldots   s s s \end{array} \right) \ , \label{value} \end{equation}
living in a $2^n$ dimensional Hilbert space. 
The sign of the $i$'th bit (in the middle expression) 
is plus for $(x)_i = 0$ and minus for
$(x)_i = 1$. 
The sign of the $j$'th term ($j = 0\ldots 2^{n-1}$)  
in the expression at the right depends on 
the parity of the string
$x \odot j$ and is equal to $(-1)^{p(x \odot j)}$.
The density matrix $\rho_x =  \psi_x   \psi_x^T$
also has for any $x$, the same terms up to the signs. 
We denote the absolute values by 
$\rho_{jk} \equiv |(\rho_x)_{jk}|$. 
The sign of each term $(\rho_x)_{jk}$ is given by
\begin{equation} 
                   (-1)^{p(x \odot j)}
                   (-1)^{p(x \odot k)} =
                   (-1)^{p[x \odot (j\oplus k)]}
\ . \label{sign} \end{equation}

A priori, all strings are equally probable 
and Eve needs to distinguish between the two density matrices
describing the parities.
%two density matrices:
%\begin{equation}
%\rho_0^{(n)}=\frac{1}{2^{n-1}}\sum_{x\,|\,p(x)=0}\!\!\!\!\!\rho_x\ ;
% \quad
%\ \rho_1^{(n)}=\frac{1}{2^{n-1}}\sum_{x\,|\,p(x)=1}\!\!\!\!\!\rho_x \ .
%\end{equation}
These matrices were calculated and analyzed in Chapter~\ref{par} 
(henceforth, the BMS work~\cite{BMS}). 
If Eve cannot obtain the error-correction code
her optimal information on the parity bit is given by (\ref{Ieven}) and
(\ref{Iodd}). 
To leading order in $\alpha$ it is given by (\ref{Iopt}) and it is
\begin{equation} I_M^{even} = {2k \choose k} \alpha^{2k} \end{equation}
for $n=2k$, and
\begin{equation} I_M^{odd} = {2k-1 \choose k-1} \frac{2}{\ln 2}
\alpha^{2k} = {2k \choose k} \frac{1}{\ln 2} \alpha^{2 k} \end{equation}
for $n = 2k -1$.
Thus, 
\begin{equation} I(n) = c {2k \choose k} \alpha^{2k} \label{BMS_info} \end{equation}
with $c=1$ for $n=2k$ and $c= 1/\ln2$ 
for $n = 2 k - 1$. 

In case Eve is being told what 
the error-correction code is,
all strings consistent with the given error-correction 
code (the $r$ sub-parities) 
are equally probable,
and Eve need to distinguish between
the two density matrices: 
\begin{equation}
\rho_0^{(n,r)}=\frac{1}{2^{n-r-1}}\!\sum_{x\,|\,{{p(x)=0}\choose {x\ \!{\rm OECC}}}}\!\!\!\!\!\!\rho_x\ \!
; \quad \!\!\!
 \rho_1^{(n,r)}=\frac{1}{2^{n-r-1}}\!\sum_{x\,|\,{{p(x)=1}\choose {x\ \! {\rm OECC}}}}\!\!\!\!\!\!\rho_x 
\label{density-matrices}  \end{equation}
where {\em OECC} is a shortcut for {\em Obeys Error-Correction Code}.

Let us look at two simple examples where $n=5$:
one with $r=1$ and the other with $r=2$. 
Suppose that 
the parity of the first 
two bits, $(x)_1$ and $(x)_2$, 
is $p_1 = 0$.
Formally, this substring is described by the $n$-bit string 
$v_1 = 24$ which is $11000$ binary; The number of $1$'s in the first
two bits of a string $x$ is given by $\hat{n}(x \odot v_1)$,
and $x$ obeys the error-correction code if $p(x \odot v_1) = p_1$.
Let $v_d$ be the binary string ($11111$ in this case) 
which describes the substring
of the desired parity.
Eve could perform the optimal attack on
the three bits which are left, or in general, on $v_1 \oplus v_d$. 
For any such case, the optimal attack is given by the BMS work and the 
optimal information depends only on   
$\hat{n}(v_1 \oplus v_d)$, the {\em Hamming distance} between the two words.
This information [using eq.~(\ref{BMS_info})] is  
\begin{equation} I(\hat{n}) = c {2k \choose k} \alpha^{2k} \label{BMS_info1} 
\end{equation}
with $c=1$ for even $\hat{n}$ (which equals to $2k$) and $c= 1/\ln2$ 
for odd $\hat{n}$ 
(which equals $\hat{n} = 2 k - 1$). 

Suppose that Eve gets another parity bit $p_2 = 1$ of the binary string
$01100$ ($v_2 = 12$). 
Now, a string $x$ obeys the error-correction code if it also obeys
$p(x \odot v_2) = p_2$.  Clearly, it also satisfies 
$p[x \odot (v_1 \oplus v_2)] = p_1 \oplus p_2$.
In the general case there are $r$ independent parity strings,
and $2^r$ parity strings in the set $\{v\}_r$.  
The BMS result cannot be directly used, but it still provides some intuition: 
For each word (i.e., each parity string) 
$v_l \in \{v\}_r$, let  
$I(\hat{n} (v_l \oplus v_d))$ be the optimal information Eve could
obtain using~(\ref{BMS_info1}). Also let 
$I_{sum}$ be the sum of these contributions from all such words.
In reality Eve cannot obtain $I_{sum}$
since each measurement changes
the state of the measured bits,
hence we expect that 
$I_{sum}$ bounds her
optimal information $I_{total}$:
\begin{description} 

\item[Conjecture]
\begin{equation} I_{total} < I_{sum} \ .
\label{conjecture} \end{equation}
\end{description}
On the other hand, Eve knows all these words at once, 
and could take advantage of it,
thus we leave this as an unproven conjecture.  

In the following we find an explicit way to calculate
the optimal information exactly.
However, this exact result requires cumbersome calculations,
thus it is used only to verify the conjecture for short strings.

The parity of the full string is also known 
since the density matrix 
$\rho^{(n,r+1)}$ corresponds to either $\rho_0^{(n,r)}$ or $\rho_1^{(n,r)}$
depending on the desired parity $p_{r+1}$,
thus we add the string $v_{r+1}=v_d$.
There are $r+1$ 
independent sub-parities altogether, hence $2^{r+1}$ parity strings
in the set $\{v\}_{r+1}$. 
A string $x$ is included in $\rho^{(n,r+1)}$ if 
\begin{equation}
p[x\odot v_l]=p_l
\end{equation}
for all given substring in $\{v\}_{r+1}$ 
In the BMS work (where $r=0$) 
the parity density matrices were put in a block diagonal
form of $2^{n-1}$ blocks of size $2 \times 2$.
This result can be generalized to the case where $r$ parities of substrings
are given.
There will be $2^{n-r-1}$ blocks of size $2^{r+1} \times 2^{r+1}$.
We shall show that the $(jk)$'th term in a density matrix $\rho^{(n,r+1)}$
of $r+1$ sub-parities
is either zero,  $\rho_{jk}$ or $-\rho_{jk}$,
that is, either all the relevant strings contribute exactly the same term, or
half of them cancels the other half.
The proof can be skipped in a first reading. 
\begin{description}
\item[Theorem] 
\quad 

The element $(\rho^{(n,r+1)})_{jk}$ is zero if 
%\hfill\newline
$j\oplus k \not\in \{ v \}_{r+1}$,
and it is $\pm \rho_{jk}$ if 
$j \oplus k \in \{ v \}_{r+1}$.
\item[Proof] 
\quad 

In case 
\begin{equation} j\oplus k \not\in  
\{ v \}_{r+1} 
\end{equation}
choose $C$ such that
\begin{equation}
p[C\odot v_l] = 0 
\end{equation}
with all $(v_l)$'s in 
$\{ v \}_{r+1}$
and  
\begin{equation}
p[C\odot (j \oplus k)] = 1
\end{equation} 
(many such $C$'s exists since $C$ has $n$ independent bits
and it needs to fulfill only $r+2$ constraints).
For such a $C$ and for any $x$ which obeys the error-correction code
there exist one (and only one) $y$, $y = x \oplus C$,
which also obeys the code (due to the first demand) but has the opposite
sign in the $jk$'th element (due to the second demand),
so $(\rho_y)_{jk} = - (\rho_x)_{jk}$ (due to eq.~\ref{sign}).
Since this is true for any relevant $x$, we obtain 
\begin{equation} (\rho^{(n,r+1)})_{jk} = 0 \ .
\end{equation}

In case 
\begin{equation} j\oplus k \in \{ v \}_{r+1}
\end{equation}  such $C$ cannot exists,
and all terms must have the same sign: Suppose that there are two terms,
$x$ and $y$ with opposite signs. Then $C=x\oplus y$ satisfies the two
demands, leading to a contradiction.

\hfill QED
\end{description}
This theorem tells us the place of all 
non-vanishing terms in the original ordering. 
The matrices can be reordered to a block-diagonal form
by exchanges of the basis vectors.
We group the vectors $s$, $s\oplus v_1$, etc., 
for all $(v_l)$'s in $\{v\}_{r+1}$
to be one after the other, so each such group 
is separated from the other groups. 
Now the theorem implies that all 
non-vanishing terms are grouped in blocks, and 
all vanishing terms are outside these blocks.
As a result, the matrix is
block-diagonal.
This forms $2^{n-r-1}$ blocks of size $2^{r+1} \times 2^{r+1}$.
All terms inside the blocks and their signs are given by eq.~(\ref{value})
and~(\ref{sign})  
respectively up to reordering.
The organization of the blocks depends only on the parity strings $v_l$ and
not on the parities $p_l$, thus,
$\rho_0^{(n,r)}$ and  
$\rho_1^{(n,r)}$ are block diagonalized in the same basis. 
The rank of a density matrix is the number of (independent) pure states which 
form it, and it is $2^{n-r-1}$ in case of the parity 
matrices [eq.~(\ref{density-matrices})].
When these matrices are put in a block diagonal form, there are $2^{n-r-1}$
(all non-zero) blocks. Thus, the rank of each block is one and the corresponding
state is pure. When a block is fully diagonalized to yield 
\begin{equation}
B^{[j]} = \left(\begin{array}{ccccc}
a_j    &     0     &     0     & \ldots    &     0     \\       
0      &     0     &     0     & \ldots    &     0     \\       
0      &     0     &     0     & \ldots    &     0     \\       
\ldots &           &           &           &           \\       
0      &     0     &     0     & \ldots    &     0            
\end{array}\right) \ , \end{equation}
the non-vanishing term $a_j$ in 
the $j$'th block is the probability that a measurement will 
result in this block. 

In the BMS work ($r=0$), 
the information, in case of small angle, 
was found to be exponentially small with the length of the string.
When each probe is in a pure state, 
this result can be generalized to $r>0$ as follows:  
The optimal mutual information carried by two pure states
(in any dimension) is well known.
The two possible pure states in the $j$'th block of $\rho_0^{(n,r)}$
and $\rho_1^{(n,r)}$ can be written as 
$\cos\beta  \choose \pm \sin \beta $.
The optimal mutual information which can be obtained from the $j$'th block
is given by the overlap (the angle $\beta_j$) 
\begin{equation}
I_j = 1 + p_j \log p_j + (1-p_j) log (1-p_j) \ ,
\end{equation} 
where 
\begin{equation} p_j = \frac{1 - \sin 2 \beta_j }{2} \ ; 
\end{equation} 
The overlap 
is calculated using 
eq.~(\ref{value}) and~(\ref{sign}).
Thus, for any given error-correction code, we can find 
the two pure states in each block, 
the optimal information $I_j$,
and finally, the total information 
\begin{equation}
I_{\rm total} = \sum_j a_j I_j \ . 
\end{equation}
We did not use the value of $v_d$ in the proof, and thus, the 
final key could be the parity of any substring. 
Moreover, we intend to develop similar methods to analyze keys
of several bits which can be formed from parities of several substrings.
 
We wrote a computer program which receives any (short)  
error-correction code 
and calculates the total information as a function of the angle
$\alpha$ between the pure states of the individual probes.
We checked many short codes (up to $n=8$) to
verify whether $I_{total} < I_{sum}$ as we conjectured.
Indeed, all our checks showed that the conjecture holds.
The information for small angle $\alpha$ is bounded 
by 
\begin{equation}
I_{sum} = C \alpha^{2k}
\end{equation}
as previously explained, where $C$ is given by 
summing the terms which contribute to the highest order of eq.~(\ref{BMS_info1}),
and the Hamming distance $\hat{n}$ (which is $2k$ or $2k -1$), 
can be increased by choosing longer codes
to provide any desired level of security.

In addition to a desirable security level, the error-correction code 
must provide
also a desirable reliability;
A complete analysis must include also estimation of the probability
$p_f$ that Alice and Bob still has wrong (i.e. different) final key.
For enabling such analysis, one must use known error-correction codes.
Random Linear Codes allow such analysis but cannot be used efficiently
by Alice and Bob.
Hamming codes~\cite{MS}, $H_r$ 
which use $r$ given parities for correcting one error in strings of length
$n=2^r - 1$, have  
an efficient decoding/encoding procedure
and a simple way to calculate $p_f$.  
A Hamming code has $2^r$ words in $\{v\}_r$, all of them,
except $00 \ldots 0$, are at the same distance
$\hat{n}=2^{r-1}-1$ from $v_d$.
Using our conjecture and eq.~(\ref{BMS_info1}) 
(with $k=\frac{\hat{n}+1}{2}=2^{r-2}$) we obtain
\begin{equation}
 I_{\rm total} < (2^r - 1) \frac{1}{\ln 2} {2^{r-1} \choose 2^{r-2} }
\alpha^{(2^{r-1})} + O\left(\alpha^{(2^r - 1)}\right) \ .
\end{equation}

For $r=3$ ($n=7$) the conjecture yields $I_{\rm total} < 60.6 \alpha^4$,
where the $2^3-1$ words
($0111001$; $1101010$; $1110100$ and their linear combinations
$1010011$; $1001101$; $0011110$; $0100111$)
are at a distance $\hat{n}=3$ from $v_d = 1111111$,
and $0000000$ at a distance $\hat{n}=7$. 
This Hamming code was presented in Section~\ref{int.qua-cry}.
For the exact calculation we use $v_d$ so there are
$4$ independent parity strings: $0111001$; $1101010$; $1110100$; $1111111$.
As a result, there are $2^4$ resulting strings where 
$0001011$; $0010101$; $0101100$; $0110010$; 
$1000110$; $1011000$; $1100001$; $1111111$ are added to the above.
Averaging over all the density matrices which are
consistent with such given parity strings, 
and organizing the matrices in a block diagonal form, we calculated
(using the computer program) the pure states in each block and the resulting
information obtained by the optimal measurement.
The exact calculation also 
gives the same
result as above, showing that the conjecture provides an
extremely tight bound in this case.

Using 
\begin{equation}
{ 2^{r-1} \choose 2^{r-2} } < \frac{2^{(2^{r-1})}}{\sqrt(\frac{\pi}{2}
2^{r-1})} 
\end{equation}
and some calculation 
we finally obtain
\begin{equation} I_{\rm total} < \left( \frac{2}{\ln2 \sqrt\frac{\pi}{2}}\right)
\sqrt{2^{r-1}} (2\alpha)^{(2^{r-1})} \ , \end{equation}
bounding $I_{\rm total}$ to be exponentially small with $n$ 
[which follows from $2^{r-1} = (n+1)/2$]:
\begin{equation} I_{\rm total} < C(n) 
 (2\alpha)^{(n+1)/2} \ , \label{BMres} \end{equation}
with $C(n) = 
\sqrt{n+1}
\frac{2}{\ln2 \sqrt{\pi}}$.
As a function of the error rate the information is 
\begin{equation} I_{\rm total} < C(n) 
 (16 p_e \tan^2 2\theta)^{(n+1)/8} \ . \label{BMer} \end{equation}

The rate of errors in the string shared by Alice and Bob
(after throwing inconclusive results) is the normalized error rate, 
$p_e^{{}^{(N)}} = p_e / (p_c + p_e)$,
where $p_c = \sin (\theta + \theta')$ is the probability of obtaining
a correct and conclusive result.
For small $\alpha$ it is 
$p_e^{{}^{(N)}} = \frac{2 p_e}{\sin^2 2 \theta} = 
\frac{2 \cos^2 2 \theta}{\sin^4 2 \theta} \alpha^4$.
The final error probability $p_f$ is bounded by
the probability to have more than one
error in the initial string, since the code corrects one error.
It is $p_f = \frac{n(n-1)}{2} (p_e^{{}^{(N)}})^2 + O[(n p_e^{{}^{(N)}})^3]$,
showing that we can use the Hamming codes as long as $n p_e^{{}^{(N)}} << 1$.
In case it is not, better codes such
as the BCH codes~\cite{MS} (which correct more than one error)
are required,  
but their analysis
is beyond the scope of this work. 
Extracting more than one final bit is also possible,
but is also beyond the scope of this work.

\section{Bounds on Information}\label{colB}

In terms of quantum information theory the result
of eq.~(\ref{BMres}) (henceforth, the 
BM result) extends the BMS result to the case where parities of
substrings are given (error-correction code).
For purposes of quantum key distribution, the 
BM result provides the first security proof
(assuming the verified conjecture~\ref{conjecture})
against a strong attack.
However, it is restricted to attacks in which Eve's probes are in a pure state.
Unfortunately, most possible translucent 
attacks on the two-state scheme~\cite{Ben92}, 
which can be used in the first step
of the collective attack, leave each of Eve's probes in a mixed state.
Also, any translucent attack on the four-state 
scheme~\cite{BB84} leaves each probe
in a mixed state (at least for two out of the four possible states).

The goal of this section is to apply the BM result to the  
case of mixed states.
We demonstrate that any type of information
which can be extracted from certain two-dimensional mixed states can be bounded,
if the solution for pure states is known.
In the next sections we use this bound and conjecture~\ref{conjecture} to 
explicitly demonstrate, via examples, 
how to bound Eve's optimal information
(for a given induced error rate).

Any state (density matrix) in 2-dimensional Hilbert space
can be written as  
\begin{equation}
\rho = \frac {\hat I + r \cdot \hat\sigma} {2}
\end{equation}
so that
\begin{equation}
\rho = \frac{1}{2} \left(\begin{array}{cc} 1 + z & x - i y \\ x + i y & 1 - z 
\end{array}\right) \ ,
\end{equation}
with $r = (x,y,z)$ being a vector in ${\cal{R}}^3$, $\hat{\sigma}
= (\hat{\sigma}_x, \hat{\sigma}_y, \hat{\sigma}_z)$ the Pauli matrices,
and $\hat I$ the unit matrix. In this spin notation, each 
state is represented by the corresponding vector $r$.
For pure states $|r| = 1$, and for mixed states $|r|<1$.  
Suppose that $\chi$ and $\zeta$ are two density matrices, 
represented by $r_{\chi}$ and $r_{\zeta}$ respectively.
It is possible to construct the density matrix 
\begin{equation}
\rho = m \zeta + (1-m) \chi 
\end{equation}
from the two matrices (where $0 \le m \le 1$ since $m$ and $1-m$ are
probabilities), and  
the geometric representation of such a density matrix 
$ \rho = \frac{\hat I + r_{\rho} \cdot \hat{\sigma}}{2} \ $ 
is \begin{equation}
r_{\rho} = m r_{\zeta} + (1-m) r_{\chi}
\ . \end{equation}
Two pure states can always be expressed as
$ | \Phi_0 \rangle = {c \choose s}   $ and  
$ | \Phi_1 \rangle = {c \choose -s}$, with $c=\cos \alpha$ and $s=\sin \alpha$.
Using the notations of density matrices 
\begin{equation}
\Phi_0 \equiv | \Phi_0 \rangle \langle \Phi_0 |
\end{equation}
etc. the two pure states
are
\begin{equation}
\Phi_p   
= \frac{1}{2} \left(\begin{array}{cc} 1 + z & x \\ x & 1 - z 
  \end{array}\right) 
\ , \end{equation}
with 
$ z = \cos 2 \alpha $ and $x =  \sin 2 \alpha$ for $p=0$ and
$x = -\sin 2 \alpha$ for $p=1$.
If $\Phi_p$ is used to describe bit $p$,
the receiver can identify the bit by distinguishing the two pure states.
Two (not necessarily pure) density matrices $\rho_p$ 
in two-dimensional Hilbert space,  
with equal determinants (which are equal to $|r|$) can also be expressed
using similar form with 
$z = |r|\cos 2 \alpha$ and $x = \pm |r| \sin 2 \alpha$.
For two such mixed states 
let us choose a state $\chi_n$, and two 
pure states $\Phi_0$, $\Phi_1$ such that
\begin{eqnarray} \rho_0 &=& m \Phi_0 + (1-m) \chi_n    \nonumber \\   
 \rho_1 &=& m \Phi_1 + (1-m) \chi_n   \label{neutral} \ .  \end{eqnarray}
Let $I$ be some (positive)
measure for the optimal distinguishability of two states,
so that {\em any operation done on them} cannot lead to a 
distinguishability better than $I$.
From the construction of~(\ref{neutral}), 
it is clear that any such measure, $I$, for optimal distinguishability
must find that the two mixed states $\rho_p$ are not more distinguishable than
the two pure states $\Phi_p$.
That is 
\begin{description}

\item[Theorem] 
\begin{equation} I(\Phi_0;\Phi_1)\ge I(\rho_0;\rho_1) \ .
\end{equation}
\item[Proof]
\quad

Suppose the contrary $I(\Phi_0;\Phi_1)<I(\rho_0;\rho_1)$.
Then, when one receives $\Phi_p$
he can mix them with $\chi_n$ and derive a better distinguishability than
$I(\Phi_0;\Phi_1)$, in contradiction to the definition of 
$I(\Phi_0;\Phi_1)$.

\hfill QED
\end{description}

We can choose any measure of an optimal information
carried by these systems 
to describe the distinguishability.
Very complicated types of information can be extracted from such systems,
as for example, the optimal 
information on the parity of an $n$-bit string of 
such quantum bits~\cite{BMS,BM1}. 
In the case~\cite{BM1}, where parities of substrings 
are given, a solution exists only for pure states with small angles
(the BM result).
We can use this known solution 
to bound the optimal information  
on mixed states which are close to each other.

\begin{figure}[t]
\vspace{7cm}
\hspace{-2.3cm}
\psfig{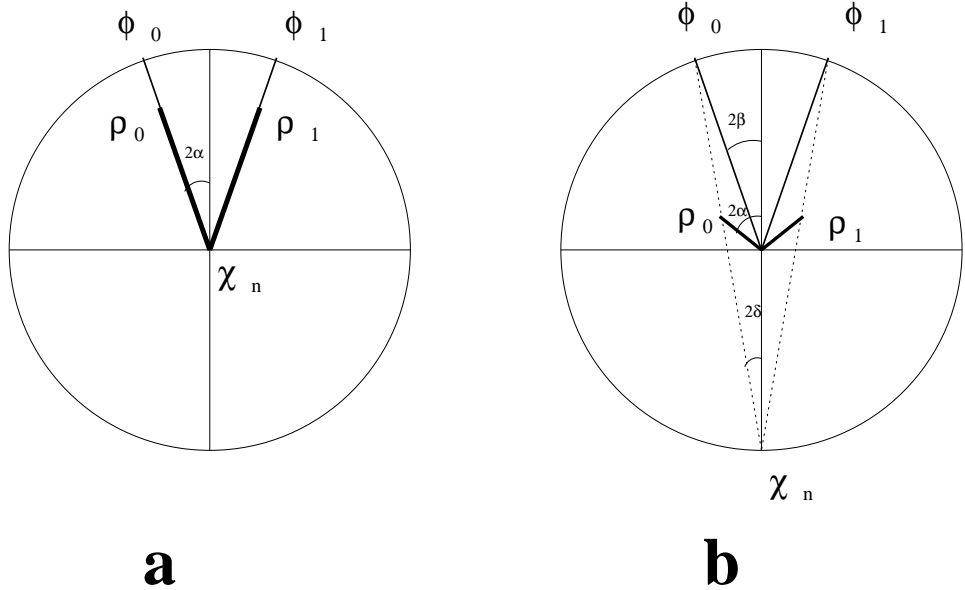}
\vspace{-7cm}
\caption[Two ways of finding 
two pure states $\Phi_p$ from which 
the two density matrices $\rho_p$ can be constructed using  
a third state $\chi_n$ 
common to both density matrices.]{
Two ways of finding 
two pure states $\Phi_p$ from which 
the two density matrices $\rho_p$ can be constructed using  
a third state $\chi_n$ 
common to both density matrices.
In a), $\chi_n = \rho_{cms}$, the completely mixed state.
In b), $\chi_n = \downarrow_z$, the ``down $z$'' pure spin state.}
\end{figure}

Let $\rho_{\rm cms}$ be the completely mixed state 
$\rho_{\rm cms} = \frac{1}{2} \hat I $.
Also let $\rho_{\downarrow}$ be the pure state of 
spin down in the $z$ direction.
Two cases of~(\ref{neutral}) are useful for our purpose:
\begin{itemize}
\item[a)] $\rho_p = m \Phi_p + (1-m) \rho_{\rm cms} $, 
where the pure states $\Phi_p$ have the same angle as $\rho_p$ (see Fig.~4.1a);
\item[b)] $\rho_p = m \Phi_p + (1-m) \rho_\downarrow $, 
where $\Phi_p$ (which are uniquely determined) are shown in Fig.~4.1b.
\end{itemize}
The first type of bound is useful if $\rho_p$ have a small angle $\alpha$
(which satisfies $\tan 2 \alpha = x/z$), so that the angle between the pure 
states $\beta = \alpha$ is also small.
The second type of bound is useful when the `distance' $2 x$ between 
the two possible mixed states is small (while $\alpha$ might be large).
In this case $x$ is small and $z$ positive hence the resulting angle 
$\beta$ between the two pure states is small
(following $\tan \beta = \tan 2 \delta = \frac{x}{z+1} \le x$).
Thus, in both cases the angle between the two pure states 
is small so that 
$ I(n,\beta) < C(n) 
 (2\beta)^{(n+1)/2}$
(with $C(n) = 
\sqrt{n+1}
\frac{2}{\ln2 \sqrt{\pi}}$)
provides an upper bound on 
Eve's information on the final key. 
Note that our final goal is to get $I(n,p_e)$, thus we need also to find 
the relation between $p_e$ and $\beta$ (through the known relation with 
$\alpha$).

\section{Security Against Symmetric Collective Attacks}\label{colC}

An explicit calculation of Eve's density matrix as a function 
of $p_e$ must be done separately 
for any suggested attack to obtain $I(n,p_e)$.
However, the fact that Eve is permitted to induce only small error rate
restricts her possible transformations at the first stage of
the collective attack, hence, the two possible states of each of her probes
must be largely overlapping (for a symmetric attack).
Concentrating on two-dimensional probes, this implies that the second 
of the two cases above can {\em always}
be used
to bound Eve's information to be exponentially small.
For certain examples -- the first case is sufficient, hence the 
angle between the two possible pure states can be calculated from Eve's
density matrix directly using $\beta = \alpha$.
Let us show two examples in details, to conclude that Eve's information
is exponentially small with the length of the string.
Both examples use the same unitary transformation
but are applied onto different quantum cryptographic schemes,
the two-state scheme~\cite{Ben92} and the
four-state scheme~\cite{BB84}. 

In our examples Eve uses a 2-dimensional probe in an initial state
${ 1 \choose 0}$. She performs a unitary transformation
\begin{equation}
U {1 \choose 0} |\phi\rangle
\end{equation}
(with $|\phi\rangle$ Alice's state), where 
\begin{equation}    
U = \left(\begin{array}{cccc} 1    &    0     &     0     &    0    \\
                              0    & c_\gamma & -s_\gamma &    0    \\ 
                              0    & s_\gamma &  c_\gamma &    0    \\ 
                              0    &    0     &     0     &    1   
  \end{array}\right) \ , 
\label{wse} \end{equation}
with $c_\gamma = \cos \gamma$, etc.
With $\gamma = \frac{\pi}{2}$ this gate swaps the particles
\begin{equation}
{1 \choose 0}|\phi\rangle \rightarrow |\phi\rangle {1 \choose 0} \ .
\end{equation}
Eve chooses a small angle $\gamma$ so that the attack is a {\em weak swap}. 
Let Alice's possible initial states be 
$|\phi_p\rangle={\cos \theta \choose \pm \sin \theta}$
in the two-state scheme, and 
$|\phi_m\rangle = 
\frac{1}{\sqrt 2}{1 \choose i^m}$ (with $i=\sqrt{-1}$ 
and $m=0\cdots 3$) in the four-state 
scheme. 
The corresponding final states are 
\begin{equation}
|\Psi_p\rangle = \left(\begin{array}{c} \cos\theta     \\
                       \pm    \sin\theta \   c_\gamma        \\ 
                       \pm    \sin\theta \   s_\gamma        \\ 
                              0    
  \end{array}\right) \ ; \quad
|\Psi_m\rangle = \frac{1}{\sqrt 2} \left(\begin{array}{c} 1  \\
                                     i^m \   c_\gamma        \\ 
                                     i^m \   s_\gamma        \\ 
                                          0    
  \end{array}\right) \ , 
\label{fin} \end{equation}
respectively.

Bob's reduced density matrices (RDMs)
are calculated from $|\Psi\rangle\langle\Psi|$
by tracing out Eve's particle. 
This operation is usually denoted by 
\begin{equation} \rho_B = {\rm Tr}_{{}_E}\ 
[|\Psi\rangle \langle \Psi|] \ ,
\end{equation}
where the full formula is given
by eq.~(5.19) in~\cite{Per93}
($\rho_{nm}=\sum_{\mu\nu} \rho_{n\nu,m\mu} \delta_{\mu\nu}
=\sum_{\mu} \rho_{n\mu,m\mu}$).
We denote this operation by
\begin{equation} \rho_B = {\rm Tr}_{{}_E}\ 
\left[(|\Psi\rangle \langle \Psi|) \hat{I}\right] \ , 
\end{equation}
where the two-dimensional $\delta_{\mu\nu}$ is written as $\hat{I}$.
From Bob's matrices we find the error rate,
that is, the probability $p_e$ that he recognizes a wrong bit value.
Calculating Eve's density matrix is more tricky; 
we need to 
take into account the additional information she obtains from the classical
data, in order to obtain an {\em information-dependent RDMs}.
This is a trivial task for the four-state scheme but a rather confusing 
task in case of the two-state scheme.

In case of the four-state scheme Bob measures his particle in one of the 
basis $x$ (corresponding to $m=0,2$) or $y$ ($m=1,3$).
Suppose that Alice and Bob use the $x$ basis; 
Bob's RDMs are
\begin{equation}
 \rho_B  = \left(\begin{array}{cc} 
 \frac{1}{2} + \frac{1}{2} (s_\gamma)^2     & \pm \frac{1}{2} c_\gamma    \\ 
 \pm  \frac{1}{2} c_\gamma     &   \frac{1}{2} - \frac{1}{2} (s_\gamma)^2     
  \end{array}\right) \ ,
\end{equation}
leading to an error rate $p_e=\sin^2(\gamma/2) $ which is the probability 
that he identifies $|\phi_2\rangle$ when $|\phi_0\rangle$ is sent.
Eve has the same knowledge of the basis, hence her 
information-dependent RDMs are 
\begin{equation}
 \rho_E  = \left(\begin{array}{cc} 
 \frac{1}{2} + \frac{1}{2} (c_\gamma)^2     & \pm \frac{1}{2} s_\gamma    \\ 
  \pm \frac{1}{2} s_\gamma     &   \frac{1}{2} - \frac{1}{2} (c_\gamma)^2     
  \end{array}\right) \ ,
\end{equation}
so that $x=s_\gamma$, $z=(c_\gamma)^2$, and the relevant angles are 
$2\beta=2\alpha = (\tan)^{-1}(s_\gamma/{c_\gamma}^2)$ (using the first
type of bounds).
For a small angle $\gamma$ we get
$p_e\approx \gamma^2 / 4  + O(\gamma^4)$, 
$\beta  \approx \gamma / 2 + O(\gamma^3)$, and thus 
$p_e \approx \beta^2 + O(\beta^4)$.  The information
is thus bounded by 
\begin{equation}
 I(n,p_e) <  C(n) 
(4 p_e)^{(n+1)/4} \end{equation}
to be exponentially small
[using~(\ref{BMres})].   

It is interesting to note that this spin-exchange 
attack against the four-state scheme 
provides the optimal information vs.\ disturbance for individual
attacks. This can be easily seen by calculating
$I_2(\gamma)$ from $\rho_{{}_E}$ and then by comparing
$I_2(p_e)$ to the optimal result calculated recently 
in~\cite{FGGNP}.
As a result, this attack also provides the maximal possible `distance' 
between Eve's possible density matrices (on the Poincar\'e sphere)
for a given error rate.
Thus, with some more work, it is possible to obtain 
a bound on Eve's optimal information
on the final key using any symmetric collective attack (on the
four-state scheme) which uses probes in 
2 dimensions. This interesting result was obtained only
while adding final corrections to this thesis, and thus, it 
shall be explicitly shown in a future review paper on the security
issue, but not in this thesis.

In case of the two-state scheme 
Bob's RDMs are
\begin{equation}  \rho_B  = \left(\begin{array}{cc} 
 (c_\theta)^2 + (s_\theta)^2 (s_\gamma)^2  & 
           \pm   c_\theta s_\theta c_\gamma    \\ 
\pm c_\theta s_\theta c_\gamma    & (s_\theta)^2 (c_\gamma)^2  
  \end{array}\right) \ .
\end{equation}
Bob chooses one of two possible measurements, $M_{0\rightarrow 1}$ or
$M_{1\rightarrow 0}$, with equal probability $p_{0\rightarrow 1}
=p_{1\rightarrow 0} = 1/2$;
In case of $M_{0\rightarrow 1}$, Bob measures the received state to distinguish 
$\phi_0$ from its orthogonal state ${\phi_0}'$ 
and finds a conclusive result `1'  
whenever he gets ${\phi_0}'$. (The conclusive result `0' is obtained
by interchanging $0$ and $1$ in the above).
The error rate is the probability of identifying
${\phi_p}'$ when $\phi_p$ is sent, and  it is 
\begin{equation}
p_e = (s_\theta)^2 (c_\theta)^2 [ 1 - c_\gamma]^2 + (s_\theta)^4 (s_\gamma)^2
\ . \end{equation}

To obtain Eve's density matrices one must take into account all 
the information she possibly has. If one ignores the classical information 
and calculates the standard RDMs (as in~\cite{FP}), then the result
is of significant importance to quantum information, while it is less
relevant to quantum cryptography.
Recall that Bob keeps only
particles identified conclusively (as 
either ${\phi_0}'$ or ${\phi_1}'$);
Bob informs Alice --- and thus Eve --- which they are, and, as a result,
Eve knows that Bob received either 
${\phi_0}'$ or ${\phi_1}'$ in his measurement, and not
${\phi_0}$ or ${\phi_1}$.
This fact influences her density matrices, 
and these are not given anymore by the simple 
tracing formula  
$\rho_E = {\rm Tr}_{{}_B}\ 
\left[(|\Psi\rangle \langle \Psi|) \hat{I}\right]$. 
In general, {\em information dependent} RDMs are obtained  
by replacing $\hat{I}$ by any other positive operator $\hat{A}$:
\begin{equation} \rho_E = {\rm Tr}_{{}_B}\ 
\left[(|\Psi\rangle \langle \Psi|) \hat{A} 
\right] \label{tracing}     \end{equation}
(This is a rather obvious conclusion from the discussions prior to 
eq.~(5.19) and also from 
page 289 in Section 9 in~\cite{Per93});
The correctness of this technique can easily be verified.
In our case 
\begin{equation}
\rho_E = {\rm Tr}_{{}_B}\left[ ( |\Psi \rangle \langle \Psi |)  
(\frac{1}{2} |{\phi_0}'\rangle \langle {\phi_0}'| + 
\frac{1}{2} |{\phi_1}'\rangle \langle {\phi_1}'|) \right]
\ , \end{equation}
where the halves result from $p_{0\rightarrow 1}$ and 
$p_{1\rightarrow 0}$.
This tracing technique leads to 
\begin{equation}    
\rho_E =  
    \left(\begin{array}{cc} 
 (s_\theta)^2 (c_\theta)^2 + (s_\theta)^2 (c_\theta)^2 (c_\gamma)^2  & 
                                 \pm    c_\theta (s_\theta)^3  s_\gamma    \\ 
 \pm c_\theta (s_\theta)^3  s_\gamma & 
                                        (s_\theta)^4 (s_\gamma)^2  
  \end{array}\right)  . 
\label{2-sE} \end{equation}
After normalization we get
\beq x = 2 s_\gamma c_\theta (s_\theta)^3 /{\rm Tr} \rho_E \eeq
and 
\beq z = \frac{1+z}{2} - \frac{1-z}{2} = 
[(c_\theta)^2 (s_\theta)^2 [1+(c_\gamma)^2] - (s_\theta)^4 (s_\gamma)^2] /
{\rm Tr} \rho_E \ . \eeq
The relevant angles are again $2\beta = 2 \alpha = \tan^{-1} (x/z)$.
For small angle $\gamma$ we get  
$p_e \approx {s_\theta}^4 \gamma^2 + O(\gamma^4) $,
$2 \beta  \approx (s_\theta / c_\theta ) \gamma + O(\gamma^3)$.
Finally we get 
\begin{equation} p_e \approx (s_\theta)^2 (c_\theta)^2 (2 \beta)^2 
+ O(\beta^4) \end{equation} 
from which we find 
\begin{equation}  I(p_e,n) < C(n) \left(\frac{p_e}{s_\theta c_\theta}
\right)^{(n+1)/4}
\ , \end{equation}
with $C(n)$ as before.

From the dependence on $p_e^{(n+1)/4}$ (that is, from $p_e \approx
\beta^2$) one might get the wrong
impression that this attack is much worse to Eve than 
the EHPP attack [see~(\ref{BMer}) where $p_e \approx \alpha^4$].
This is indeed the case since the EHPP attack is especially designed
to attack the two-state scheme. 
However, Alice and Bob can easily overcome this problem
if they use other (additional) states in the protocol.
Such additional states are used only for error estimation
and are much more sensitive to the EHPP attack.
For instance, if Alice sends the additional state
${1 \choose 0}$ to Bob, then Bob's RDM is
\begin{equation} 
\left( \begin{array}{cc}
\frac{{c_\alpha}^2 {c_{\theta'}}^2 }{{c_\theta}^2} & 0 \\
 0 & \frac{{s_\alpha}^2 {s_{\theta'}}^2 }{{c_\theta}^2} 
\end{array}\right)   
 \ , \end{equation}
as a result of the EHPP transformation~(\ref{EHPPtrans}).
The induced error rate is 
\begin{equation} p_e = 
 \frac{{s_\alpha}^2 {s_{\theta'}}^2 }{{c_\theta}^2} \approx \tan^2 \theta 
 \  \alpha^2 \end{equation}
proving that $p_e \approx \alpha^2$ as in the attacks presented
in this section
(actually, Alice and Bob could even identify that Eve is performing 
an attack specially designed against the two-state scheme if $p_e$
is so much different for different states).
Thus, if Alice and Bob use such verification states, Eve must take this into
account and decrease $\alpha$ in an appropriate way.
As a result of the existence of such unique attacks against
the two states of the two-state scheme, and the recent proof that the
optimal attack on the four-state scheme also yields $p_e \approx
\beta^2$~\cite{FGGNP} it is quite clear that it is better 
to abandon the two-state scheme as it is 
today and to modify it as we just suggested.
Whether the modified two-state scheme still deserves being called
a two-state scheme is a matter of taste, but clearly it loses
one of its main advantages over the four-state scheme, namely using
two states instead of four for secure transmission.

\section{Security Against Non-Symmetric Collective Attacks}\label{colD}

While writing this thesis we extended the previous result a bit
further, analyzing also a typical example of a non-symmetric collective 
attack. These results are yet partial, but due to their importance, we
prefer to add them in an incomplete form rather than to ignore them.

In this section we present the gate of {\em weak measurement}
and use it to perform a non-symmetric attack on the four-state scheme.
This attack generalized the simple attack presented in 
Section~\ref{int.qua-cry},
where Eve performed standard measurements in one of the two
basis. In the original attack Eve can only attack a small potion of the
particles, $\eta$, or else she induces too much noise.
Using the weak measurement gate presented here, she can attack all
particles (for any permitted error rate).

A gate for standard measurement performs the transformation
\begin{equation} {1\choose 0}_E 
{1/\sqrt2 \choose \pm 1/\sqrt2}_A
\longrightarrow 
{1/\sqrt2 \choose \pm 1/\sqrt2}_E
{1/\sqrt2 \choose \pm 1/\sqrt2}_B \ .
\end{equation}
We generalize this gate to 
\begin{equation} {1\choose 0}_E 
{1/\sqrt2 \choose \pm 1/\sqrt2}_A
\longrightarrow 
{\cos \gamma \choose \pm \sin \gamma }_E
{1/\sqrt2 \choose \pm 1/\sqrt2}_B \ ,
\end{equation}
which yields the previous gate for $\gamma = \pi/4$ and yields 
a weak measurement gate for small $\gamma$.
This unitary transformation stands for a weak measurement in the $x$ basis.

Let us apply the unitary transformation to ${1 \choose 0}_A$
and ${0 \choose 1}_A$:
\begin{equation}
\left(\begin{array}{c} 1 \\ 0 \\ 0 \\ 0
\end{array}\right)  
\rightarrow 
\left(\begin{array}{c} c_\gamma \\ 0 \\ 0 \\ s_\gamma
\end{array}\right)  
\quad ; \quad     
\left(\begin{array}{c} 0 \\ 1 \\ 0 \\ 0
\end{array}\right)  
\rightarrow 
\left(\begin{array}{c} 0 \\ c_\gamma \\ s_\gamma \\ 0
\end{array}\right)  
\ . \end{equation} 
This unitary transformation can be chosen as:
\begin{equation}
\left(\begin{array}{cccc}
                  c_\gamma  &     0     &     0     & -s_\gamma  \\
                      0     & c_\gamma  & -s_\gamma &     0      \\
                      0     & s_\gamma  & c_\gamma  &     0      \\
                  s_\gamma  &     0     &     0     &  c_\gamma  \\
\end{array}\right)  
\label{measur-gate}\end{equation}
Applying this ($x$ basis) 
weak measurement transformation to $|\phi_p \rangle$ and $|\phi_m \rangle$ we
get
\begin{equation}
\left(\begin{array}{c} c_\theta \\ \pm s_\theta \\ 0 \\ 0
\end{array}\right)  
\rightarrow 
\left(\begin{array}{c} c_\gamma c_\theta \\ \pm c_\gamma s_\theta \\ 
 \pm s_\gamma s_\theta \\ s_\gamma c_\theta
\end{array}\right)  
\quad ; \quad     
\frac{1}{\sqrt 2}\left(\begin{array}{c} 1 \\ i^m \\ 0 \\ 0
\end{array}\right)  
\rightarrow 
\left(\begin{array}{c} c_\gamma \\ i^m c_\gamma \\ i^m s_\gamma \\
s_\gamma 
\end{array}\right)  \ .
\end{equation} 
One can verify that for $\gamma = \pi/4$ and $m=0,2$ this is the
standard measurement gate (in the $x$ basis) we started with.

When Eve applies this transformation onto the two-state scheme as above, 
she actually applies another
symmetric collective attack. 
However, when Eve applies this gate to attack the four-state scheme
by choosing randomly in which basis ($x$ or $y$, in our case) to 
measure\footnote{ The gate presented above can be easily generalized
to give a weak measurement in any other direction.}, she applies 
a very enlightening example of a non-symmetric collective
attack.

We consider the case where Eve performs
a weak measurement in the $x$ direction.
If Alice and Bob also use the $x$ basis
Eve induces no errors and the angle for her probe
is $\alpha = \gamma$.
If they use the $y$ basis
the reduced density matrices in Bob's hands are
\begin{equation}
 \rho_B  = \left(\begin{array}{cc} 
 \frac{1}{2}      & \mp \frac{i}{2} c_{2\gamma}    \\ 
 \pm  \frac{i}{2} c_{2\gamma}   &   \frac{1}{2}     
  \end{array}\right) \ ,
\end{equation}
leading to an error rate 
$p_e=\sin^2 \gamma$.
The average error rate is thus,
$p_e=(\sin^2 \gamma)/2$, and this bounds $\gamma$ to be small.
For small angles $\gamma = \sqrt{2 p_e}$.

Eve has the same knowledge of the basis, hence her 
information-dependent RDMs are 
\begin{equation}
 \rho_E  = \left(\begin{array}{cc} 
 \frac{1}{2} + \frac{1}{2} (c_{2\gamma})   & 0    \\ 
        0    &   \frac{1}{2} - \frac{1}{2} (c_{2\gamma})     
  \end{array}\right) \ .
\end{equation}
These RDM's are independent of the state sent by Alice.
Thus, Eve's density matrix provides no information at all.

Although Eve's average information on each bit does not differ
much from the previous example (using the spin-exchange gate),
her average information on the parity bit is much reduced.
In most cases it is exactly zero since 
the probability that Eve guesses the basis of all the bits
in an $n$-bit string correctly is $2^{-n}$.
Even when she does guess it correctly, her information is still 
exponentially small 
\begin{equation} 
I < C(n) (8 p_e)^{(n+1)/4} \ , 
\end{equation}
but larger than the information Eve obtains using the spin-exchange gate
by a factor of $2^{(n+1)/4}$. 
On average, her information is reduced by a factor of
$2^{[(n+1)/4] - n} \approx 2^{-3 n/4}$ compared to that symmetric attack.

To gain a better comparison, one should calculate Eve's information
in a symmetric collective attack using 
a weak measurement in the Breidbart basis (the basis symmetric to 
the $x$ and the $y$ bases), or even better --- to calculate
Eve's information in a non-symmetric attacks where
Eve measures in some chosen angle from the Breidbart basis.

\section{Security Against Any Collective Attack}\label{colE}

We have seen simple symmetric collective attacks in Section~\ref{colC}.
The information available to Eve when she performs any other
symmetric collective
attack with two-dimensional probes can also be calculated using our method.
In particular, when the optimal attack of that type will be found,
our method can be used to prove security against it; 
For small enough
error rate Eve's probes must have small angles between them, 
and thus, our proof can be applied using the first type of bounds (Fig.~4.1a). 

The second type of bounds 
(Fig.~4.1b) is usually
irrelevant when the 
attack is given 
(since Eve's initial state is usually in a pure state), 
but it can be very useful for {\em finding}
the optimal attack, requiring only to find the maximal `distance', 
$2 x$, between
the two possible states of the probe. For the attacks on the two-state scheme
the optimal distance can be found as in~\cite{FP}. Note that it will not
provide the optimal attack, but it shall provide a bound on Eve's 
optimal information.
For the attacks on the four-states scheme we can already perform the
calculation as previously explained, and shall do it very soon.

More general collective attacks can use non-symmetric translucent attacks
and/or can use probes in higher dimensions
in the first step of the collective attack.
Methods similar to ours can be used for proving security against various 
non-symmetric collective attacks (in 2 dimensions), 
as is done in Section~\ref{colD}.
But the calculation
becomes more complicated, and we are not sure if all cases can be 
solved. 
However, the case solved in the previous section provides some intuition
why non-symmetric attacks
are not good for Eve.

Our bounds cannot be used when 
Eve uses higher dimensional probes.
On one hand, the two possible states of
each probe in this case are still highly overlapping, and the same intuition
which holds in our work shall still hold.  However, 
extending the information bounds we found to three or four dimensions
might be a difficult task since there no simple geometrical
representations as for two dimensions.
Note that analysis of
dimensions higher than four is not required since they cannot help
the attacker due to  
reasons shown in~\cite{FP}.

\chapter{Quantum Networks for Quantum Cryptography}\label{net}

Quantum correlations between two particles 
show non-classical properties 
which can be used for providing secure transmission 
of information.
In this chapter we present a quantum cryptographic system, 
in which users store particles
in a transmission center, where their quantum states
are preserved using quantum memories.
Correlations between the particles stored by two users 
are created upon 
request by projecting their product state  
onto a fully entangled state. 
Our system secures communication between
any pair of users who have particles in the same center.
Unlike other quantum cryptographic systems, it can work 
without quantum channels and is suitable for  
building a quantum cryptographic network.  
We also present a modified system with many centers.
This part was done together with Eli Biham and Bruno Huttner~\cite{BHM}.

The new developments in quantum computation
started a wide surge of interest in the field
of quantum computing gates and devices.  
However, building such computing devices
is a difficult task, and
quantum computing
is only 
doing its first experimental steps.
The building blocks of future quantum computers are one-bit and 
two-bit quantum logical
gates~\cite{Deutsch89,DiV,BDEJ,CZ},
which are currently under intensive 
development~\cite{Haroche,Kimble,Wineland}. 
Building quantum computing 
devices to factor large numbers 
does not seem to be practical in the foreseeable future since it 
requires combining many one-bit and 
two-bit gates. However, 
a single two-bit gate also have intriguing uses in
information processing
and quantum communication such as 
teleporting a quantum state~\cite{BBCJPW}, and dense coding in
quantum cryptography~\cite{BW}. 
We shall show in this chapter
that the use of quantum gates together with quantum
memory
opens new directions in quantum cryptography.
Our system may be practical long before quantum computers are, 
hence it provides a short-term application for quantum gates.

One of the main disadvantages of quantum cryptography is its
restriction to relatively short channels. This is due to the fact
that, in contrast to classical channels, a quantum channel cannot use
repeaters\footnote
{In parallel, we investigated this problem as well and find a surprising
way of defining quantum repeaters. See Section~\ref{repeaters} 
and Section~\ref{QECforQPA}.}
to amplify the signal without loss of coherence.
Currently, working prototypes enable transmission to distances of
many kilometers (see the general introduction in Chapter~\ref{int}).
Commercial systems may become available in the
near future~\cite{LosAl},
so that two users will be able to communicate securely 
(if they are not too far).    
However, building quantum cryptographic {\em networks} based on the
existing 
schemes\footnote{For a suggestion of a quantum cryptographic
networks based
on the existing schemes see~\cite{Townsend}.}
seems to cause severe difficulties (which may even 
make it impractical):    
\begin{enumerate}
\item Quantum communication requires 
any pair of users to
have a common quantum channel, or alternatively a
center (or a telephone-like switching network) connected by 
quantum channels to all the users, which should match any pair of
channels
upon request;
enhancing the security of 
the current world-wide telephone network (which contains about $N
\approx
10^9$ users (telephones) )
using quantum cryptography requires huge investments in quantum
channels and devices.
\item Any user must have the financial and technological abilities
to operate complicated quantum devices.
\item  The keys must be transmitted {\em online}, or else one would
need to transmit $O(N^2)$ keys in advance to enable any pairs of
users to communicate in secrecy.
\item The network must assure authenticity of the users.
\end{enumerate}
It is important to have quantum cryptographic networks  
not suffering from these problems. 

In this chapter we suggest a new cryptographic scheme 
in which users store quantum states in 
quantum memories, kept in a transmission  center. Upon request from
two users, the center uses two-bit gates to project the product state
of two non correlated particles (one from each user)
onto a fully entangled state. As a result, 
the two users can share a secret
bit, which is unknown {\em even to the center}. 
Our scheme can operate
without quantum channels, if the
quantum states are ``programmed'' at the center. 
In that case, the scheme does not suffer from the four problems just
mentioned, 
and can operate at any distance.
Hence,  
it is especially appropriate
for building a quantum
cryptographic
network of many users.  
Such a system actually shows some of the useful  
properties of the public key cryptosystems,
but yet, it doesn't require computation assumptions.

In Section~\ref{EPR}, we present 
a new two-party quantum cryptographic scheme
which is a time-reversed EPR-scheme.
In Section~\ref{network}, we present a quantum network based on 
this scheme
with the addition of quantum
memories.
In Section~\ref{implementation2}, we discuss the possibilities of
implementing our scheme 
in practice.
In Section~\ref{www}, we present a more advanced network, based on
quantum
teleportation, where users can store their states
in different centers and the centers teleport states upon request.
This network uses quantum channels;
however, it requires quantum channels only between the centers,
so that the 4 problems
stated above do not arise.
In Section~\ref{conclusion} we summarize this chapter.

\section{A Time-Reversed EPR-Scheme for Quantum Cryptography}
\label{EPR}

The first goal of this chapter is to suggest another scheme for
quantum cryptography, which we shall call {\em the time-reversed EPR-scheme}:
in the EPR scheme, presented in Section~\ref{int.qua-cry},
a singlet state is prepared and, 
later on, it is projected on
the states of the four-state scheme; 
in our time-reversed EPR-scheme, each user prepares
one of the four states of the
four-state scheme, and, 
later on, the product state is 
projected upon a basis which includes the 
singlet state.
Let both Alice and Bob send one of the four states,
$|\! \uparrow \rangle$, $|\! \downarrow \rangle$, 
$|\! \leftarrow \rangle$ or $|\! \rightarrow \rangle$ 
to a third party 
whom we refer to as {\em the center} (the purpose of using this name
shall
be clarified in Section~\ref{network}).  
The center measures their qubits together to find 
whether the two particles are in a singlet state. 
This can be done
by measuring the total-spin operator $(\hat{S}_{total})^2$. 
If the result of the measurement is $s=0$, then the two particles are
projected onto
the singlet state. In that case Eq.~(\ref{singlet-pair})
ensures that,
if the two spins were \underline{prepared} along the same axis,
then they necessarily had opposite values (the projection of the
states with identical spins on the singlet state is zero). 
As a result, Bob knows Alice's bit and vice versa. However, from 
Eq.~(\ref{singlet-pair}), an honest center, who follows the protocol
and projects onto the singlet state, has absolutely no knowledge on
these bits. For example, when Alice and Bob both used the vertical
axis, the center does not know whether Alice had the up state and Bob
the down state, or vice versa. 
If the measurement result is $s=1$, Alice and Bob cannot infer
anything about the value of each other's bit, and they discard the
transmission.
The probability of obtaining the singlet state is zero when Alice
and Bob sent the same state (e.g., $\uparrow\uparrow$), 
and is half in case they sent opposite
states. Taking into account the case where Alice and Bob use
different axes (which will also be discarded), we find that the
overall
probability to obtain a usable state is only one eighth.

To create a key with many bits,   
Alice and Bob send strings of quantum states 
($L'>8(L+l)$ qubits) to the center.
The center must be able to keep them for a while (in case the states
do not
arrive at the same time from Alice and Bob), and then to measure the
first pair,
the second pair etc.
The center tells Alice and Bob all cases in which the 
result of the measurement is a singlet,
which happens in one fourth of the cases.
Alice and Bob then compare their axes. 
When they used the same axis
(which happens about half of the time), they know that their
spins are necessarily opposite, and thus Bob can calculate
Alice's bits to share a key with her.
As in the BB84 scheme and the EPR schemes Alice and Bob use the
classical discussion channel to estimate the error rate.
If it is tolerable they perform error correction 
and privacy amplification, to derive a final $L$-bit key.

The security of our protocol derives from the security of the EPR
protocol, and relies on the fact that the singlet state is the only
state for which the two spins are anticorrelated both in the 
$\hat{S}_z$ and in the $\hat{S}_x$ bases. However, as explained
previously, if the center projects on the singlet state, he does not
get any information on Alice's and Bob's bits. Therefore, a cheating
center needs to project onto a different state (possibly entangled
with his own system), which cannot
give perfect anticorrelations along both  $\hat{S}_z$ and
$\hat{S}_x$ axes. Since the center cannot
know in advance which basis was used by Alice and by Bob (the two
density matrices corresponding to using $\hat{S}_z$ or $\hat{S}_x$
are identical),  he will unavoidably introduce errors, which 
Alice and Bob shall identify  during the discussion. 

\begin{figure}[t]
\vspace{8cm}
\caption[Two processes, which we use to prove 
the security of the protocol.]{
Two processes, which we use to prove 
the security of the protocol.
In both figures, one 
particle of each EPR correlated pair (denoted by dashed lines)
is sent to the center, 
who performs a Bell measurement. We consider only the case
where the result of the measurement is a singlet state.
The second
particles are sent to Alice and to Bob respectively, 
who project them 
onto the BB84 states. 
In a), the first measurement is done by
the center. The particles arriving to Alice and Bob 
are therefore in the singlet state like in the EPR based 
protocol. In b), the first measurement is performed by 
Alice and Bob. Each particle sent to the center is 
therefore in one of the BB84 states. This is similar to
our protocol.}
\end{figure}

In fact, in
terms of eavesdropping possibilities, our protocol and the EPR
protocol are equivalent, as we show using the scheme presented
in Fig.~5.1.  In this scheme, two
EPR pairs are created, one particle of each pair is sent to the
center, and the
second one to Alice and to Bob. In Fig.~5.1a, the center performs a
measurement on his two particles first.
An honest center, who follows the agreed protocol, projects the
particles 
onto the singlet
state. The two particles sent to Alice and to Bob are now in the
singlet state as well. This is therefore equivalent to the EPR
scheme. The only difference is that the projection onto the singlet
state performed by the center succeeds with probability 1/4 only.
This means that the center will ask to discard 3/4 of the
transmission, but this does not affect the eavesdropping issue.
A cheating center
can send to Alice and Bob any state he wants, 
including any desired entanglement with his own system, 
by choosing an appropriate unitary transformation 
and the correct state on which to project his own particles. 
To show that, we 
start with the two singlet pairs 
and let the center introduce an ancilla in a state $A_{\rm init}$.
The state of the whole system is:
\begin{equation}  
\Phi_{ABC} =   \frac{1}{2} \Bigl(|\! \uparrow  \downarrow \rangle 
    -  |\!  \downarrow \uparrow \rangle) \otimes
              (|\! \uparrow \downarrow \rangle 
    -   |\! \downarrow \uparrow \rangle \Bigr) \otimes A_{\rm init}
\; .
\end{equation} 
The first particle of each singlet pair is sent to Alice and to Bob
respectively, while the center keeps the second, together with his
ancilla. The state $\Phi_{ABC}$ can thus be rearranged as:
\begin{equation}
\Phi_{ABC} = 
 \frac{1}{2} \Bigl( |\! \uparrow \uparrow \rangle_{AB} \otimes
              |\! \downarrow \downarrow \rangle_{C} 
       +      |\! \downarrow \downarrow \rangle_{AB} \otimes
              |\! \uparrow \uparrow \rangle_{C} 
       -      |\! \uparrow \downarrow \rangle_{AB} \otimes
              |\! \downarrow \uparrow \rangle_{C}
       -      |\! \downarrow \uparrow \rangle_{AB} \otimes
              |\! \uparrow \downarrow \rangle_{C}           \Bigr)
              \otimes A_{\rm init} \; , 
\end{equation}
where the index $AB$ refers to the particles sent to Alice and Bob,
and the index $C$ refers to the particles kept by the center. 
The center now applies a unitary transformation $U$ to entangle his
particles with the ancilla in the following way:
\begin{eqnarray} 
U \Phi_{ABC}  =  \frac{1}{2} &\Bigl(&
               |\! \uparrow \uparrow \rangle_{AB} \otimes
              |\! \downarrow \downarrow \rangle_C  \otimes A_1
       +      |\! \downarrow \downarrow \rangle_{AB} \otimes
              |\! \uparrow \uparrow \rangle_C \otimes A_2 
\nonumber \\               
     & & \quad 
      -        |\! \uparrow \downarrow \rangle_{AB} \otimes
              |\! \downarrow \uparrow \rangle_C  \otimes A_3
       -      |\! \downarrow \uparrow \rangle_{AB} \otimes
              |\! \uparrow \downarrow  \rangle_C \otimes  A_4 \Bigr)
\; ,
\end{eqnarray}
with $A_i$ any normalized states of the ancilla which are (in general)  
not orthogonal to one another.

By projecting his state onto 
$ \psi =   
   \alpha |\! \downarrow \downarrow \rangle_C +
   \beta  |\! \uparrow \uparrow \rangle_C - 
   \gamma  |\! \downarrow \uparrow \rangle_C - 
   \delta  |\! \uparrow \downarrow \rangle_C \ \  $
(this projection succeeds with probability 1/4 on average), the center creates
the state
\begin{equation}  \Psi_{ABC} =  \frac{1}{2} \Bigl(
          \alpha^*  |\! \uparrow \uparrow \rangle_{AB} \otimes A_1
        + \beta^*   |\! \downarrow \downarrow \rangle_{AB} \otimes A_2
        + \gamma^*  |\! \uparrow \downarrow \rangle_{AB} \otimes A_3 
        + \delta^*  |\! \downarrow \uparrow \rangle_{AB} \otimes A_4  
\Bigr) \; , 
\end{equation}
which is the most general state the center could create when cheating
the EPR scheme~\cite{BBM}. This demonstrate the equivalence between
Fig.~5.1a and the EPR scheme.

In Fig.\ 5.1b, the first measurement is performed by Alice and
Bob, who project the particles onto the BB84 states. Therefore, the
particles arriving at the center are also in the BB84 states, and
this scheme is identical to ours. Since the relative time of the
measurements cannot influence the outcome, all these schemes are
equivalent.
Following the same reasoning, but in two steps (first letting only
Alice measure before the center) it is also possible to show that 
the security of the BB84 scheme implies the security of our scheme.
Since the security of the EPR scheme implies
the security of the BB84 scheme~\cite{BBM}, our proof actually 
shows that 
the security of the three
schemes is equivalent.

Using only the total spin measurement, less than one eighth of the 
qubits could be used. 
A better choice, although 
possibly more difficult to implement in practice, 
is to measure the Bell operator
(defined in~\cite{BMR}) whose eigenstates (the {\em Bell states})
are the singlet state, 
 $\Psi^{(-)}$ [equation~(\ref{singlet-pair})],
and the three other states:
\begin{equation}    
 \Phi^{(+)} 
 = \sqrt{\frac{1}{2}}
 \left(|\!  \uparrow \uparrow \rangle +  
|\!\downarrow \downarrow \rangle
\right)  
= \sqrt{ \frac{1}{2} }
 \left(|\! \leftarrow \leftarrow \rangle + 
|\! \rightarrow \rightarrow \rangle \right) 
\; , \label{phiplus} \end{equation}   
\begin{equation} 
\Psi^{(+)} = \sqrt{ \frac{1}{2} }
 \left(|\!  \uparrow \downarrow \rangle +  
|\! \downarrow \uparrow \rangle \right) 
= - \sqrt{ \frac{1}{2} }
 \left(|\! \leftarrow \leftarrow \rangle - 
|\! \rightarrow \rightarrow \rangle \right) 
   \label{psiplus} \end{equation} 
and
\begin{equation}
\label{phiminus}
 \Phi^{(-)} = \sqrt{\frac{1}{2}}
 \left(|\!  \uparrow \uparrow \rangle -  
\downarrow \downarrow \rangle \right)  
= \sqrt{ \frac{1}{2} }
 \left(|\! \leftarrow \rightarrow \rangle + 
|\! \rightarrow \leftarrow \rangle \right)  \ ,
\end{equation}   
where the second expression for each of the Bell states is derived by
expanding them (as was done for the singlet state)
into the 
$(|\! \leftarrow \rangle \, , \, |\! \rightarrow \rangle )$ basis.
Consider a case where Alice and Bob used the same basis.
According to the result of the measurement, and to the
choice of axes by Alice and Bob, their prepared states are known to
be either
correlated (e.g. if the result is $\Phi^{(-)}$ and they both used the
$z$ axis), or anticorrelated (e.g. if the result is still
$\Phi^{(-)}$ but they both used the $x$ axis).

The protocol is:
\begin{itemize}
\item
The center retrieves the particles from Alice and Bob and 
measures the Bell operator on each pair. He gets 
one of the 
four above states, and tells his result to Alice and Bob. 
\item  Alice and Bob 
tell each other the axis they used (but not the bit value). When they
used different axes, they discard the transmission. Whenever
they used 
the same axis, they know if their bits are correlated 
or anticorrelated. In this case
half of the quantum states are used 
to derive the desired key, and $L'>2(L+l)$ qubits are required. 
\end{itemize}

The proof of security for this case is similar to the proof 
in the singlet case. An honest center, who projects the states onto
the allowed states, cannot get any information on the bits. For
example, if the center obtains the state $\Phi^{(+)}$, and Alice and
Bob 
announce later that they used the horizontal axis, the center only
knows that
either both Alice and Bob have the left state, or both have the right
state. But he cannot know which of these two possibilities occurred,
hence has no information on the bit values. Moreover, similarly to
the singlet case, $\Phi^{(+)}$ is the only state for which Alice's
and Bob's states have such correlations along both $x$ and $z$ axes.
Therefore, for each bit on which a cheating center attempts to eavesdrop, he 
needs to create a different state in order to gain information, 
which shall be detected with finite probability. By
checking a large number of bits, Alice and Bob will therefore detect the 
cheating with probability exponentially close to one.

\section{A Quantum Cryptographic Network}
\label{network}
In this section we combine the reversed EPR scheme  
and the use of quantum memories into a classical network to
present a {\em quantum cryptographic network}.
The classical protocol for a network uses a {\em hidden file}
managed by a communication center. 
Any user is permitted to put data (secret keys) in the file, under his
name, but 
only the center has access to the data. 
Let there be $N$ users, and let each of them store many $L$-bits
strings.
Upon request from two users, the center uses their data and creates 
a secret key for them, which is 
shared by both of them:
the center calculates the XOR of one 
string of the first user (say, $a_1 \ldots a_L$ of Alice),
and one string of the second user (say, $b_1 \ldots b_L$ of Bob); 
the XOR of a string is calculated bit by bit using  
$c_j = a_j \oplus b_j$
(the parity of the two bits),
and the resulting string, $C = c_1 \ldots c_L$, is transmitted 
(via a classical  unprotected channel) to Alice;
Alice rederives Bob's string by calculating the XOR of  
her string with the received 
string, and can use Bob's string as their common key.
Secure transmissions from each user to the center can be done either
by personal delivery, trusted couriers
or quantum key distribution.
Such a classical key distribution scheme  
is perfectly secure if we 
assume that the center holding them  
is perfectly safe and trusted. 
No other person (except 
the center) can have any information on their key.
Even a powerful eavesdropper who can impersonate the center and all
the  
users cannot eavesdrop, since the center and each of 
the legitimate users can use 
some of the secret bits for authenticating each other.
Alice and Bob need to trust the center for two different purposes:
\begin{enumerate}
\item To ``forget'' their secret key (and not trying to listen to the
messages
transmitted using the key);
\item To authenticate one to the other in case they 
have no other way of authentication. 
This is a new possibility of authentication, added to the two
options, previously 
mentioned in Section~\ref{EPR}.
Thus the assumption
of having classical channels which cannot be modified 
can be completely removed, 
even if the users have no other way to authenticate each other.
\end{enumerate}
The main reason why this simple scheme is
not satisfactory in practice is that it concentrates too much power
in the distribution
center. Indeed the center can understand all the secret
communications going
through 
its distribution web, or connect Alice to an adversary instead of to
Bob. 
Even if we assume that the center 
is trusted,
any eavesdropper who manages to get access to it could decipher all
the 
communications.  

Using a quantum memory instead of a classical memory is the key-point
in deriving the quantum network. 
We now present a quantum key 
distribution network which uses 
a {\em quantum file} instead of the classical hidden file,
and {\em removes} the
requirement of a trusted center. 
Alternatively, we can release the usual
assumption of quantum cryptography \cite{BBBSS}
 -- that classical channels cannot be 
modified by Eve -- if we are willing 
to trust the center
for authentication (without trusting him for ``forgetting'' their
qubits).
Instead of storing $L$ classical bits to make a future
key, each user shall store $L'$ quantum states (qubits) in specially
devised
quantum memories kept in a center. 
Upon request from two users, the center 
performs the time reversed EPR scheme described in the previous
Section and creates correlations between the bits. 
The resulting
string C, which holds the correlation data is sent 
to Alice and Bob via a classical channel.
As in the classical case, using string C, Alice can calculate Bob's
string
to derive a final common key of $L$ bits.
If Alice and Bob compare bases {\em after} deriving the data from the
center, 
then, as explained in Section~\ref{EPR}, any attempt by the
center to obtain the value of these bits will create errors and be
discovered by Alice and Bob. The center therefore does not need to be
trusted anymore. 
Unlike other quantum schemes, the actual (online) distribution of the
secret
keys is performed on classical channels. First, the center let Alice
and Bob
know the state he got. Then, Alice and Bob continue as in the other
two schemes
previously described to obtain the final key.
All the quantum communication is done in advance,
when the users ``deposit'' their quantum strings in the center
(preferably in a personal meeting).

When $L$-bit strings are stored in a classical hidden file, 
two users derive $L$-bit strings of correlated bits. 
Using quantum states for representing the bits, longer strings of
length $L'>L+l$ are required, 
since some bits will be used for error estimation,
error correction and 
privacy amplification.
The exact ratio depends on the expected error rate in the channel.
Only the bits which are encoded in the same basis by both users 
can be used, therefore $L' > 2(L+l)$ bits are actually required 
(we assume that the more 
efficient scheme of measuring the Bell basis is used).

Let us summarize the protocol as follows:
\begin{itemize}
\item In the preparation step the user sends (gives) 
\mbox{$L'$-qubit} strings
to the center, each qubit is   
one of the four states of the BB84 protocol.
The center keeps these quantum states in a quantum file without
measuring them. It is important that the system used for keeping the
quantum states will preserve them for a long time (as long as
required
until the actual key distribution is performed). 
\item 
When Alice and Bob wish to obtain a common secret key, they ask the
center to 
create correlations between two strings, one of Alice and one of Bob.
The center performs the Bell operator measurement 
on each pair of qubits, which  projects them onto one of the Bell
states, and tells Alice and Bob the result he obtained. 
After Alice gets the results from 
the center (and not before that),  
Alice and Bob compare the basis they used and keep only 
the bits for which they used the same basis. 
In this case, and according to the state obtained, the  states of
Alice and Bob are either correlated or anticorrelated. So, Alice for
example inverts all her bits which should be anticorrelated with
Bob's. The remaining string should be identical with Bob's, apart
from possible errors.
\item An honest center, who performed the correct projections on the
Bell states does not get any information on the string.
\item A cheating center (or any other eavesdropper who might have had
access to the quantum files), who modifies the allowed states,
unavoidably introduces errors between the two strings.
\item Alice and Bob perform error estimation, error correction
and privacy amplification to derive a final key.
\end{itemize}

The quantum channel 
is used only as a preparation step between each user and the center, 
and all the online communication 
is done via a classical channel.
Yet, only $O(N)$ keys 
are required 
to enable secret communication between 
any pair of the $N$ users.
The conventional quantum key distribution schemes require 
$O(N^2)$ keys, or else require online quantum communication.
In fact, our scheme does not require quantum {\em channels} at all.
As in old implementations of quantum cryptography~\cite{BBBW}, 
the four quantum states can be chosen in any 2-dimensional 
Hilbert space.
Instead of sending them, each user could arrive once in a while to 
the center, and ``program'' his states into the quantum file.
If the memory can keep the states unperturbed long enough then each
user
can put as many strings as he needs till his next visit to the
center.
By using personal delivery of the quantum states we
replace the distance
limitation of all other schemes by a time limit, 
and solve the problems
of a quantum cryptographic network which were described in the
introduction to this chapter.
All the technologically difficult 
steps, such as storing qubits and 
performing Bell measurements
occur only at the center.

\section{Implementations}
\label{implementation2}
Our scheme requires the possibility to program, 
store and manipulate quantum bits rather than to transmit them.
Therefore, the discussion in Section~\ref{int.mem} 
perfectly applies to our demands.
In our case the quantum bits
are subjected only once to a unitary operation of calculating the Bell state.

We estimate\footnote{The following suggestions were investigated
with the help of D. DiVincenzo~\cite{DD}.} that 
it may be possible to implement a working prototype of 
our scheme within a few years, 
with small modifications of existing technology of ion traps.
Such a prototype shall be able to keep quantum states of several users for
a few minutes 
and to allow to perform two-bit operations on qubits of two users.
To do so, the center should hold $L'$ ion traps, and
each of the two (or more) users puts one qubit in each ion trap. 
Upon request of two users the center shall be able to 
measure the $i$'th qubit of Alice and the $i$'th qubit of Bob 
together in the $i$'th ion trap.
It may well be that such a prototype will not enable 
an operation on the qubits of other users,
since the state of the other qubits in the ion trap might be destroyed.

The way to establish a real working scheme with a center and 
with many users is still long.
The main obstacle is that it is currently infeasible to transfer
a quantum state from one ion trap to another.
A real network should enable each user to program his string of quantum states
in a register (say, at least one separated ion trap for each user).
Then, upon request of two users, the center should be able to move
one qubit of each of the two users to another ion trap where he can
perform
the Bell measurement without disturbing the other quantum states.
As discussed in Section~\ref{int.mem},
this might be possible using `cavitraps'
but it is difficult to estimate now whether 
it will become practical in a few
years. It may well be that other candidates based on nuclear spins
will become more favorable.

\section{World-Wide Network of Many Centers}
\label{www}

The network of Section~\ref{network} is well suited for 
communication
among users who
are not far away and can arrive to the center.
For two users who are far away from each other 
and cannot come to the same center
our network may not be appropriate.
We now show that it can be modified to be useful also in this case.
Let there be many centers, and many users registered in each center.
Every two centers should share many EPR singlet pairs.
Upon request of users of the two centers, one center would
teleport \cite{BBCJPW} the
qubits of his user to the other center. This operation can be done
with 100\% efficiency (but it may, however, increase the error rate).
The singlet pairs can be transmitted using any quantum cryptography 
scheme or even using a teleportation scheme and
a {\em supercenter} who shares EPR pairs with all centers. 
However, the transmission and distribution of singlet pairs require
quantum channels even if done in advance\footnote{And transporting 
quantum states to deliver them by a personal meeting is similar
to transmitting them through channels.}.

Do we lose all the benefits we gained before?
Certainly not.
Still, only the centers need to have ability of performing quantum
operations.
All quantum transmission is done in advance, and yet,
there is no need for $O(N^2)$ strings, since the number of centers
is much smaller than the number of users.
Authentication is still simple.
One problem of any quantum channel is the limit on its length.
Our first scheme (with one center) replaced it by  a limit on time. 
It would be bad to have both problems in a network of many centers.
However, the suggestion of purifying singlets~\cite{purification} 
enables transmitting signals to longer distances.
Our scheme can make an excellent use of it, since only the
centers
need to have the technological ability of purifying singlets.
Moreover, several transmission stations can be put in between to
improve 
transmission (this idea was suggested by DiVincenzo~\cite{DD}) by
performing purification of singlets between any two neighboring
stations
and then use teleportation from one station to the next to derive
purified singlets shared by the centers.
   
\section{Summary and Discussion}
\label{conclusion}

In this chapter we introduced a new scheme for quantum cryptography
based on a time-reversed EPR scheme.
We suggested two new types of networks:
a classical one based on hidden files, and one based on quantum
files.
The security of key distribution protocols in these networks 
does not rely on computational complexity assumptions.
Both networks can be used 
to distribute keys in a secure way among any two users using 
simple online communication via classical channels.
In the case of hidden files it is done with the help
of a trusted center who can have access to all the information
exchanged through its lines. 
Using quantum files, the center need not be trusted.
Users have the means to
check whether the center, or any other eavesdropper, tried to obtain
information on the transmitted messages. 

The one-center quantum network we suggested
does not require any quantum channel 
at all, and can be implemented in a center where 
each user ``programs'' 
his states into a quantum memory.
We estimate that a working prototype may be built in the near future
using ion-trap technology.
A real network can be built when the problem of transmitting 
a quantum state from one trap to another is solved (perhaps
using
cavitrap with polarization states).
The machinery required for our scheme is also required for 
much more complicated tasks such as purification and quantum
computing.
In this respect, our scheme may represent a first practical 
application for these new devices, which are now being planned 
in various laboratories. 
We hope that our work will motivate more research for systems which
can both
keep a quantum state for a long time, and enable the desired
programming
and measurements.
We didn't pay much attention to the delicate problem of programming
the states.
The programming requires a simple equipment to make sure that the
center does not
eavesdrop on the preparation step. While cavitraps are very
complicated,
programming the polarization
state of each of the photons may be quite simple in the future.

A future system of secure communication based on the protocol
of Section~\ref{www}
would involve a number of large transmission 
centers, which can exchange EPR correlated particles and 
store them, together with the qubits deposited at the centers
by various users. Secure communication between any pair
of users would then requires teleportation of the states to 
the same center, followed by the creation of correlations
between the two strings.

\chapter{Quantum Error Correction and Quantum Privacy Amplification
}\label{QECQPA}

While doing this research, the 
feasibility of quantum memory was much increased (at least conceptually) 
due to the suggestions for quantum error correction~\cite{Shor1}.
Furthermore, additional fascinating applications of quantum memories 
were suggested
by other people:
\begin{enumerate}
\item breaking quantum bit commitment~\cite{DM4} 
\item entanglement purification~\cite{purification}
\item quantum privacy amplification~\cite{oxford}
\end{enumerate}
The first issue was mentioned in Chapter~\ref{int}, 
and we will not discuss it further.
Let us explain the second issue in brief:
Alice and Bob exchange many singlet pairs 
through a noisy channel.  Because of the natural noise (or any other reason), 
the resulting entangled states are imperfect.
Using an entanglement purification process~\cite{purification}, they can  
distill (if the initial error rate is not too large) nearly perfect
singlet states. 
This is impressive since the 
purification is done by local operations
performed at Alice's site and Bob's site,
and using only classical discussion.

In this chapter we present our research on quantum error correction and 
quantum privacy amplification.
Quantum error correction and quantum privacy amplification generalize
the analogous classical concepts to the quantum case.
In the classical case, error correction 
is useful for improving communication, computation and memory,
and privacy amplification is useful for cryptography.
We use the term quantum privacy amplification literally.

Quantum computation, communication, and memory are very sensitive to noise.
Unlike the classical cases, errors cannot be removed using simple
redundancy if non-orthogonal quantum states are used,
due to the no-cloning theorem.
Quantum error correction (QEC)
was suggested~\cite{Shor1} as a way to solve these problems by 
using new type of redundancy.
Several schemes have been recently suggested to reduce these problems, 
mainly focusing on quantum computation, trying to minimize the redundancy.
In Section~\ref{QEC} we investigate the role of QEC in 
quantum communication and memory. 
We focus on the fidelity of the 
the recovered state, an issue ignored by most of the works on this subject.
To improve the fidelity, we suggest to replace the error-correction codes
by {\em error-reduction} codes, and we analyze the ways to calculate the
fidelity. 
For $n$, a large integer, we suggest a code which
uses $N=n^2$ quantum bits to encode one quantum bit, and reduces 
the error exponentially with $n$~\cite{QEC}. 
Our result suggests the possibility
of sending quantum states over very long distances, and maintaining
quantum states for very long times. 
This part was done with the help of Eli Biham, 
Gilles Brassard, Lior Goldenberg and Lev Vaidman.

Privacy amplification is an important step in any quantum
key distribution protocol.
While we (and also Yao~\cite{Yao} and Mayers~\cite{DM2}) 
investigated the efficiency
of classical privacy amplification against a quantum eavesdropper,
the possibility of using quantum privacy amplification (QPA) instead 
of classical privacy amplification 
was suggested~\cite{oxford}.
Ignoring the problem of practice, this idea is very interesting.
They suggested to perform a QPA which is based upon the entanglement 
purification (EP) developed in~\cite{purification}. 
We call this process {\em EP-based QPA}, for convenience. 
In this process Alice and Bob use purified singlets to execute
the EPR scheme or the four-state scheme (via teleportation). 
They can even choose a fixed basis and use orthogonal states,
if they previously verified that the quantum privacy amplification
works to yield the desired entanglement purification.
The EP-based QPA provides a proof of security of quantum key distribution
under the assumption that Alice and Bob have perfect devices\footnote{
The assumption appears on page 2819, col.\ 1, line 18.}~\cite{oxford}.
In Section~\ref{QPA} we criticize the possibility that 
EP-based QPA provides secure quantum key distribution.

One of the main surprises of the idea of~\cite{oxford} is that, 
unlike the classical case (where error correction and privacy amplification
are usually considered separately), the suggested
QPA process already contains the 
error correction as well!   
This was a hint that QEC itself could be used 
for suggesting another QPA process.
In Section~\ref{QECforQPA} we show that QEC codes can also serve for QPA
(e.g., by encoding the qubits of the four-state scheme), 
if a simple randomization step is applied.

In the previous chapter and in the literature~\cite{BBM}
it is usually claimed that the EPR scheme provides no significant
benefit over the four-state scheme. 
The EP-based QPA suggests a  
genuinely different quantum key distribution protocol, where the
use of shared entangled states  
yields a protocol that has no analogue
along the lines of the four-state scheme.
Our QEC-based QPA shows similar features to the EP-based QPA, 
and it can work with any scheme.
Due to the use of randomization of the code qubits, 
Alice could even start with
only two known orthogonal states, encoded using the QEC code, to execute
the key distribution. This aspect also presents a 
similarity to the use of the EP-based QEC. 
Finally, 
we explain that
QPA might still be crucial to the security of
quantum key distribution over large scale distances or times.
This is due to quite different reasons 
from those suggested in~\cite{oxford}.

\section{Quantum Error-Correction}\label{QEC} 

\subsection{Error-Correction and Error-Reduction}

The ability to correct errors in a quantum bit (qubit) 
is crucial to the success of quantum computing,
and it is very important to the implementations
of quantum communication systems and quantum memories.
Motivated by quantum computing, 
Shor~\cite{Shor1} shows that quantum errors can be corrected
(in some analogy to classical error correction~\cite{MS}).
The many works which follow Shor's idea focus 
on improving his result~\cite{BDSW,LMPZ,Steane1,CS} to better fit 
the requirements of quantum computing, or to provide a 
better understanding of the properties of the error-correction 
codes (the previous works and also~\cite{Steane2,EM,Shor2}).  
In this section we apply this idea to quantum memories and 
quantum communication, where improving the fidelity of the recovered state
is the main goal.
The {\em error-reduction}
scheme we suggest here enables, in principle, to reduce the noise in
a transmission channel or in a quantum memory
to any desirable level, while using much less qubits than the analogous
error-correction schemes.

Classical error correction is based 
on redundant encoding which uses more than one bit (on average) to  
encode one bit.
The simplest scheme is the $1\rightarrow 3$ repetition code 
in which
each bit is repeated three times in the encoding, and a majority
vote is chosen for decoding.
This code corrects one error, thus if the 3-bit string suffers
from up to one error, the original bit is recovered with perfect fidelity.
In a realistic analysis one must also take into account the possibility
of more than one error.
For instance, if a single bit contains an error with probability $p$
(where we assume that $p$ is small), 
then $p_l={3 \choose l} p^l (1-p)^{3-l}$ 
is the probability of having 
exactly $l$ errors. Thus,
the probability of having 
an error in the recovered
bit is
$ P= p_2 + p_3 = 3 (1-p) p^2 + p^3 = 3 p^2 - 2 p^3 $. 
We call the probability $P$ of having an error in the recovered bit
the {\em remainder error}, and the fidelity is 
$1-P$.

The analogous quantum error correction~\cite{Shor1}
uses 9 qubits\footnote{
Other codes were found later on which use less qubits.} to encode a single qubit 
(to perfectly correct a single error) using the following procedure:
a $1\rightarrow 3$ repetition code 
in the $z$ basis
$  |0\rangle \rightarrow |000\rangle $ and 
$  |1\rangle \rightarrow |111\rangle $
(where $|000\rangle$ stands for the tensor product
$|0\rangle |0\rangle |0\rangle$
of three qubits);
a transformation to the $x$ basis
$|0\rangle \rightarrow(1/\sqrt 2)(|0\rangle + |1\rangle)$ and
$|1\rangle \rightarrow(1/\sqrt 2)(|0\rangle - |1\rangle)$
for each qubit;
and finally, again a $1\rightarrow 3$
repetition code in the (new) $z$ basis.
All together, the encoding is:
\begin{eqnarray}
|0\rangle &\rightarrow& 
\frac{1}{\sqrt8} (|000\rangle+|111\rangle)(|000\rangle+|111\rangle)
(|000\rangle+|111\rangle) \ ,
\nonumber \\
|1\rangle &\rightarrow& 
\frac{1}{\sqrt8} (|000\rangle-|111\rangle)(|000\rangle-|111\rangle)
(|000\rangle-|111\rangle) \ . 
\label{Shor-scheme}
\end{eqnarray}
We denote it as $R_3  U R_3$ where $R_n$ 
stands for $1\rightarrow n$ 
repetition code, and $U$ for transformation from the $z$ basis to the $x$ basis.
From now on, $N$ denotes the total number of code
qubits which encode one original qubit. 
In an $R_n U R_n$ code (which is the obvious generalization
of the code suggested by Shor)
$N=n^2$
qubits are used.

A qubit is described by a two-dimensional Hilbert space (e.g., spin of
a spin-half particle) and it generalizes the classical bit
to the quantum state 
$ | \psi\rangle = \alpha |0 \rangle + \beta | 1 \rangle $
with $|\alpha|^2 + |\beta|^2 =1$.
When it is encoded using $R_nUR_n$ we get the entangled state
\begin{equation} |\psi\rangle \rightarrow
 |\Psi_{_{RUR}}\rangle = \alpha |0_{_{RUR}}\rangle
+ \beta |1_{_{RUR}}\rangle \ . \end{equation}
This state is made of $N=n^2$ qubits, and lives
in a $2^{(n^2)}$ dimensional Hilbert space,
with
\begin{eqnarray}
|0_{_{RUR}}\rangle =
\left(\frac{1}{(\sqrt2)^n}\right)
(|0_1 \cdots 0_n\rangle +
                     |1_1 \cdots 1_n\rangle)
% \quad \quad \quad \nonumber \\
% (|0_1 \cdots 0_n\rangle +
%                      |1_1 \cdots 1_n\rangle)  
\quad \cdots\quad  
(| 0_1 \cdots 0_n\rangle +
                     |1_1 \cdots 1_n\rangle)  
                                                     \nonumber \\ 
|1_{_{RUR}}\rangle =
\left(\frac{1}{(\sqrt2)^n}\right)
(|0_1 \cdots 0_n\rangle -
                     |1_1 \cdots 1_n\rangle)
% \quad \quad \quad \nonumber \\
%(|0_1 \cdots 0_n\rangle -
%                     |1_1 \cdots 1_n\rangle)  
\quad \cdots\quad  
(| 0_1 \cdots 0_n\rangle -
                     |1_1 \cdots 1_n\rangle)  
                                                    \label{my-code} 
\end{eqnarray}
where there are $n$ multiplets of $n$ bits each.
In the decoding process, the disturbed state is projected on the
desirable 2-dimensional subspace spanned by the two states
$ |0_{_{RUR}}\rangle $ and
$ |1_{_{RUR}}\rangle $.

An error-correction code is not supposed to 
correct arbitrary types of errors.
In all the discussions on quantum error correction 
transformations of many particles together are dismissed since
in reality such errors are 
much smaller than errors in individual bits.
As a result, the interaction of the qubits can be written as a tensor 
product of the interactions of the individual bits
(each with an independent environment).
We adopt this assumption and refer to it as {\em independent disturbance}.

To simplify the analysis even further, usually an additional assumption
is made, 
that most of the bits are completely unaffected by the transformations,
while few bits are strongly affected (and suffer from a bit-flip,
a phase-flip or both\footnote{  
The analysis in~\cite{BDSW,EM} might have shown 
that resistance against such errors
promises resistance against any quantum error.
We are not fully convinced that this is indeed true.}).
In such a case, a code provides a perfect fidelity if it can correct
at least as many bits as were affected.
Unfortunately, this assumption is not natural, and it
contradicts the assumption 
of independent disturbance which says that 
errors occur independently in any of the qubits.  
We shall not assume that only some qubits are affected.
All bits are affected hence a code cannot
provide perfect fidelity; there is always a remainder 
error $P$ in the recovered qubit.

We can estimate 
the average error rate $p$ for a qubit transmitted through the channel,
and calculate the remainder error $P$ which results from it.
For a small $p$, an error-correction code usually yields a much smaller $P$.
However, such an approach is somewhat {\em classical}
and it is not clear 
that it provides a correct calculation for $P$.
To the best of our knowledge, it is not completely proven that the calculation 
of $P$ can be based upon $p$, since the qubits are not measured separately
in the QEC process. 

In the next subsection we shall try to remove this assumption (for a very 
simple case)
and follow an alternative approach 
which is fully {\em quantum}.
Instead of choosing $p$ as a parameter, we shall choose the 
``average'' transformation\footnote{ 
In order to yield a small $p$ the transformation must be close
to unity.}
which acts on each qubit.
It may even be more realistic to assume that some of the transformations 
are not weak, but we shall be satisfied 
with analyzing an average transformation.
Usually it is assumed that it suffices to consider 
the classical probability $p$ instead of the average transformation.

In this subsection we focus on analyzing the connection between 
$p$ and $P$ for $R_n$ codes, assuming that such calculation is relevant
to the quantum case of $R_n U R_n$ as well.

For communication purposes, we are mainly interested in 
the remainder error.
It can be reduced if we use longer codes (or more efficient codes)
which correct more errors.
Alternatively, for a given code, it is possible to reduce the remainder
error if we perform error reduction instead of error correction!
Let us explain the difference between them (in the classical case).
We previously presented the classical error-correction scheme $R_3$. 
This code improves the remainder error (to $3 p^2 - 2 p^3$) only on average:
it improves it much when no errors are found,
but it does not improve it when one error is found; 
in case we know that one error was identified and corrected
[which happens with probability $ p_2 + p_1 = 3 p^2 (1-p)+3 p (1-p)^2$]
the probability of having a remainder error is exactly $p_2/(p_1+p_2)=p$,  
and we gain 
no error reduction at all!
In case we throw away cases were an error is found, instead of
correcting it, the remainder error is $p_3/(p_0 + p_3)= p^3/(p^3 +(1-p)^3)$.
For a small $p$ the remainder error is reduced from 
$3 p^2 - 2 p^3$ to $p^3 + 3 p^4 + O(p^5)$.
The probability of success (the probability of not throwing the bit)
is $Q = p_0 + p_3 = 1 - 3p +O(p^2)$.
The possibility of using error reduction was already considered in~\cite{VGW},
who suggested the $n=2$ error reduction as an alternative to the $n=3$
error correction.
The $n=2$ (classical) error-reduction code 
provides a remainder error $P \approx  p^2 + 3p^3$ which is better than 
$3 p^2 - 2 p^3$ 
for a small $p$.

In any error-correction code, the remainder error increases with the 
number of corrected errors.
If we throw away cases where the scheme suffers from large number of errors
the average remainder error can be much improved.
Therefore, in our error-reduction code 
the majority vote is replaced by
a unanimous decision; in case of a disagreement 
the bit is thrown away.
The classical ($1\rightarrow n$) repetition code $R_n$ with $n=2 t + 1$
provides successful 
unanimous decision with probability $Q=(1-p)^n + p^n$, and the remainder
error in this case is $P=p^n/Q$.
For a small $p$ 
\begin{equation}
P\approx \left(\frac{p}{1-p}\right)^n \ ,
\end{equation}
and (for a large $n$) $Q$ is given by  
\begin{equation} Q \approx (1-p)^n \approx 1 - np + \frac{(np)^2}{2!}
-\frac{(np)^3}{3!} \ldots 
\ . \end{equation}
For a large $n$ but small $np$ we obtain $ Q \approx 1 - np$.

The same code can also be used to correct up to $t$ errors 
(since $n = 2t+1$),
but with a much higher (average) remainder error, which can be calculated
from the binomial expansion of $[ p + (1-p) ]^n$.
As in the previous example, if exactly 
$t$ errors are identified and corrected, the probability 
that there were actually $t+1$ errors 
(hence, there is a remainder error) is $p$.
If $np>1$ (or even $np \approx 1$) and $Q$ becomes too small,
error correction and error reduction can be combined together 
for the price of increasing the remainder error $P$.
This is done by using a code which can correct up to $t$ errors and use it
to correct $t'$ errors where $0 < t' < t$.
The case of $t' = 0$ reduces to an error-reduction code,
and the case of $t=t'$ reduces to an error-correction code.
For simplicity we shall consider only ``pure''  
error-reduction schemes,
but our calculations can be generalized to combine the correction of
few bits as well.

The quantum error-reduction scheme 
$R_2 U R_2$ (suggested in~\cite{VGW})
improves the remainder error compared to Shor's scheme, 
while using only 4 qubits instead of 9 for the encoding.
The error-reduction process is done by projecting the state of the code's
qubits on a desirable subspace; for instance, in case of the
$R_n U R_n$ quantum error-reduction code,
it is projected on the subspace spanned by the two states
of eq.~(\ref{my-code}).     
If the projection fails, the
qubit is not corrected but is thrown away.
Clearly, this does not complicate the implementation, and would
probably even simplify it.
Unless $Q$ is very small, throwing away the bits has 
only small influence on a quantum key distribution protocol
since the legitimate users throw away most of the bits anyhow
due to other reasons.
However, in quantum computing 
throwing one bit
destroys the computation.

In a scheme which combines error reduction and $t'$-errors error correction, 
one has to check into which subspace the state is projected,
and if this subspace corresponds to $t'$ errors (or less) the state 
is corrected by simple transformations (see~\cite{Shor1}).

Suppose that the average error rate in a quantum error-reduction scheme is $p$.
Following~\cite{Shor1,VGW} 
and other works on this 
subject we calculate the classical remainder error $P$
(from the analogous classical code) 
given that errors 
occur with probability $p$. 
Let us use the error-reduction code $R_n U R_n$ with
large $n$.
This code encodes one qubit into $N=n^2$ qubits.
It reduces the remainder error to 
\begin{equation} P=\frac{p^n}{p^n + (1-p)^n} \ , \end{equation} 
that is, to be
exponentially with $n$.
More efficient codes could be used as well,
based, for instance, on~\cite{Steane1,Steane2,CS}.

\subsection{A Realistic Calculation of the Remainder Error in a Simple Case}

To the best of our knowledge, 
a calculation of the remainder error which takes into account
the fact that each bit is subjected to a transformation
was done
only in~\cite{KL}.
It was done using {\em interaction operators}
and {\em recovery operators}, but only the case
where one of the operators is proportional to the identity
was considered, 
and it is not clear if it covers all possible transformations.

For the simple special case of the code
$R_2 U R_2$ 
we wish to calculate explicitly the remainder error $P$,
and the probability of success $Q$, to verify
that they are of the same order as $P$ and $Q$ of the classical error
reduction code $R_2$.
We calculate the remainder error without using the classical analogous 
code.
Let each qubit in the code 
be transformed arbitrarily
(but independently).
In general, the transformation is not unitary since an ancilla
(e.g., environment) might be involved.
However, for any given initial state
we can still consider unitary transformations
and add the effect of decoherence (non-unitary transformations) 
by averaging over several different 
unitary transformations with
appropriate probabilities. 
(We believe that a similar argument is provided
in~\cite{EM}). 

For instance, when the weak-measurement gate (eq.~\ref{measur-gate}) acts on 
a code qubit and an environment qubit 
in a state ${1 \choose 0}{1 \choose 0}$,
it yields the same 
RDM (for the code qubit)
as the average of the two transformations
\begin{equation}
\left(\begin{array}{cc} c_\gamma & s_\gamma \\
                        s_\gamma & c_\gamma
\end{array}\right)  \end{equation}
and 
\begin{equation}
\left(\begin{array}{cc} c_\gamma & - s_\gamma \\
                       - s_\gamma & c_\gamma
\end{array}\right) \ . \end{equation}

The use of unitary transformation provides the 
order of magnitude of the resulting 
$P$ and $Q$, for verifying that the frequently used
classical analog is not strictly wrong.

Let each qubit $j$ in the code  
be exposed to the most general
one-particle transformation
\begin{equation}
U_j = \left(\begin{array}{cc}
                \cos \theta_j    &  \sin \theta_j e^{ i \phi_j} \\
 - \sin \theta_j e^{ i \eta_j}  &  \cos \theta e^{ i(\phi_j + \eta_j)} 
\end{array}\right) 
\end{equation}
(up to an irrelevant overall phase),
where all angles are smaller than some small angle $\chi$,
so that $p \approx \chi^2$ if a single qubit is measured.

In the quantum error-reduction process $R_2 U R_2$
the original qubit is encoded into  
$|\Psi_{_{RUR}}\rangle = \alpha |0_{_{RUR}}\rangle + \beta |1_{_{RUR}}\rangle$
with 
$ |0_{_{RUR}}\rangle = (1/2)( 
|0000\rangle + |0011\rangle + |1100\rangle + |1111\rangle)$ and 
$ |1_{_{RUR}}\rangle = (1/2)( 
|0000\rangle - |0011\rangle - |1100\rangle + |1111\rangle)$.
The original state is 
in a $2^{(2^2)} = 16$ dimensional
Hilbert space, and can be written as   
\begin{equation} 
|\Psi_{_{RUR}}\rangle = \frac{1}{2} [ (\alpha + \beta) |0000\rangle +
(\alpha - \beta) |0011\rangle +
(\alpha - \beta) |1100\rangle +
(\alpha + \beta) |1111\rangle ] \ . \end{equation}
This state 
is transformed (due to  
$\langle U_1 U_2 U_3 U_4| 0011\rangle$, etc.) into:
\begin{equation}
\left(\begin{array}{c} 
\alpha +\beta \\
0 \\ 
0 \\ 
\alpha -\beta \\
0 \\ 
\cdot \\ 
\alpha -\beta \\
0 \\ 
0 \\ 
\alpha +\beta \\
\end{array}\right)   \rightarrow 
\left(\begin{array}{c} 
x_{0000} \\
x_{0001} \\ 
x_{0010} \\ 
x_{0011} \\ 
x_{0101} \\ 
\cdot \\ 
x_{1100} \\ 
x_{1101} \\ 
x_{1110} \\ 
x_{1111}  
\end{array}\right) \ .   \end{equation}
Using $c_\theta = \cos \theta$ and $s_\theta = \sin \theta$, the terms are  
$x_{0000} = (\alpha+\beta) c_{\theta_1} c_{\theta_2} c_{\theta_3} c_{\theta_4}
          + (\alpha-\beta) c_{\theta_1} c_{\theta_2} s_{\theta_3} s_{\theta_4}
                    e^{i \phi_3} e^{i \phi_4}
          + (\alpha-\beta) s_{\theta_1} s_{\theta_2} c_{\theta_3} c_{\theta_4}
                    e^{i \phi_1} e^{i \phi_2}
          + (\alpha+\beta) s_{\theta_1} s_{\theta_2} s_{\theta_3} s_{\theta_4}
                    e^{i \phi_1} e^{i \phi_2} e^{i \phi_3} e^{i \phi_4}  $, 
etc.
In the quantum error-reduction process the state is projected
onto the subspace spanned by 
$ |0_{_{RUR}}\rangle $ and
$ |1_{_{RUR}}\rangle $.

\noindent 
Let $C=
                 (\cos \theta_1 \cdots \cos \theta_4\  \cos \phi_1
                  \cdots \cos\phi_4\  \cos \eta_1 \cdots \cos \eta_4)$,
thus, $C \approx 1 - O(\chi^2)$.
After a lengthy calculation we obtain the (unnormalized)
final state
$| \Phi_{_{RUR}}\rangle = C
                \left[ {\alpha \choose \beta} + O(\chi^2) \right] $ .
This final state, when normalized, is almost identical to the initial state
$|\Psi_{_{RUR}}\rangle$,
and the corrections are of the type 
$\sin \theta_1 \sin \theta_2$; $\sin \eta_3 \sin \eta_4$, 
etc., all of order $O(\chi^2) $ or smaller.
Thus, the remainder error probability is indeed 
$O(\chi^4) \approx O(p^2) $, 
with probability of success $Q \approx C^2 \approx 1-O(\chi^2) \approx 1-O(p)$.

This simple example (which is already quite complicated) shows us that
the classical use of $p$ could be correct. Yet, a more general proof is
certainly needed.

\subsection{Application to Quantum Communication 
and \mbox{Memory}}\label{repeaters}

The main problem of a scheme which performs only error reduction is that, 
when $n$ is increased, 
the probability of successful projection diminishes 
as $(1-p)^n$.
We could combine it with some (small-$t'$) error correction
as previously explained, but there is also a different solution,
which should be preferable in case the noise changes in time
as
$\theta \approx wt$.
In this case, the probability of success can be much improved
using the Zeno effect
(see discussion in~\cite{VGW,ChuaYam}) by performing $M$ projections in between,
at equal time steps,
reducing $p$ to $p/(M^2)$, and $Q$ to $ (1-\frac{P}{M^2})^{nM}\approx
1-np/M$. 
The remainder error is also much improved by this process.
Performing $M$ projections is rather simple 
when enhancing a quantum memory is considered (meaning
that it does not add any further complication).
When transmission to long distances is considered, Alice and
Bob need to have {\em projection stations} between them.
We shall explain later in Section~\ref{QECforQPA} that 
(contrary to the common belief)
the usage of such repeater stations
is possible 
even when used for secure transmission of quantum information
(as in quantum key distribution).

Analysis of the effectiveness of this scheme when real devices are
used is much more complicated and (to our knowledge) is still missing.
It might be partially or completely solved if 
fault-tolerant quantum error
correction~\cite{Shor2} is used.

\section{Quantum Privacy Amplification}\label{QPA}

Quantum privacy amplification (QPA) is a scheme for privacy
amplification which uses quantum gates and memory.

Purification~\cite{purification} uses quantum gates and classical communication
to distill purified singlet states from 
non-pure singlet states. 
Suppose that the two non-pure singlets are in a tensor product and that 
Alice and Bob have perfect devices. 
Then, if the purification succeeds, 
the obtained singlet state have a better fidelity than 
the average fidelity of the two original state.
When a supply of such non-pure singlets is given, and 
this process is repeated many times it converges\footnote{
If the original singlets are not too far from being pure.} to provide
singlets that are arbitrarily close to being perfect.
Thus, purification can be used to provide Alice and Bob perfect
singlets under these assumptions.

Recently, a wonderful idea was raised in~\cite{oxford} 
to use purification as a QPA
for the EPR scheme (but it can be useful to any other scheme
due to teleportation).
This paper provides an (almost complete) proof of security of quantum 
key distribution under the assumption that Alice and Bob have perfect devices.
It also creates the impression that purification based QPA can provide
a proof for the ultimate security of quantum key distribution.
In this section we criticizes this possibility.
It is  
based on discussions with Charles Bennett, Eli Biham, Gilles Brassard
and John Smolin.

When Eve performs a collective attack, the original
pairs are in a tensor product with each other, even if entangled with 
Eve's probes.
In this case, and with the additional demand that 
Alice and Bob have perfect devices, this QPA provides an
easy proof of security.
[Yet, it does not provides a calculation of Eve's information as a function 
of the original average fidelity and the number
of singlets required (which might be exponentially large with some parameter
if not proven otherwise). 
Only an incomplete numerical analysis is provided].

Altogether, there are three assumptions which make the proof correct
and complete:
(1) Alice and Bob have exponentially large number of particles.
(2) Alice and Bob have perfect devices.
(3) Eve performs a collective attack (this results from 
eq. (9) in ~\cite{oxford}).
Here, we ignore the efficiency problem, although this problem  
must be solved if one wishes to use the scheme. 

Let us try to remove assumptions (2) and (3) to see
whether the purification based QPA can provide a proof of security
in principle.
We conclude that it might provide it, but the task might 
be much more difficult
than proving security using classical privacy amplification.

If Alice and Bob have perfect devices but Eve performs a 
(non-collective) joint attack
their proof is almost complete:
Suppose that Eve knows which pairs shall be used at each step
(including the pairs which shall be used for verification, but not the basis
used for each verification)
Eve could prepare a state
which yields an error only in certain qubits, since all the transformations
are unitary (this observation is due to Charles Bennett and John Smolin).
Indeed, in this case, it is quite clear that randomizing the pairs
solves the problem. However, 
an analysis of this attack and a proof of security against it are missing
from~\cite{oxford}, 
and it is not clear if changing the intuition into a proof is trivial.
Note that we use a similar randomization argument in
Chapter~\ref{secur}, where we provide an intuition that randomization 
makes a collective attack the strongest joint attack.

The proof in~\cite{oxford} considers noisy channels, 
but assumes non-realistic perfect devices.
The case of perfect devices is not of much interest unless the 
proof can be extended to real devices.
We analyze both the security against collective attacks and against
joint attacks.

If Eve performs a collective attack and Alice and Bob have real devices 
it is still possible to generalize the proof of~\cite{oxford}. 
On one hand, Alice and Bob cannot check 
whether the purification converges.
Perfect devices are required to enable 
Alice and Bob to derive any level of confidence that the remaining 
pairs are in the desired pure state, and quit the protocol if this level
is not reached.
On the other hand,
repeating the purification many times and randomizing the pairs
probably reduces the entanglement with Eve to zero
even if the states are not
purified. 
It is not clear how to prove security and
how to find Eve's information as a function
of the number of iterations in this case.

While assuming either perfect devices or collective attacks seems to cause
problems which are solvable in principle,
removing both assumptions might leave an unsolvable problem.

Let the RDM of one such pair be $\rho$, 
and the RDM of all the pairs together be $\chi$.
Also let the state of the entire system
(pairs and Eve's probes) be $\psi$.
With $\rho$ close enough to the desired singlet
state 
$| \phi_- \rangle \langle \phi_- |$, Alice and Bob cannot distinguish 
$\rho$ from the desired state
due to their non-perfect devices.
Let us present various attacks Eve could perform in such a case:
\begin{enumerate}
\item 
Eve can prepare a joint state $\chi$ for the pairs such 
that the purification does not converge due to 
her special choice of the entanglement of the pairs.
\item
Eve can prepare a different state $\chi$ for the pairs which is not changed
by randomization.
\item
Eve could prepare a state $\psi$ for the system and her probes, such
that she obtains information from it.
\end{enumerate}
Furthermore, consider the following attack:
\begin{itemize}
\item
Eve might be able to combine the above
three techniques to create a situation where
purification does not converge, randomization does not work, and 
yet she gets some information on some final bit, or collective property
of some bits.
\end{itemize}
We were not able to prove that such an attack is impossible.
Furthermore, it is not even clear how to find 
an intuitive argument for security in this case.

Thus the EP-based QPA only provides the possibility for 
a very restricted security proof, assuming either
perfect devices or the collective attack.
Note that improving their proof is an extremely difficult task when we recall 
that Eve has access to all information regarding the purification process,
thus can follow the changes in the state of her probes throughout the process.
Therefore, we believe that proving security using the much simpler
classical privacy amplification is an easier task.  

\section{Quantum Error Correction as Quantum Privacy Amplification}\label{QECforQPA}

In this section we suggest another QPA which is based on QEC. 
After reading the previous section,
one may wonder whether QPA (of any type) is still interesting. 
We believe it is, for reasons which are shown at the end of this section.

One of the effects of QPA is error correction.
This raises the question of whether QEC 
can serve as QPA.
Assuming Alice and Bob have perfect devices we suggest the following 
(simple) QEC-based QPA:
Alice prepares $M$ qubits that she wishes to send to Bob. 
She chooses the qubits according to the four-state (or the 
two-state) scheme of quantum cryptography.
She encodes each qubit 
using $N$ qubits following any QEC scheme
and sends the $MN$ qubits to Bob.
Bob receives the particles and keeps them in a quantum memory
till receiving all qubits which encode an original qubit. 
Since perfect devices are used, 
QEC can reduce a small error rate in a quantum memory
or in a quantum channel to any desired level due to analysis
of the remainder error (apart for accumulating many-bits error).

We can use projection stations to improve this QEC-based QPA much further,
due to the Zeno effect.
We call them quantum repeaters.
When such quantum repeaters are placed in appropriate predetermined
distances, they keep the error rate at low levels {\em without}
reducing security.
Note that, when real devices are used, 
the use
of fault-tolerant QEC~\cite{Shor2} might still enable reducing the errors
to any desired level (although the
use of projection stations becomes more subtle).

Let us now analyze the case where an eavesdropper attempts
to eavesdrop on the code qubits.
Eve can either get the qubits at some intermediate stage, or even control the 
projection stations used to improve the probability of successful
projection $Q$. 
Indeed, since 
each such station is only required to perform the desired projection,
it can even be controlled by the eavesdropper;
if Eve tries to do anything other than the required projections ---
she increases the error rate and will be detected.
Eve could even control the encoding and decoding process
of the original qubit.

The only assumption required for the success of any error-correction
or error-reduction scheme is that each code bit is disturbed independently 
of the others. 
If real noise causes many-particle transformations
the scheme will fail, but for bits stored or transmitted separately,
such effects are
expected to be negligible.
Thus, the legitimate users can use error-reduction
schemes to decrease the noise, 
and expect much less errors
when comparing a portion of the data. 
Eve is still permitted to do whatever she likes,
including creating many-particle coherence.
She could clearly benefit from performing a coherent attack in this case,
but she is restricted to induce a much smaller error rate on the original qubit.

If perfect devices are used, Alice and Bob can reduce the 
permitted error rate exponentially to zero.
In case of real devices a proof is still missing.
As previously said, analyzing this aspect of QPA 
is rather complicated and we believe that a security proof
based on classical privacy amplification is simpler.

If Eve deviates from the protocols and the error rate is larger than expected,
Alice and Bob quit the transmission.
If she deviates from the protocol but provides the final state with
the permitted error rate, Alice and Bob do not care which operations she has
done. 
Then, the permitted small error rate (which is verified), promises
them that her information is limited as desired.

The errors due to the frequent projections in a ``many-stations'' system
were not considered here. As in the case of a 
fault-tolerant calculation~\cite{Shor2}, it may well be that there is some
optimal number of stations $M$ such that a larger number of stations
causes an increase of the remainder error.
Note also that some errors are due to the creation and measurement of the 
code state
in the labs of Alice and Bob, and for the time being
these limit our ability to reduce $P$.
We believe that error correction by symmetrization is the right way
to handle these problems, but we are not aware of works 
which are showing this.
However, the main limitations on quantum cryptography are maintaining
coherence over long distances and long times 
and these limitations are solved efficiently
using the scheme we suggest.

Suppose now that 
Alice sends the $MN$ qubits in a random order 
(so that Eve will not know which $N$ qubits
encode each original qubit). 
In this case, our
QPA scheme probably becomes more secure:
Without having the knowledge regarding the random permutation  
we conjecture that she could benefit from a coherent eavesdropping. 
Thus, the randomization of the code qubits 
probably improves the QPA.
However, as for all other uses of the randomization argument there is
no proof for this intuition.

The EP-based and the QEC-based QPA processes 
do not assure security.
They only allow Alice and Bob to work with a much lower permitted error rate.
Even though we do not believe that QPA improves our ability
to prove security in a direct way, it might still help it in other
ways. 
a) It can be used to reduce the error rate very much. Such a reduction might be 
crucial for other security proofs such as the one we presented 
in Chapters~\ref{secur} and~\ref{col}, which relies
on various approximations of small angles.
b) It enables implementing quantum cryptography over large scale distances
and times 
since the final error rate  
depends only on the devices' errors and not on the 
channel/memory's error.
When we combine both arguments we see that QPA might be crucial for proving
security using classical privacy amplification when large scale
cryptography is performed, by reducing the errors to levels
which can be dealt with by the BM approach.
   
\chapter{Conclusions}\label{conc}

Quantum memory shall sooner or later become a common tool. In 
this thesis we provided the first thorough analysis of the impacts
of using quantum memories for both improving quantum cryptography and for
attacking it.
In the hands of an eavesdropper a quantum memory is a crucial tool:
without it classical privacy amplification provides a proven security of
quantum key distribution (although an explicit bound is still missing).
In this work we provided strong evidences that classical privacy amplification
still works when Eve holds a quantum memory, and we suggested a way
to establish a full proof.
In the hands of the legitimate users, a quantum memory provides 
new ways of establishing a network for many users.
It also provides the ability to improve large scale quantum key 
distribution.

\end{document}